\documentclass{aa}
\usepackage{graphicx}
\usepackage{natbib}
\usepackage{enumerate}
\usepackage{enumitem}
\usepackage{float}
\usepackage[section]{placeins}
\usepackage{txfonts}
\begin{document}
%
   \title{Oscillation mode linewidths and heights of 23 main-sequence stars observed by \textit{Kepler}}

   \author{T.~Appourchaux
          \inst{1}
          \and
          H.~M.~Antia\inst{2}
          \and
          O.~Benomar\inst{3,4}
          \and
          T.~L.~Campante\inst{5,9}
          \and
          G.~R.~Davies\inst{5,10}
          \and
          R.~Handberg\inst{5,9}
          \and
          R.~Howe\inst{5}
          \and
          C.~R\'egulo\inst{6,7}
            \and
          K.~Belkacem\inst{8}
          \and
          G.~Houdek\inst{9}
          \and
          R.~A.~Garc\'\i a\inst{10}
          \and
          W.~J.~Chaplin\inst{5}
          }

   \institute{Univ. Paris-Sud, Institut d'Astrophysique Spatiale, UMR 8617, CNRS, B\^atiment 121, 91405 Orsay Cedex, France
           \and
          Tata Institute of Fundamental Research, Homi Bhabha Road, Mumbai 400005, India
          \and
          Sydney Institute for Astronomy (SIfA), School of Physics, University of Sydney, New South Wales 2006, Australia
          \and
          Department of Astronomy, The University of Tokyo, 113-033, Japan
          \and
          School of Physics and Astronomy, University of Birmingham, Edgbaston, Birmingham B15 2TT, United Kingdom
          \and
          Instituto de Astrof\'isica de Canarias, 38205 La Laguna, Tenerife, Spain
          \and
          Universidad de La Laguna, Dpto. de Astrof\'isica, 38206 La Laguna, Tenerife, Spain
          \and
          LESIA, Observatoire de Paris, CNRS UMR 8109, UPMC, Universit\'e Denis Diderot, 5 place Jules Janssen, 92195, Meudon Cedex, France
                  \and
          Stellar Astrophysics Centre, Department of Physics and Astronomy, Aarhus University, DK-8000 Aarhus C, Denmark
                   \and
          Laboratoire AIM, CEA/DSM-CNRS-Universit\'e Paris Diderot, IRFU/SAp, Centre de Saclay, 91191 Gif-sur-Yvette Cedex, France
          }

   \date{Version 10.0. 17 March 2014. Accepted}

 
  \abstract
  {Solar-like oscillations have been observed by {{\it Kepler}} and CoRoT in many solar-type stars, thereby providing a way to probe the stars using asteroseismology.}
   {We provide the mode linewidths and mode heights of the oscillations of various stars as a function of frequency and of effective temperature.}
   {We used a time series of nearly two years of data for each star.  The 23 stars observed belong to the simple or F-like category.  The power spectra of the 23 main-sequence stars were analysed using both maximum likelihood estimators and Bayesian estimators, providing individual mode characteristics such as frequencies, linewidths, and mode heights.  We study the source of systematic errors in the mode linewidths and mode heights, and we present a way to correct these errors with respect to a common reference fit.}
   {Using the correction, we could explain all sources of systematic errors, which could be reduced to less than $\pm$15\% for mode linewidths and heights, and less than $\pm$5\% for amplitude, when compared to the reference fit.  The effect of a different estimated stellar background and a different estimated splitting will provide frequency-dependent systematic errors that might affect the comparison with theoretical mode linewidth and mode height, therefore affecting the understanding of the physical nature of these parameters.  All other sources of relative systematic errors are less dependent upon frequency.  We also provide the dependence of the so-called linewidth dip, in the middle of the observed frequency range, as a function of effective temperature.  We show that the depth of the dip decreases with increasing effective temperature.  The dependence of the dip on effective temperature may imply that the mixing length parameter $\alpha$ or the convective flux may increase with effective temperature.}
   {}

   \maketitle
%

\section{Introduction}
\label{sec:intro}

Stellar physics is undergoing a revolution thanks to the wealth of asteroseismic data that have been made available by space missions such as CoRoT \citep{Baglin2006} and \textit{Kepler} \citep{Gilliland2010}.
With the seismic analyses of these stars providing the frequencies of the stellar eigenmodes, asteroseismology is rapidly becoming a tool for understanding stellar physics.  
 
Solar-type stars have been 
observed over periods exceeding
six months using CoRoT and \textit{Kepler} providing many lists of mode frequencies required for seismic analysis \citep[See][and references therein]{Appourchaux2012a}.  Additional invaluable information about the evolution of stars is provided by the study of the internal structure of red giants  \citep{Bedding2011a, Beck2011, Beck2012, Mosser2012,Mosser2012a} and of sub giants  \citep{Deheuvels2012, Benomar2013}.  The large asteroseismic database of \textit{Kepler} allowed us to estimate the properties of an ensemble of solar-type stars that is large enough to perform statistical studies \citep{Chaplin2013a}.  

Solar-like oscillations are stochastically excited and damped by the convection.  Thus measurements of mode linewidths and mode heights provide information about how the stellar modes are excited and damped.  The processes involved are related to the generation of the acoustic noise and the dissipation of energy at the surface of the star \citep[See][]{GH99,Samadi2011}.  


For solar-like stars, a scaling relation for mode linewidth related to the stellar effective temperature has been proposed by \citet{Chaplin2009} using ground-based observations, \citet{Baudin2011} using CoRoT data and Appourchaux et al. (2012a) using \textit{Kepler} data. Those relations are based upon the linewidth measured at the frequency of maximum mode height and have been found by \citet{Belkacem2012} to be in qualitatively good agreement with the theoretical predictions. The relation was extended to lower effective temperature for red giants \citep{Corsaro2012}.

These previous scaling studies do not provide the frequency dependence of the linewidth.  Using a simple modelling approach, \citet{Gough1980} suggested that solar linewidths might have a local decrease, or dip, at the frequency of maximum power.  With more accurate and detailed modelling, \citet{Balmforth1992} showed that there is indeed such a linewidth dip for the Sun.  \citet{GH99} found that stellar mode linewidths show either a dip or {\it plateau} close to the maximum of mode height.  The plateau is located at the frequency of the maximum of the mode height as shown by \citet{Belkacem2011}, which is also
related to the Mach number (${\cal M}_a$), the ratio of convective velocity to the sound speed.  This dip was first observed but not acknowledged in the solar p-mode linewidths by \citet{Libbrecht1988}, while a small dip or plateau was indeed observed by \citet{Chaplin1997}.    The dip is caused by a resonance between the thermal adjustment time of the superadiabatic boundary layer and the mode frequency \citep{Balmforth1992}.  Since the thermal adjustment time is proportional to the acoustic cut-off frequency $\nu_{c}$, the two frequencies follow a scaling relation as shown by \citet{Belkacem2011}.   \citet{Claus1997} observed a very pronounced dip during solar minimum, hypothesizing that the depth of the dip may be modulated by solar activity, 
as confirmed later by \citet{Komm2000}.  These variations of the solar mode damping with solar activity were thought to be related to the change of the solar granule properties with the increasing magnetic field \citep{Houdek2001, Muller2007}, but these changes were not confirmed using space-based data \citep{Muller2011}.  The variations are likely to be affected by the change in the global magnetic field during a solar cycle.  Very recently, \citet{Benomar2013} studied the frequency dependence of mode linewidth of 4 sub-giant stars having mixed modes.  They found that the linewidth of $l=0$ modes showed a clear dip at the location of the maximum of mode power.

 It was shown by \citet{Appourchaux2012} that different fits of the same data could provide significantly different results for stellar linewidths.  Understanding the source of systematic errors  will result in a better understanding of how physics operate in stars.  \citet{Appourchaux2012} provided some insight on the various sources of systematic errors  related to stellar background estimation and the mode height ratio.   \citet{Chaplin2008a} showed that biased linewidths are also obtained when measuring mode linewidth of the order of 1 to 7 times larger than the frequency resolution.   Apart from these two papers, the understanding of the origin of systematic errors on  mode linewidth and height has not been widely studied.


This paper aims at providing the frequency dependence of mode linewidth $\Gamma$, mode height $H$ and mode amplitude $A$ ($A=\sqrt{\pi H \Gamma /2}$) for 23 \textit{Kepler} main-sequence observed for nearly 2 years by \textit{Kepler}, as well as an understanding of the source of systematic errors  affecting these parameters.  The paper also aims at providing the dependence of the linewidth dip as a function of effective temperature.

Section 2 describes how the time series and power spectra were obtained.  Section 3 describes the peak fitting procedure.  Section 4 details the sources of systematic errors on the mode linewidth and mode heights provided by the fitters.  Section 5 provides a procedure for correcting the systematic errors with respect to a reference fit.    We then discuss the detection of the dip as a function of effective temperature and the implications for stellar physics. The paper includes two examples of mode linewidth and mode height and an example of systematic error correction, while tables of the parameters of the 23 stars and correction for 22 stars are available online.  




\section{Time series and power spectra}
\textit{Kepler} observations are obtained in two different operating modes: long cadence (LC) and short cadence (SC) \citep{Gilliland2010,Jenkins2010}.  This work is based on SC data. For the brightest stars (down to \textit{Kepler} magnitude $Kp \approx12$), SC observations can be obtained for a limited number of stars  (up to 512 at any given time)  with a faster sampling cadence of 58.84876~s (Nyquist frequency of $\sim$ 8.5~mHz), which permits a more precise exoplanet transit timing and improves the performance of asteroseismology.  \textit{Kepler} observations are divided into three-month-long {\it quarters} (Q).  A subset of 23 stars from the 61 stars analyzed by \citet{Appourchaux2012a} has been used in the present analysis.  The subset of stars, observed during quarters Q5 to Q12 (March  22, 2010 to March 22, 2012), were chosen because they have oscillation modes spanning more than 10 radial orders. Therefore, the longest length of data gives a frequency resolution of about 16~nHz.  The stars used in this study are listed in Table~\ref{table1}.

To maximise the signal-to-noise ratio for asteroseismology, the time series were corrected for outliers, occasional jumps, and drifts \citep[see][]{RAG2011}, and the mean level for each quarter was normalised.  Finally, the resulting light curves were high-pass filtered using a triangular smoothing with full-width-at-half-maximum of one day, to minimise the effects of the long-period instrumental drifts.   The amount of data missing from the time series ranges from 3\% to 7\%, depending on the star.  All of the power spectra were produced by one of the co-authors using the Lomb-Scargle periodogram \citep{Scargle82}, properly calibrated to comply with
Parseval's theorem \citep[see][]{Appourchaux2011}.  All power spectra are single sided.





\section{Mode parameter extraction}
\subsection{Power spectrum model}


The mode-parameter extraction was performed by eight teams of fitters whose leaders are listed in Table~\ref{methods}.  The power spectra were modelled over a frequency range typically covering 10 to 20 large separations ($\Delta\nu$) between successive radial orders.  The stellar background was modelled using a multi-component Harvey model \citep{Harvey85}, each component with up to three parameters,  and a white noise component.  The Harvey model is a modified Lorentzian profile with a different exponent.  The number of components for the stellar background is given in Table~\ref{methods}.  The stellar background was fitted prior to the extraction of the mode parameters and then held at a fixed value. For each radial order, the model parameters were mode frequencies (one for each degree, $l=0,1,2$), a single mode height (with assumed ratios between degrees given in Table~\ref{methods2}), and a single mode linewidth for all degrees; a total of 5 parameters per order.  Since we have no stars with mixed modes, we choose to set a common mode linewidth and mode height to reduce the number of fitted parameters.  Other choices could be implemented depending on hypotheses.  The relative heights $h_{(l,m)}$ (where $m$ is the azimuthal order) of the rotationally split components of the modes depend on the stellar inclination angle, as given by \citet{Gizon2003}.  For 
each star, the rotational splitting and stellar inclination angle were chosen to be common for all of the 
modes; it adds 2 additional free parameters.
The mode profile was assumed to be Lorentzian.  In total, the number of free parameters for 15 orders was at least $5 \times 15+2=77$.  The potential misidentification of the pair $l=0-2$ for the $l=1-3$ pair as in by \citet{Appourchaux2008} was avoided by having the fitters using the same correct identification.

The model described above was used to fit the parameters of the 23 stars using maximum likelihood estimators (MLE) and with a Bayesian approach.  For the MLE, formal uncertainties in each parameter were derived from the inverse of the Hessian matrix \citep[for more details on MLE, significance, and formal errors, see][]{Appourchaux2011}.  For the Bayesian approach, the uncertainties (or credible intervals) were derived from the marginal posterior distribution of each parameter \citep[for more details on credible intervals, see][]{Benomar2009,Handberg2011}. 


Tables~\ref{methods} and~\ref{methods2} provides a summary of the model used by each fitter team.

\subsection{Guess parameter and fitting procedures}
The procedure for the initial guess of the parameters is described in \citet{Appourchaux2012a}, in which the steps of the fitting procedure are also described. These steps are repeated here for completeness:
\begin{enumerate}
\item We fit the power spectrum as the sum of a stellar background made up of a combination of modified Lorentzian profiles (one or two) and white noise, as well as a Gaussian oscillation mode envelope with three parameters (the frequency of the maximum mode power, the maximum power, and the width of the mode power).  Some fitters chose to exclude the p-mode region and fit the stellar background alone.
\item We fit the power spectrum with $n$ orders using the mode profile model described above, with no splitting and the stellar background fixed as determined in step 1
\item We follow step 2 but leave the rotational splitting and the stellar inclination angle as free parameters, and then apply a likelihood ratio test to assess the significance of the fitted splitting and inclination angle.  For the Bayesian fit, step 2 {\ is by-passed then step 3 is directly} applied; {\ this is because} credible intervals are always provided for any parameter, hence the significance test is only applied by MLE fitters.
\end{enumerate}
The steps above were sometimes varied slightly depending on the assumptions that were made.  For instance, the mode height ratio could instead be defined as a free parameter to study the impact of its variations on the derivation of the mode linewidth and mode height.

\section{Sources of mode-linewidth and mode-height systematic errors and their correction}
\begin{figure*}[htbp]
\centering
\includegraphics[width=10.75 cm,angle=90]{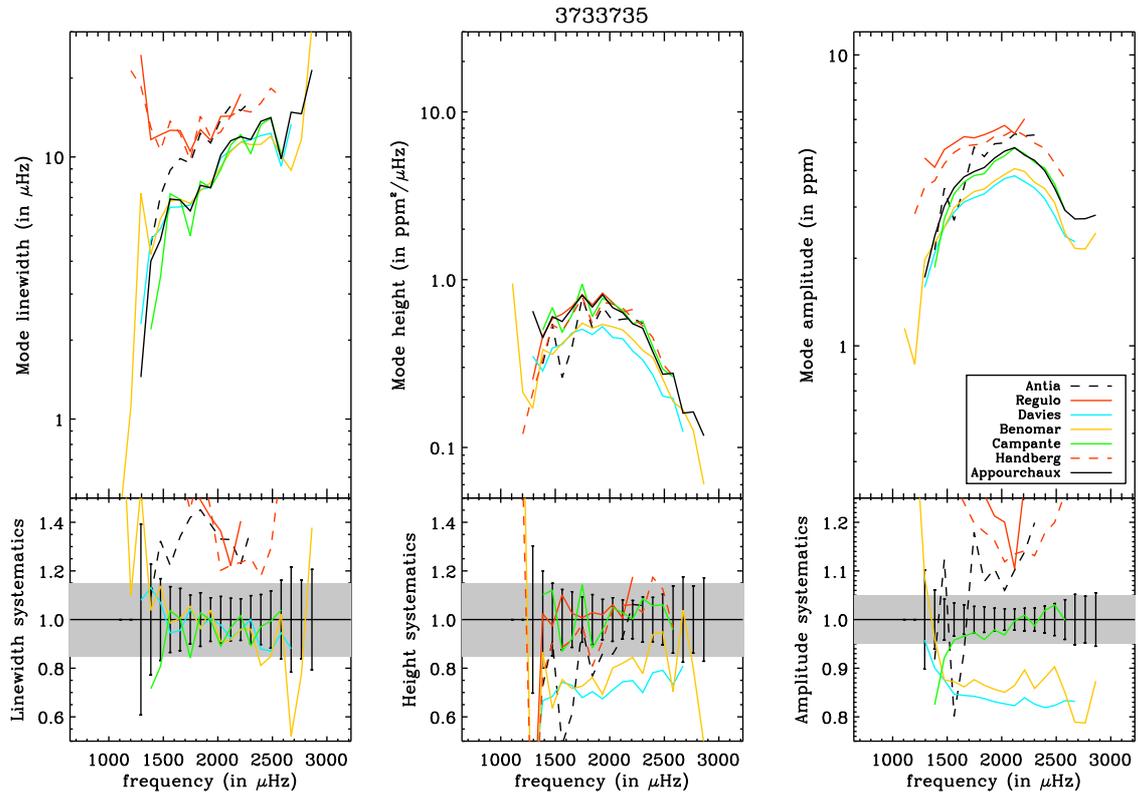}
\caption{Mode linewidth, mode height and mode amplitude (Top) and relative values of these parameters with respect to the reference fit (Bottom) as a function of mode frequency for various fitters for KIC 3733735.  The grey band indicates the range of systematic error around the reference fit values of $\pm$ 15\% for mode linewidth and mode height, of $\pm$ 5\% for mode amplitude. The 1-$\sigma$ error bars are those of Appourchaux.}
\label{fig_3733735}
\end{figure*}

When comparing the results that we obtained, it was clear that there are large differences in the mode linewidth and mode height measured by the different fitters.  Figure~\ref{fig_3733735} provides a typical example of large discrepancies between the fitters.
The different sources of bias leading to the systematic errors observed are listed in Appendix A.  The understanding of the different sources of bias provides a way to correct the systematic errors with respect to a {\it reference} fit.  In short, the main sources of systematic error in the measurement of the mode linewidth and mode height are: 1) splitting bias and 2) stellar background bias.  There are additional sources of systematic errors such as a different mode-height ratio and a different mean frequency which have been solved by assuming that the fitters use the same mode height ratio and mean frequency (see Appendix A for a full description of the various contributors).  

In order to check that all of the sources of systematic errors encountered when comparing the different results can be understood and compensated for, we have implemented a correction scheme based upon the one-fit correction of \citet{TT2005b}.  The procedure that they devised is based on fitting a spectrum model with no realization noise (i.e. the limit spectrum) with a different spectrum model.  The assumption behind that procedure is that the systematic errors are the same with the one-fit approach compared to a full Monte-Carlo simulation of the systematic errors resulting from using two different spectrum models.  The correction scheme was applied for comparing with the results of the reference fit, which were provided by Appourchaux.  The reference fit is simply used as a basis for understanding the source of the systematic errors.  The reference fit value is in no way the {\it best} set but simply a data set that can provide a basis for understanding the impact of using a different fitting model due to different theoretical or observational constraints.  The correction scheme is detailed in Appendix A.


\begin{figure*}[htbp]
\centering
\includegraphics[width=10.75 cm,angle=90]{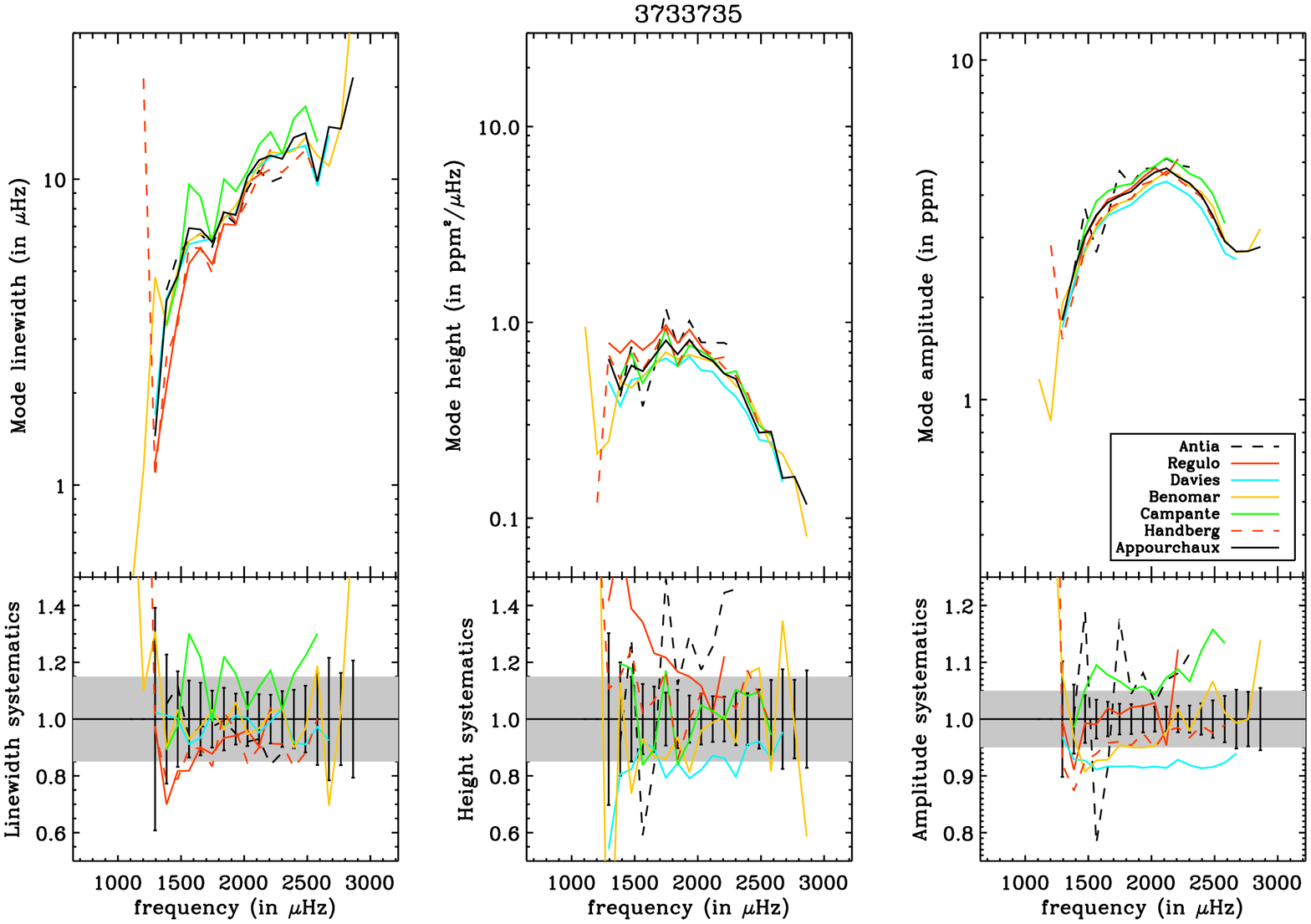}
\caption{Corrected mode linewidth, mode height and mode amplitude (Top) and relative values of these parameters with respect to the reference fit (Bottom) as a function of mode frequency for KIC 3733735. The grey band indicates the range of systematic error around the reference fit values of $\pm$ 15\% for mode linewidth and mode height, of $\pm$ 5\% for mode amplitude.  The error bars are those of the reference fit.}
\label{bias_comparison_3733735_new}
\end{figure*}





The correction has been implemented for all fitters who provided the information on the stellar background, the splitting and inclination angle and compared to a reference fit.  Figure~\ref{bias_comparison_3733735_new} shows the correction results for the largest discrepancies already shown in Fig.~\ref{fig_3733735}.  Figures B.1 to B.22 of the online material provide details of how the correction operates for all other stars on the mode linewidth and mode height.  For 17 out of 23 stars, the correction scheme provides mode linewidths, mode heights and mode amplitudes agreeing with the reference fit within $\pm$15\% for the first two parameters, and within $\pm$5\% for the latter.  As we showed above, the systematic error on the mode amplitude is bound to be smaller than for the mode linewidth or mode height alone.  The correction is really effective especially when there are large discrepancies (Antia, R\'egulo) with respect to the reference fit for which the different stellar background is the major source, with in addition a different splitting in some cases.  The remaining discrepancies (not larger than $\pm$10\%) mainly occur for the amplitude, especially when the mode height ratio used in the correction is far from the one used by the fitter.  This is the case for KIC1435467, KIC3733735, KIC3424541 and KIC12258514 for some fitters (Benomar, Davies, R\'egulo) for which even the modified ratios of 1.0 / 2.0 / 1.0 / 0.0 were underestimating their fitted mode height ratio.  Similar discrepancies (not larger than +30\%) occur for mode linewidth when using a far-from-nominal mode height ratio for KIC2837475 and KIC7103006 (R\'egulo).  In the latter case, the nominal mode height ratios of 1.0 / 1.5 / 0.5 / 0.05 overestimate the fitted mode height ratio.  For Antia's results, there are also large discrepancies for 13 stars that are traced back to a mode height ratio varying largely from mode to mode and to the use of flat stellar background.  In that latter case, neither a varying mode height ratio nor a flat stellar background are realistic assumptions regarding the fitting model.

Given the success encountered with the {\it simple} correction scheme, we chose not to make the correction model more detailed.  Therefore, we assume that the reference fit values given in this paper are corrected for known systematic errors.


\section{Results and discussion}
\begin{figure*}[htbp]
\centering
\includegraphics[width=9 cm,angle=90]{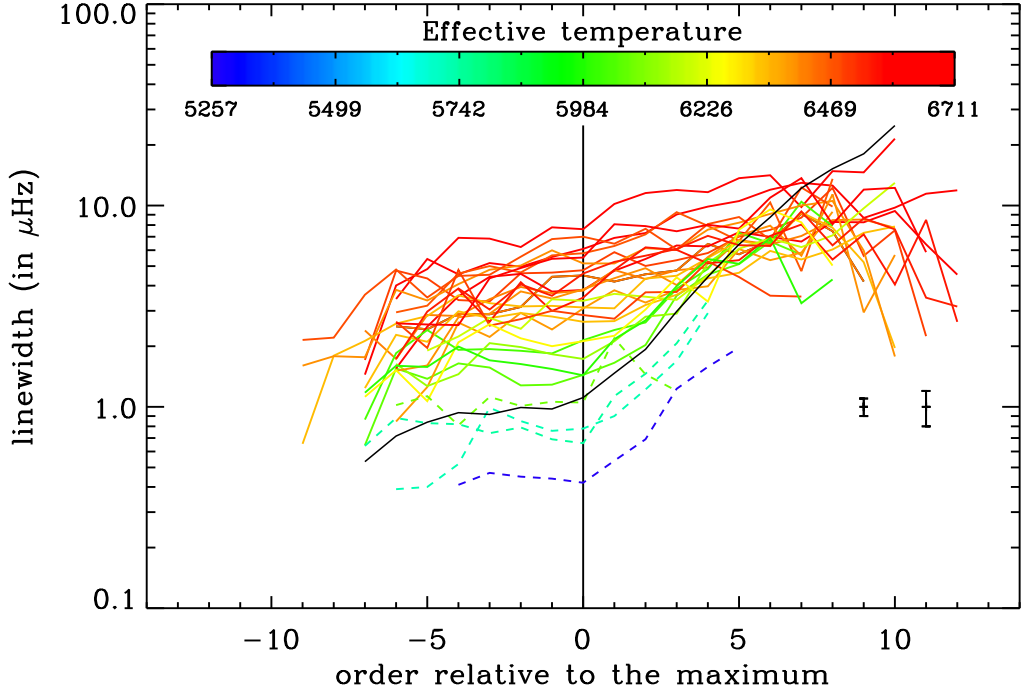}
\caption{Mode linewidth as a function of the mode order for the 23 stars with their colour identified by their effective temperature, for the $l=0$ mode linewidths of 4 sub-giant stars of \citet{Benomar2013} (dashed lines) and for the solar LOI data (solid black line).  For the 23 stars of this study the relative 1-$\sigma$ error bars range from 5\% to 10\%, while those of \citet{Benomar2013} range from 10\% to 20\%.  The 10\% and 20\% error bars are shown as an example.  The longer LOI data set has relative error bars of about 2\%.} 
\label{modeh}
\end{figure*}

\begin{figure*}[htbp]
\centering
\hbox{
\hspace{+1.2 cm}
\includegraphics[width=5.5 cm,angle=90]{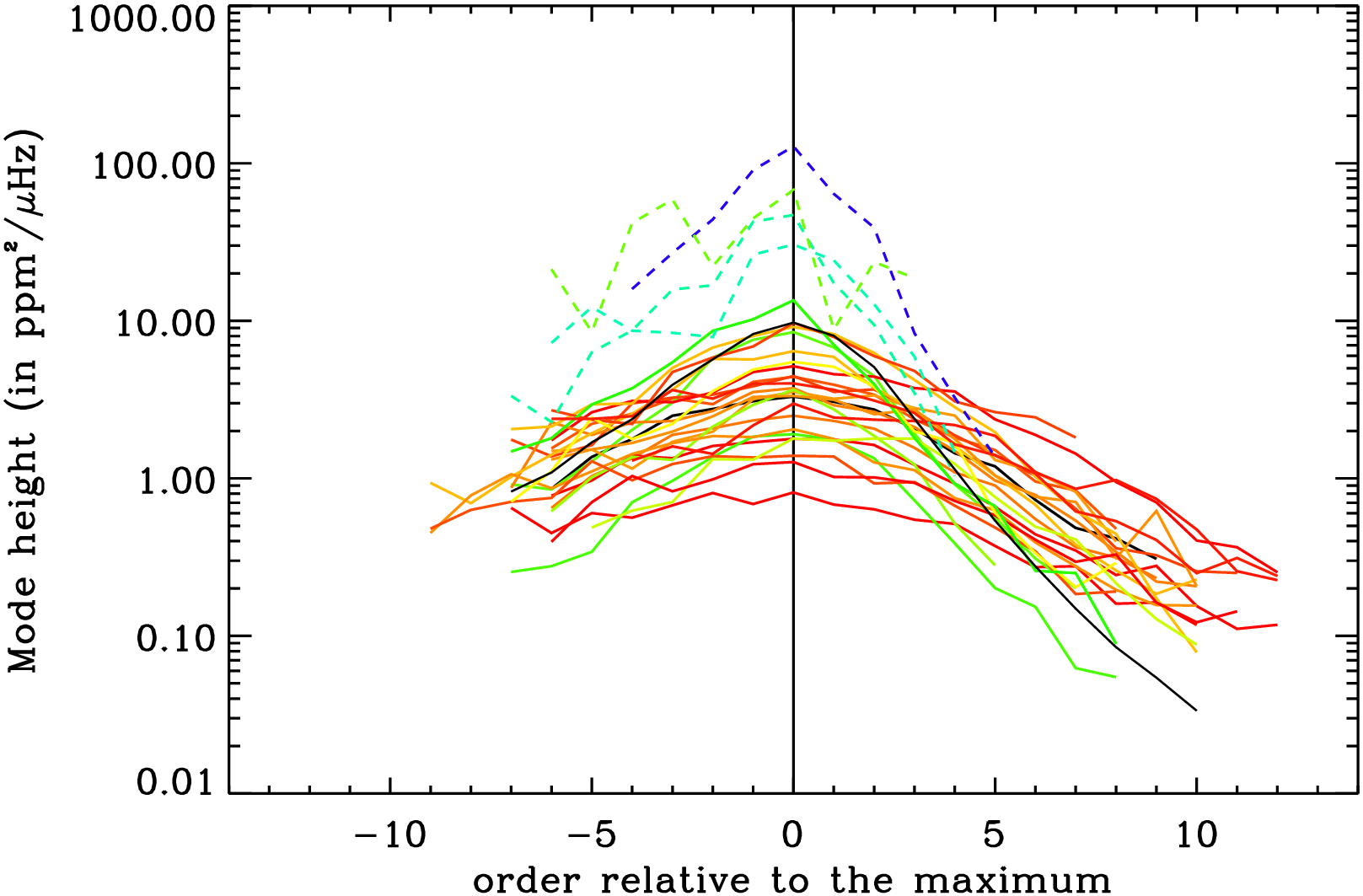}
\hspace{+1.2 cm}
\includegraphics[width=5.5 cm,angle=90]{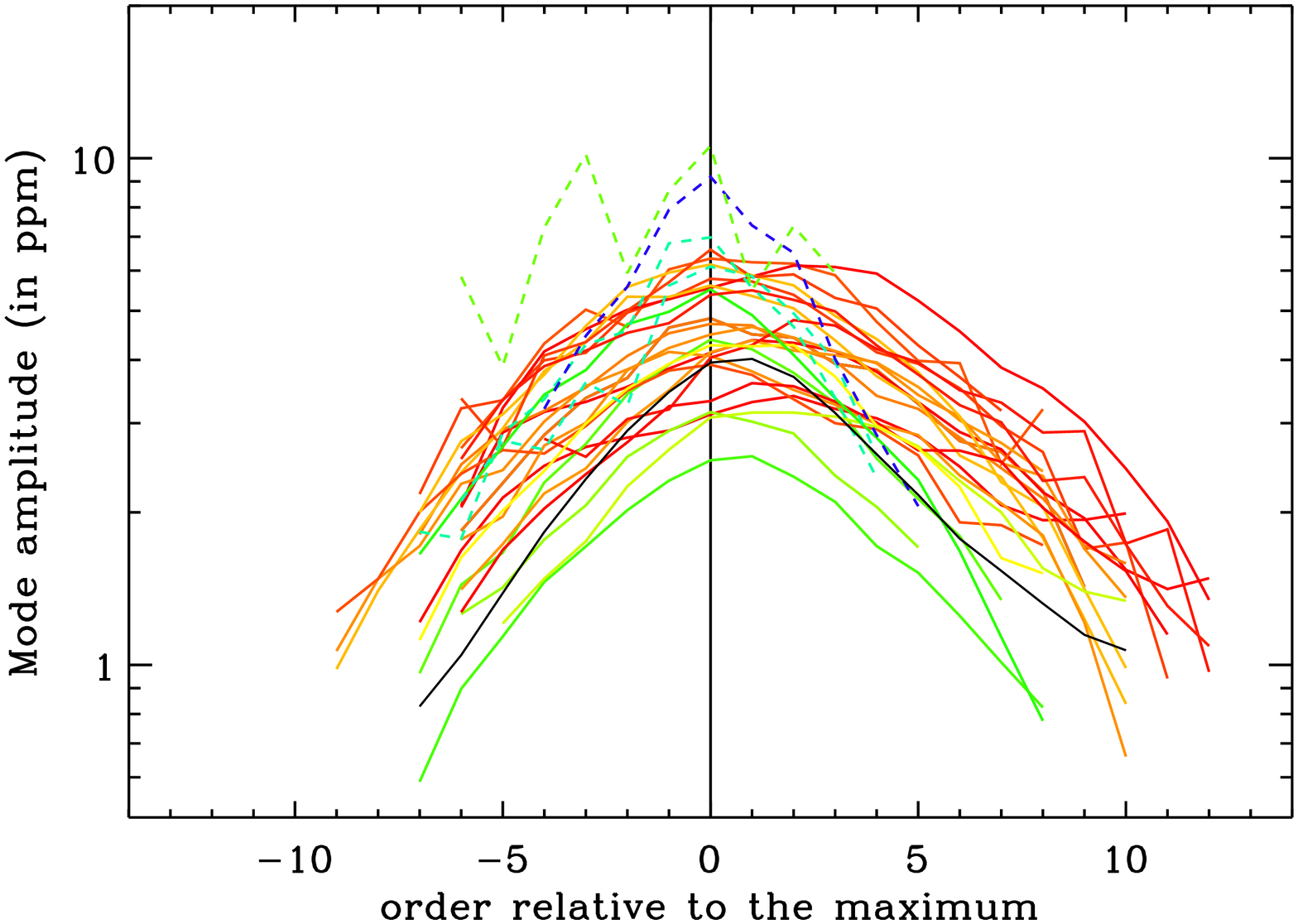}
}
\caption{Mode height and amplitude as a function of the mode order for the 23 stars with their color identified by their effective temperature (as in Fig. \ref{modeh}), for the 4 sub-giant stars of \citet{Benomar2013} (dashed lines) and for the solar LOI data (solid black line).  The relative 1-$\sigma$ error bars for mode height are similar to those of the mode linewidth, while they are typically 3 times smaller for mode amplitude.}
\label{modeh1}
\end{figure*}

Tables 4 to 26 provide the mode linewidth, mode height and mode amplitude provided by the reference fit for all 23 stars of this study and the associated error bars.  Figure~\ref{modeh} shows the dependence of the mode linewidth for the 23 stars of this study together with the $l=0$ mode linewidths of 4 sub-giant stars of \citet{Benomar2013} and that of the Sun measured by the Luminosity Oscillations Imager (LOI) of VIRGO \citep{CFJR95,TABA97}.  We computed the mean power spectra of 16 years of full-disk time series of the LOI; the data were then analysed in a similar manner as for the 23 stars but using a dual modified Harvey profile for the background.   For the data of \citet{Benomar2013}, we divided the published amplitude by $\sqrt{2}$ because the values given in their paper  were incorrectly quoted as $A=\sqrt{\pi \Gamma H}$ (for a double-sided power spectrum), instead of $A=\sqrt{\pi \Gamma H/2}$ (for a single-sided power spectrum as here); the mode height is unaffected by the correction since the former fit was performed on a single-sided spectrum.  The figures are centred with respect to the frequency of the maximum of mode height.  Figure~\ref{modeh} shows clearly that the linewidth dip disappears as the effective temperature increases.  Figure~\ref{modeh1} shows also that the maximum mode height and amplitude decrease with the effective temperature, as already shown by \citet{HK1995}.  It is also interesting to note that the location of maximum mode height also corresponds to the location of the linewidth dip as anticipated by \citet{Belkacem2011}, a location which does not coincide with the frequency of maximum amplitude, which is about half a large separation higher than the frequency of maximum mode height.


\begin{figure}[htbp]
\centering
\includegraphics[width=6.0 cm,angle=90]{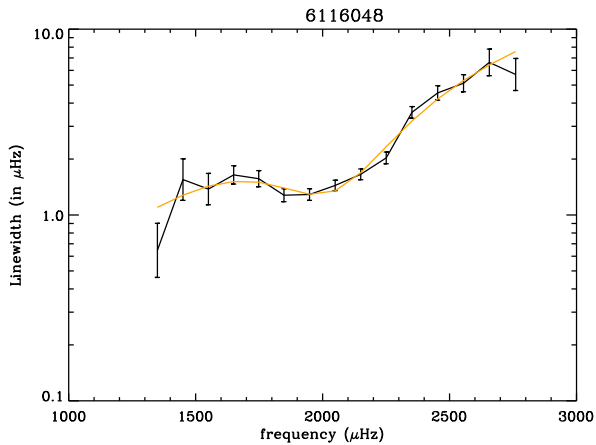}
\caption{Mode linewidth as a function of frequency for KIC 6116048 (solid line) with its 1-$\sigma$ error bars compared with the fit following Eq.~(\ref{Eq_fit}) (Orange line)}
\label{fit}
\end{figure}

\begin{figure*}[htbp]
\centering
\hbox{
\includegraphics[width=7 cm,angle=90]{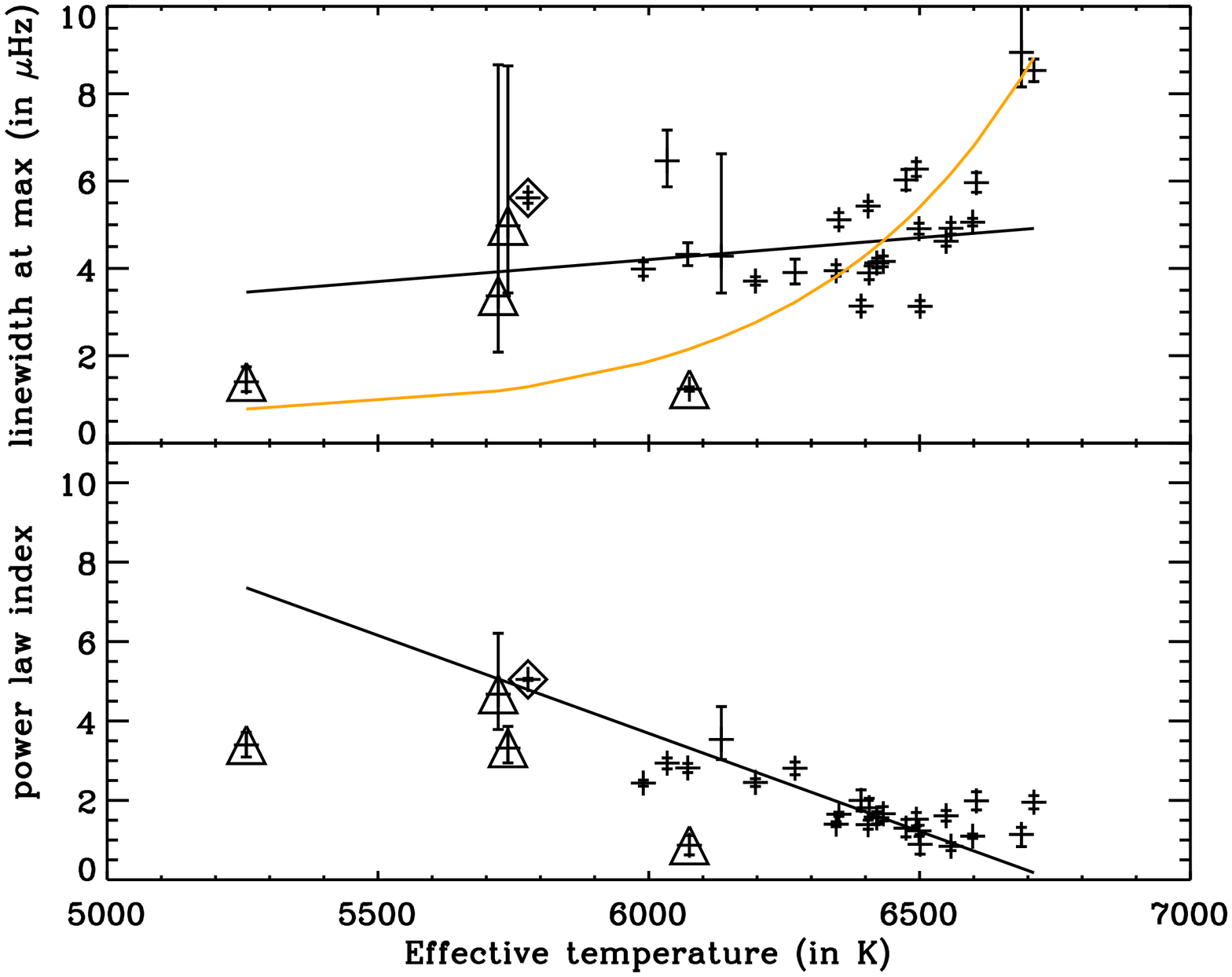}
\includegraphics[width=7 cm,angle=90]{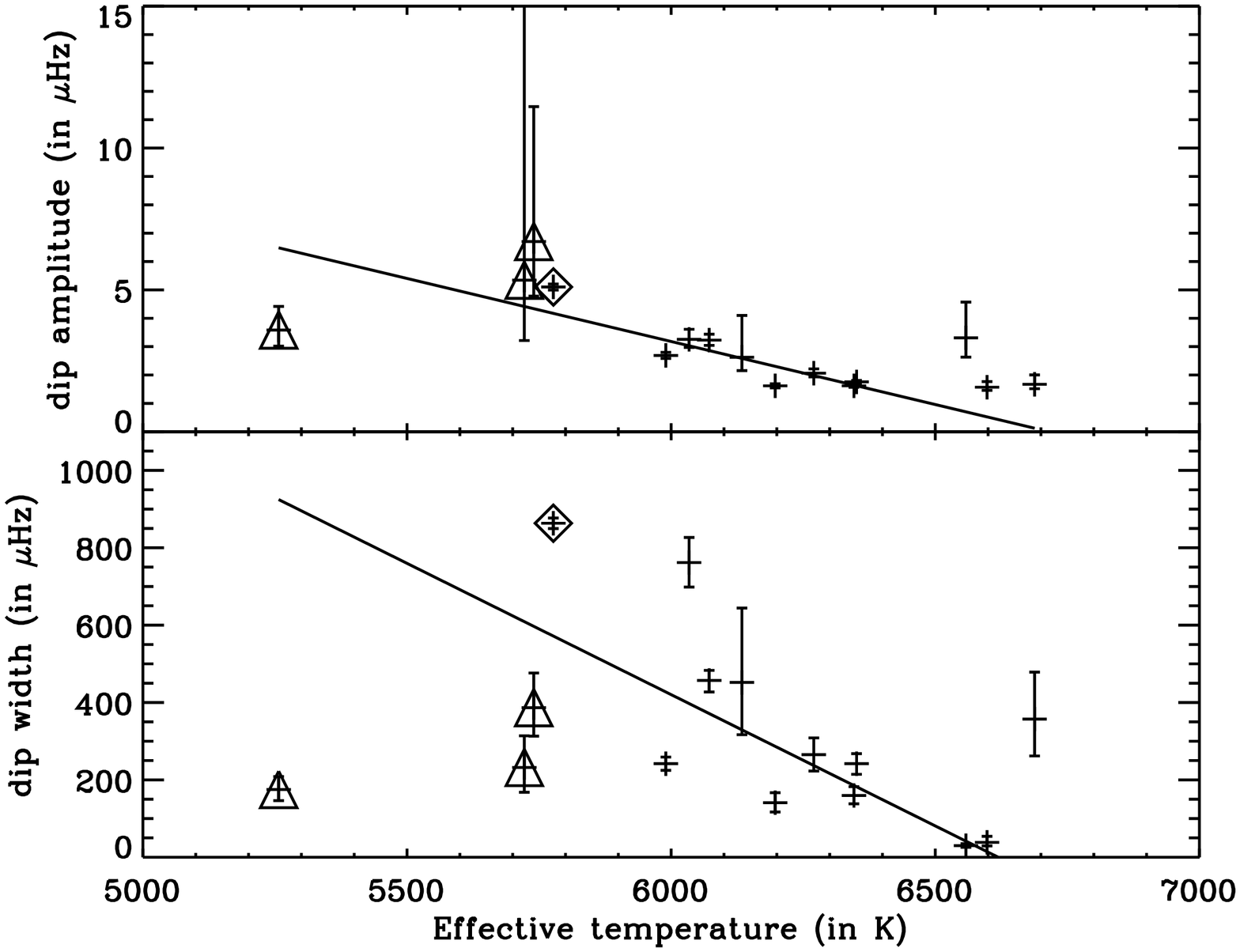}
}
\caption{Parameters of Eq.~(\ref{Eq_fit}) as a function of the effective temperature for all 28 stars, for the power law dependence (Left) and for the Lorentzian fit (Right).  The median value together with the credible intervals at 33\% and 66\% were derived from a Monte-Carlo simulation of the fit.  
The orange line shows the temperature dependence of the linewidth at the frequency of maximum mode height as derived by \citet{Appourchaux2012}.  The open diamond is the result of the fit for the solar data of the LOI.  The open triangles are the results of the fit for the sub-giant stars of \citet{Benomar2013}.  The solid lines show a linear fit of the parameters with respect to the effective temperature.  The Lorentzian parameters for stars for which the Lorentzian fit is not significant are not plotted.}
\label{dip}
\end{figure*}

In order to understand the dependence of the linewidth dip on effective temperature, we modelled the frequency dependence of the mode linewidth $\Gamma$ as a combination of the power law dependence and the dip modelled as a Lorentzian profile in $\ln \nu$:
\begin{equation}
\ln(\Gamma)=(\alpha \ln(\nu/\nu_{\max})+\ln \Gamma_{\alpha})+\left(\frac{\ln \Delta \Gamma_{\rm dip}}{1+\left(\frac{2\ln(\nu/\nu_{\rm dip})}{\ln (W_{\rm dip}/\nu_{\rm max})}\right)^2}\right)
\label{Eq_fit}
\end{equation}
where $\nu$ is the mode frequency, $\nu_{\rm max}$ is the frequency of maximum mode height, $\alpha$ is the power law index, $\Gamma_{\alpha}$ is the factor of the power law, $\Delta \Gamma_{\rm dip}$ is the depth of the dip, $W_{\rm dip}$ is the width of the dip and $\nu_{\rm dip}$ is the frequency of the dip.  Figure~\ref{fit} shows an example of a fitted linewidth using Eq.~(\ref{Eq_fit}), clearly showing a dip.  Figure~\ref{dip} shows the result of the fit of the linewidth for all the stars of Fig. \ref{modeh} and their values are given in Table~\ref{table_last}.  The left-hand panels of Fig.~\ref{dip} show the fit of the power-law only, for all 28 stars.   We checked whether the additional parameters of the Lorentzian profile of Eq.~(\ref{Eq_fit}) were significant by examining the decrease of the reduced $\chi^2$ with respect to a $\chi^2$ law with 3 degrees of freedom; the probability cut was $\frac{1}{23}$ ensuring that on average only one fit would be due to noise.  In total there were 11 out of the 23 stars for which the 5-parameter fit was significant.  In order to check the gaussianity of the probability distribution of the parameters, we used Monte-Carlo simulations of the fit for returning the median value and its associated credible intervals.  The non-gaussianity of the distribution occurs only for two stars but the Monte-Carlo scheme is applied to all stars for coherence.  The right-hand panels of Fig.~\ref{dip} show the Lorentzian fit of the dip for 15 stars for which the dip was significant.

\begin{figure*}[htbp]
\centering
\hbox{
\includegraphics[width=7 cm,angle=90]{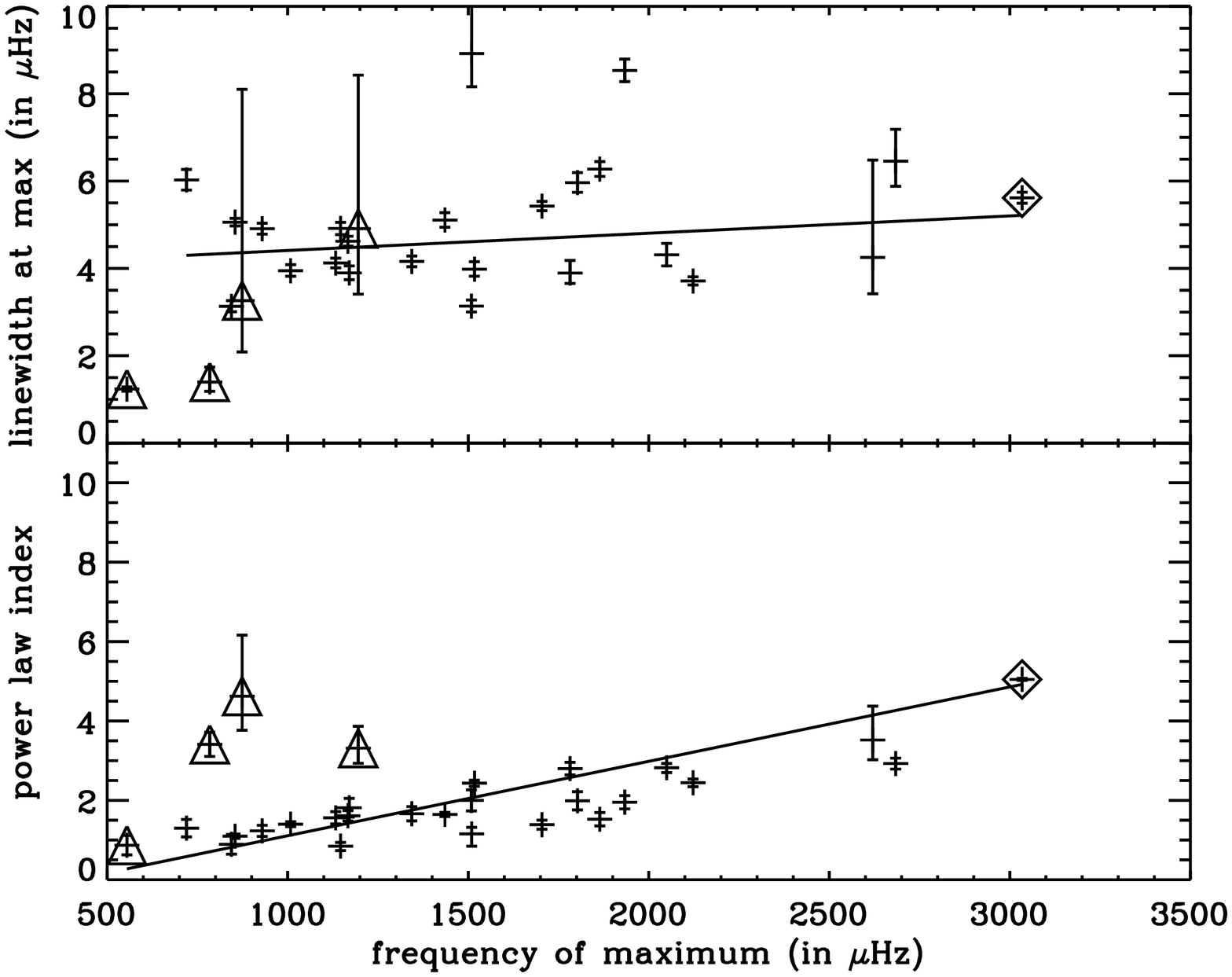}
\includegraphics[width=7 cm,angle=90]{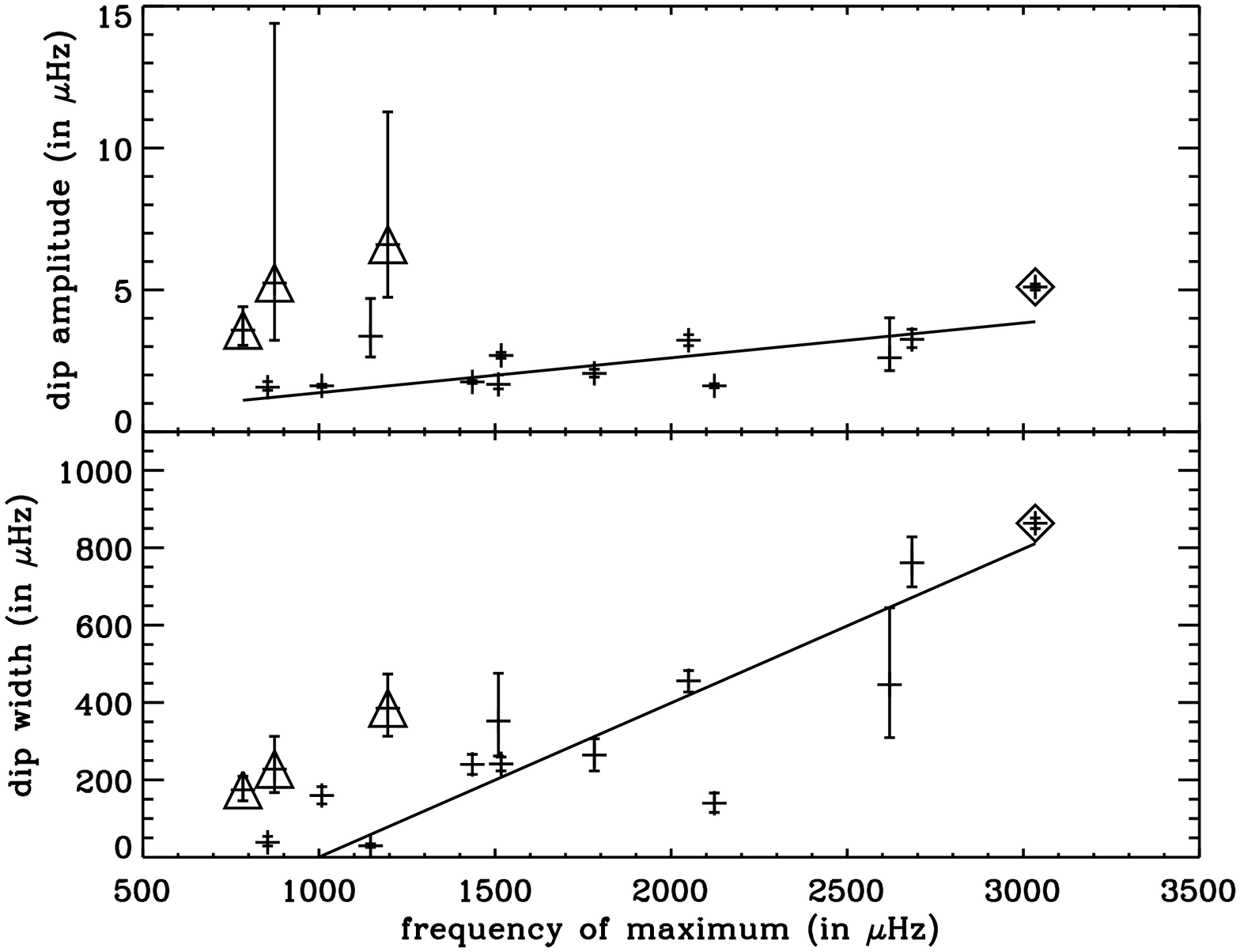}
}
\caption{Parameters of Eq.~(\ref{Eq_fit}) as a function of the frequency of maximum mode height for all 28 stars, for the power law dependence (Left) and for the Lorentzian fit (Right).  The median value together with the credible intervals of 33\% and 66\% were derived from a Monte-Carlo simulation of the fit.  The open diamond is the result of the fit for the solar LOI data.  The open triangles are the result of the fit for the sub-giant stars of \citet{Benomar2013}.   The error bars are derived from a Monte-Carlo simulation using credible intervals of 33\% and 66\%.  The solid lines show a linear fit of the parameters with respect to the frequency.  The Lorentzian parameters for stars for which the Lorentzian fit is not significant are not plotted.}
\label{dip1}
\end{figure*}

Fig.~\ref{dip1} shows the result of the fit as a function of the frequency of maximum power, $\nu_{\rm max}$, which is approximately proportional to $g/\sqrt{T_{\rm eff}}$  \citep{Brown1991, Belkacem2011}, where $g$ is the surface gravity and $T_{\rm eff}$ is the effective temperature .  

\citet{Chaplin1997} found that the solar mode linewidth follows a power law of 7$\pm 1.5$, at frequencies below 1800 $\mu$Hz in agreement with the theoretical result of \citet{Balmforth1992} and \citet{PG1994a}.  \citet{Komm2000} found a power law index of about 3.3 for mode frequency excluding the linewidth dip extending from 2400 $\mu$Hz to 3750 $\mu$Hz for the Sun.  In our case, the solar power law index is closer to that of \citet{Komm2000} because we made a global fit over a larger frequency range than that of the fit performed by \citet{Chaplin1997}.  

The width of the solar dip as measured by \citet{Komm2000} is about 1350 $\mu$Hz wide (full width) which is about twice as large as the full width at half maximum that we found for the Sun.  The right-hand panels of Fig.~\ref{dip} clearly shows that the amplitude and the width of the dip decrease with the effective temperature.  On the left top panel of Fig.~\ref{dip}, the dip also manifests itself as a deviation with respect to the linewidth measured at the maximum mode height (the orange line); the deviation becomes very small at high effective temperature.  

It was shown by \citet{Komm2000} that the solar linewidth dip is reduced for increasing solar activity.  The reduction was suggested to be due to the fact that radiative processes occurring in the upper superadiabatic boundary layer of the
convection zone that are locally destabilizing, would be in turn less unstable because of an increasing magnetic field \citep{Komm2000}.  But for other stars what is the source of the vanishing linewidth dip?  Using data from a high-resolution spectrograph, \citet{Karoff2013} measured the flux of the chromospheric lines Ca II H\&K of 11 stars in our study having a level of activity similar to that of the Sun.  All of these stars fall in the category IV (low activity stars) of \citet{Vaughan1980}.  KIC3733735 and KIC8379927 show a very high level of activity compared to the others of \citet{Karoff2013}.  KIC3733735 has the highest effective temperature of our sample, and shows a dip that is at the limit of detection with very large error bars (see Fig.~\ref{bias_comparison_3733735_new} and Fig.~\ref{dip}).  On the other hand KIC8379927 shows a measurable dip that might imply that this star is at minimum activity (see Fig.~\ref{8379927}).  Therefore the reason for the absence of a dip in some active stars could be due to the stars being at their maximum activity.  This can only be confirmed by measuring the activity of these stars over longer period of time.

\citet{Balmforth1992} showed that the linewidth dip would disappear with an increasing mixing length parameter ($\alpha$).  Numerical simulations performed by \citet{Ludwig1999}, \citet{Freytag1999} and \citet{Trampedach2011} showed that the mixing length parameter decreases with effective temperature, and therefore a smaller mixing length parameter at high effective temperature would increase the depth of  the linewidth dip contrary to what is observed in Fig.~\ref{dip}.  Results based on stellar modelling by \citet{Pinheiro2013} show an opposite dependence to that of the numerical simulations, e.g. the mixing length parameter increases with the effective temperature, being thus consistent with the findings of Fig.~\ref{dip}.  

\citet{GH96} showed that the linewidth dip becomes more pronounced with decreasing surface density if the mixing-length parameter and anisotropy of the turbulent velocity field are kept constant in the model computations.  For the stars studied in this paper, the surface density variations are dominated by changes in the effective temperature, i.e. the surface density decreases with increasing effective temperature provided the surface gravity stays approximately constant.  For that case the linewidth dip would increase with increasing surface temperature, according to \citet{GH96}, which is the opposite to what we found in Fig.~\ref{dip}.  However, the properties of the linewidth dip also depend crucially on the anisotropy of the turbulent velocity field which calibration may lead to better agreements between model computations and observations.

These contradictory findings may have important impact for the modelling of convection and turbulence in stars.

\section{Conclusions}
We have analysed the oscillation power spectra of 23 main-sequence stars for which we obtained the mode linewidths, mode heights and mode amplitude parameters.  The parameters were obtained by a team of 8 independent fitters.  We found large systematic errors between the parameters that could be traced to the way that the stellar background of the power spectrum was modelled; and to the fitted values of the rotational splitting and the stellar inclination angle.  Other sources of systematic errors related to the mean frequency definition and to the mode height ratio were also studied.  Finally using  a correction scheme derived from the one-fit approach of \citet{TT2005b}, we could explain all sources of systematic errors, which could be reduced to less than $\pm$15\% for mode linewidth and mode height, and to less than $\pm$5\% for amplitude, when compared to a reference fit value.  A different stellar background will give rise to frequency-dependent systematic errors that might affect the comparison with theoretical mode linewidth and mode height, therefore affecting the understanding of the physical nature of these parameters.  All other sources of relative systematic errors are independent of frequency.

Using the 23 stars of this study, 4 additional sub-giant stars of \citet{Benomar2013} and solar data, we also derived that the amplitude of the linewidth dip close to the maximum of frequency decreases with effective temperature.  The dependence of the dip with effective temperature is linked to the behaviour of convection in the stellar atmosphere, implying that either the mixing length or the level of activity may increase with effective temperature.



\begin{acknowledgements}
We are grateful to the referee for their comments which considerably improved the structure and readability to the paper.  The authors wish to thank the entire \textit{Kepler} team, without whom
these results would not have been possible. Funding for this Discovery mission is provided by NASA's
Science Mission Directorate.  We also thank all funding
councils and agencies that have supported the activities
of KASC Working Group~1, as well as the International Space Science
Institute (ISSI).  TA gratefully acknowledges the financial support of the Centre National d'Etudes Spatiales (CNES) under a PLATO grant.  TLC, WJC and GRD acknowledge financial support from the UK Science and Technology Facilities Council (STFC).  Funding for the Stellar Astrophysics Centre is provided by The Danish National Research Foundation. The research is supported by the ASTERISK project (ASTERoseismic Investigations with SONG and \textit{Kepler}) funded by the European Research Council (Grant agreement no.: 267864).  RAG and GRD has received funding from the European Community's Seventh Framework Programme (FP7/2007-2013) under grant agreement no. 269194.  RH acknowledges computing support from the National Solar Observatory.  SOHO is a mission of international collaboration between ESA and NASA.  We thank J\'er\^ome Ballot for useful discussions.
\end{acknowledgements}

\bibliographystyle{aa}
\bibliography{thierrya}

\begin{table*}[htbp]
\caption{Table of key stellar parameters.  The first three columns provide the KIC, HIP and HD numbers.  The fourth column provides the effective temperature and its error bar from \citet{Pins2011}.  The fifth column provides the \textit{Kepler} magnitude.  The sixth column gives the median of the large separation as measured using the mode frequencies given in this paper.  The seventh column provides the frequency of the maximum of oscillation power.  The eighth column gives the star category as provided by \citet{Appourchaux2012}.  The  penultimate column gives the duration of observation.  The last column provides the start and end quarters of the observation.}
\centering
\begin{tabular}{r r r c r r r c c c} 
\hline
\hline
KIC&HIP&HD&$T_{\rm eff}$ (K)&Kp& $\Delta\nu$ ($\mu$Hz)&$\nu_{\rm max}$($\mu$Hz) &Star category&Duration (in months)&Quarters\\
\hline
\hline
1435467&-&-&        6433 $\pm$         86&8.9&70.9&1324&F-like&24&Q5-Q12\\
2837475&-&-&       6688 $\pm$          57&8.5&76.0&1522&F-like&24&Q5-Q12\\
3424541&-&-&        6475 $\pm$          66&9.7&41.3&712&F-like&24&Q5-Q12\\
3733735&94071&178971&        6711 $\pm$          66&8.4&92.4&2041&F-like&24&Q5-Q12\\
6116048&-&-&        6072 $\pm$          49&8.4&100.7&2020&simple&24&Q5-Q12\\
6508366&-&-&        6499 $\pm$          46&9.0&51.6&959&F-like&24&Q5-Q12\\
6679371&-&-&        6598 $\pm$          59&8.7&50.4&908&F-like&24&Q5-Q12\\
7103006&-&-&        6421 $\pm$          51&8.9&59.9&1072&F-like&24&Q5-Q12\\
7206837&-&-&        6392 $\pm$          59&9.8&79.0&1556&simple&24&Q5-Q12\\
8379927&97321&187160&        6034 $\pm$          74&7.0&120.4&2669&simple&24&Q5-Q12\\
8694723&-&-&        6351 $\pm$          62&8.9&75.1&1384&simple&24&Q5-Q12\\
9139151&92961&-&        6134 $\pm$          48&9.2&117.0&2610&simple&24&Q5-Q12\\
9139163&92962&-&        6405 $\pm$          44&8.3&81.4&1649&simple&24&Q5-Q12\\
9206432&93607&-&        6494 $\pm$          46&9.1&85.1&1822&F-like&24&Q5-Q12$\,^{a}$\\
9812850&-&-&        6407 $\pm$          47&9.5&65.3&1186&F-like&24&Q5-Q12\\
10162436&97992&-&        6346 $\pm$          108&8.6&55.9&1004&simple&24&Q5-Q12$\,^{b}$\\
10355856&-&-&        6558 $\pm$          56&9.2&68.3&1210&F-like&15&Q5-Q9$\,^{c}$\\
10454113&92983&-&        6197 $\pm$          45&8.6&105.2&2313&simple&24&Q5-Q12\\
10909629&-&-&        6501 $\pm$          61&10.9&49.7&813&F-like&21&Q5-Q11$\,^{d}$\\
11081729&-&-&        6605 $\pm$          51&9.0&90.2&1820&F-like&24&Q5-Q12\\
12009504&-&-&        6270 $\pm$          61&9.3&88.1&1768&simple&24&Q5-Q12\\
12258514&95568&183298&        5990 $\pm$          85&8.1&74.8&1449&simple&24&Q5-Q12\\
12317678&97316&-&        6549 $\pm$          48&8.7&64.1&1162&F-like&24&Q5-Q12\\
\hline
\end{tabular}
\begin{list}{}{}
\item[$^{\rm a}$] Q7 missing.
\item[$^{\rm b}$] Q7, Q10, Q11 missing.
\item[$^{\rm c}$] Q7 missing.
\item[$^{\rm d}$] Q8, Q9, Q10 missing.
\end{list}
\label{table1}
\end{table*}

\begin{table*}[htbp]
\caption{Characteristics of the fit performed by each fitting group.  The first column provides the fitter name.  The second column provides methods used by the fitters.  The third column provides the number of fitted stars.  The  fourth column provides the number of parameters used per order.  The fifth column provides the number of additional parameters that are common to all the modes and plus those describing the stellar background.  The sixth column provides the number of fitted radial orders.  The last column provides the range over which the fit is performed.  MLE stands for maximum likelihood estimators.  MCMC stands for Monte Carlo Markov chain.}             
\label{tab_summary1}      
\centering                          
\begin{tabular}{c c c c c c c c c}        
\hline                 
\hline     
Fitter&Method&Number of&Param.& Add.&Orders& Window\\
&&stars fitted&per order& parameters&&size\\
\hline  
\hline                           
Appourchaux, IAS&MLE Global$^{b}$&23&5&5& $\le$20&$\le 20 \Delta \nu$\\
Howe, BIR&MLE Global$^{f}$&10&5&5&$\le$15&$\le 15 \Delta \nu$\\
Davies, BIR&MLE Global$^{d}$&23&5&4& $\le$20&$\le 20 \Delta \nu$\\
Antia, TAT&MLE Global$^{b}$&23&7&5& $\le$15&$\Delta \nu$\\
R\'egulo, IAC&MLE Local$^{a}$&23&12&None& $\le$15&$\Delta \nu$\\
Campante, BIR&MLE Global$^{b}$&23&5&6& $\le$20&$\le 20\Delta \nu$\\
Benomar, SYD&Bayesian MCMC$^{c}$&19&5&10& $>$10&$> 10 \Delta \nu$\\
Handberg, AAU&Bayesian MCMC$^{e}$&21&5&5& $>$10&$> 10 \Delta \nu$\\
\hline  
\hline  
\end{tabular}
\begin{list}{}{}
\item[$^{a}$] \citet{EA90}
\item[$^{b}$] \citet{Appourchaux2008}
\item[$^{c}$] \citet{Benomar2009}
\item[$^{d}$] \citet{Fletcher2009} 
\item[$^{e}$] \citet{Handberg2011}
\item[$^{f}$] derived from \citet{Howe1998}

\end{list}
\label{methods}
\end{table*}

\begin{table*}[htbp]
\caption{Table of linewidth parameters as per Eq.~(\ref{Eq_fit}) for the 23 stars in this study, the 4 stars of  \citet{Benomar2013} and of the Sun-as-a-star.  The first column provide the KIC numbers.  The second column provides the effective temperature.  The third and fourth columns provide the power law index and the 66-\% credible error.  The fifth and sixth columns provides the linewidth factor and their uncertainties.  The seventh and eighth columns provides the width of the linewidth dip and the and their uncertainties.  The ninth and tenth columns provides the depth of the linewidth dip and and their uncertainties.}
\centering
\begin{tabular}{r r c c c c c c c c} 
\hline
\hline
 KIC&$T_{\rm eff}$& $\alpha$ & 1-$\sigma$ uncertainty& $\Gamma_{\alpha}$ & 1-$\sigma$ uncertainty& $W_{\rm dip}$ & 1-$\sigma$ uncertainty& $\Delta\Gamma_{\rm dip}$ & 1-$\sigma$ uncertainty\\
  & (K) & & &(in $\mu$Hz) & (in $\mu$Hz)&(in $\mu$Hz) & (in $\mu$Hz)&(in $\mu$Hz) & (in $\mu$Hz)\\
\hline
\hline
  1435467 & 6433 & 1.66 & +0.18 / -0.18 & 4.16 & +0.12 / -0.12 & - & - & - & - \\
  2837475 & 6688 & 1.13 & +0.17 / -0.31 & 9.01 & +1.63 / -0.80 & 365. & +124. / -97. & 1.68 & +0.36 / -0.16 \\
  3424541 & 6475 & 1.30 & +0.21 / -0.21 & 6.03 & +0.23 / -0.23 & - & - & - & - \\
  3733735 & 6711 & 1.95 & +0.16 / -0.16 & 8.53 & +0.26 / -0.25 & - & - & - & - \\
  6116048 & 6072 & 2.82 & +0.11 / -0.11 & 4.32 & +0.28 / -0.24 & 458. & +28. / -29. & 3.23 & +0.20 / -0.19\\
  6508366 & 6499 & 1.23 & +0.13 / -0.13 & 4.91 & +0.12 / -0.12 & - & - & - & - \\
  6679371 & 6598 & 1.10 & +0.05 / -0.05 & 5.06 & +0.09 / -0.09 & 39. & +16. / -9. & 1.57 & +0.19 / -0.12\\
  7103006 & 6421 & 1.56 & +0.15 / -0.15 & 4.12 & +0.11 / -0.11 & - & - & - & -  \\
  7206837 & 6392 & 2.00 & +0.26 / -0.26 & 3.14 & +0.14 / -0.13 & - & - & - & -  \\
  8379927 & 6034 & 2.93 & +0.13 / -0.14 & 6.45 & +0.69 / -0.58 & 763. & +63. / -64. & 3.25 & +0.35 / -0.29\\
  8694723 & 6351 & 1.65 & +0.05 / -0.05 & 5.11 & +0.16 / -0.15 & 241. & +25. / -5.0 & 1.76 & +0.05 / -0.05\\
  9139151 & 6134 & 3.53 & +0.87 / -0.51 & 4.31 & +2.42 / -0.89 & 451. & +193. / -139. & 2.64 & +1.53 / -0.49 \\
  9139163 & 6405 & 1.39 & +0.11 / -0.11 & 5.43 & +0.10 / -0.10 & - & - & - & - \\
  9206432 & 6494 & 1.52 & +0.16 / -0.16 & 6.28 & +0.17 / -0.16 & - & - & - & - \\
  9812850 & 6407 & 1.81 & +0.23 / -0.23 & 3.90 & +0.16 / -0.15 & - & - & - & - \\
 10162436 & 6346 & 1.41 & +0.06 / -0.06 & 3.95 & +0.13 / -0.12 & 161. & +22. / -22. & 1.62 & +0.05 / -0.05\\
 10355856 & 6558 & 0.85 & +0.09 / -0.10 & 4.91 & +0.13 / -0.13 & 30. & +5. / -4. & 3.36 & +1.31 / -0.73\\
 10454113 & 6197 & 2.45 & +0.09 / -0.09 & 3.71 & +0.10 / -0.10 & 140. & +25. / -23. & 1.62 & +0.07 / -0.07\\
 10909629 & 6501 & 0.90 & +0.24 / -0.24 & 3.13 & +0.12 / -0.12 & - & - & - & - \\
 11081729 & 6605 & 1.99 & +0.22 / -0.22 & 5.96 & +0.23 / -0.22 & - & - & - & - \\
 12009504 & 6270 & 2.81 & +0.16 / -0.15 & 3.91 & +0.30 / -0.25 & 267. & +44. / -42. & 2.05 & +0.15 / -0.13\\
 12258514 & 5990 & 2.43 & +0.07 / -0.07 & 3.99 & +0.16 / -0.16 & 243. & +18. / -19. & 2.69 & +0.10 / -0.10\\
 12317678 & 6549 & 1.61 & +0.13 / -0.13 & 4.62 & +0.11 / -0.11 & - & - & - & - \\
\hline
  6442183 & 5740 & 3.31 & +0.55 / -0.36 & 5.00 & +3.40 / -1.57 & 387. & +88. / -73. & 6.72 & +4.41 / -1.91\\
 11026764 & 5722 & 4.73 & +1.57 / -0.92 & 3.40 & +5.60 / -1.28 & 233. & +85. / -64. & 5.44 & +11.0 / -2.20\\
 12508433 & 5257 & 3.40 & +0.29 / -0.29 & 1.41 & +0.34 / -0.22 & 176. & +34. / -29. & 3.60 & +0.84 / -0.54\\
 11771760 & 6075 & 0.87 & +0.24 / -0.24 & 1.24 & +0.06 / -0.05 & - & - & - & - \\
\hline 
 Sun & 5777 &  4.97  &  +0.03 / -0.03  &    4.65 &    +0.11 / -0.11 &     824.  &    +14. / -15. &     4.66 &    +0.10 / -0.10\\
 \hline
 \hline
\end{tabular}
\label{table_last}
\end{table*}



\begin{table*}[!]
\centering
\caption{{\bf Mode heights, mode linewidths and mode amplitude with their associated uncertainties} for KIC 1435467}
\begin{tabular}{c c c c c c c} 
\hline
\hline
Frequency&Mode height & 1-$\sigma$ uncertainty & Linewidth & 1-$\sigma$ uncertainty&Amplitude&1-$\sigma$ uncertainty\\
(in $\mu$Hz)&(in ppm$^2 \mu$Hz$^{-1}$) & (in ppm$^2 \mu$Hz$^{-1}$)& (in $\mu$Hz) & (in $\mu$Hz)&(in ppm)&(in ppm)\\
\hline
\hline
921.50&0.86&+1.81 / -0.58&2.51&+9.26 / -1.97&1.84&+0.46 / -0.37\\
995.28&1.37&+0.14 / -0.12&2.43&+0.38 / -0.32&2.28&+0.13 / -0.12\\
1064.75&1.78&+0.30 / -0.26&2.90&+0.53 / -0.45&2.84&+0.11 / -0.11\\
1136.22&2.50&+0.34 / -0.30&2.88&+0.40 / -0.35&3.36&+0.10 / -0.10\\
1206.38&2.76&+0.33 / -0.29&3.14&+0.38 / -0.34&3.68&+0.10 / -0.10\\
1275.16&3.09&+0.27 / -0.25&4.43&+0.37 / -0.35&4.63&+0.10 / -0.09\\
1343.74&3.31&+0.29 / -0.27&4.49&+0.38 / -0.35&4.83&+0.09 / -0.09\\
1414.21&3.06&+0.27 / -0.25&4.19&+0.35 / -0.32&4.49&+0.09 / -0.09\\
1484.44&2.74&+0.24 / -0.22&4.56&+0.38 / -0.35&4.42&+0.09 / -0.09\\
1556.47&2.07&+0.20 / -0.19&4.75&+0.46 / -0.42&3.93&+0.09 / -0.09\\
1626.85&1.44&+0.15 / -0.13&6.44&+0.66 / -0.60&3.82&+0.09 / -0.09\\
1696.93&1.19&+0.13 / -0.12&5.73&+0.69 / -0.61&3.27&+0.10 / -0.09\\
1771.67&0.73&+0.11 / -0.10&6.67&+1.15 / -0.98&2.77&+0.11 / -0.10\\
1840.59&0.48&+0.07 / -0.06&9.10&+1.47 / -1.27&2.63&+0.11 / -0.11\\
1911.83&0.42&+0.07 / -0.06&7.46&+1.47 / -1.23&2.21&+0.11 / -0.11\\
1984.87&0.31&+0.11 / -0.08&4.19&+1.86 / -1.29&1.43&+0.14 / -0.13\\
\hline
\hline
\end{tabular}
\end{table*}

\begin{table*}[!]
\centering
\caption{{\bf Mode heights, mode linewidths and mode amplitude with their associated uncertainties} for KIC 2837475}
\begin{tabular}{c c c c c c c} 
\hline
\hline
Frequency&Mode height & 1-$\sigma$ uncertainty & Linewidth & 1-$\sigma$ uncertainty&Amplitude&1-$\sigma$ uncertainty\\
(in $\mu$Hz)&(in ppm$^2 \mu$Hz$^{-1}$) & (in ppm$^2 \mu$Hz$^{-1}$)& (in $\mu$Hz) & (in $\mu$Hz)&(in ppm)&(in ppm)\\
\hline
\hline
1060.53&0.78&+0.26 / -0.19&3.43&+1.36 / -0.97&2.05&+0.13 / -0.12\\
1132.64&0.97&+0.10 / -0.09&5.43&+0.66 / -0.59&2.88&+0.10 / -0.10\\
1205.92&1.39&+0.18 / -0.16&4.55&+0.64 / -0.56&3.16&+0.09 / -0.09\\
1280.70&1.34&+0.15 / -0.13&5.18&+0.58 / -0.53&3.30&+0.09 / -0.09\\
1356.21&1.61&+0.17 / -0.16&4.96&+0.55 / -0.50&3.54&+0.09 / -0.08\\
1432.77&1.70&+0.17 / -0.15&5.55&+0.56 / -0.51&3.85&+0.08 / -0.08\\
1509.59&1.78&+0.16 / -0.15&6.09&+0.56 / -0.51&4.13&+0.08 / -0.08\\
1585.90&1.76&+0.14 / -0.13&6.91&+0.57 / -0.52&4.38&+0.08 / -0.08\\
1660.35&1.63&+0.13 / -0.12&7.30&+0.56 / -0.52&4.32&+0.08 / -0.08\\
1734.13&1.22&+0.09 / -0.08&9.03&+0.69 / -0.64&4.16&+0.08 / -0.08\\
1810.46&0.93&+0.07 / -0.06&9.88&+0.78 / -0.72&3.79&+0.08 / -0.08\\
1886.98&0.66&+0.06 / -0.06&10.52&+1.12 / -1.01&3.30&+0.08 / -0.08\\
1963.29&0.44&+0.05 / -0.04&11.98&+1.44 / -1.28&2.88&+0.09 / -0.09\\
2037.87&0.35&+0.04 / -0.03&12.96&+1.65 / -1.46&2.66&+0.09 / -0.09\\
2117.33&0.24&+0.04 / -0.03&12.56&+2.56 / -2.12&2.19&+0.11 / -0.10\\
2194.94&0.28&+0.05 / -0.04&8.67&+2.01 / -1.63&1.95&+0.11 / -0.10\\
2267.55&0.15&+0.04 / -0.03&9.76&+3.51 / -2.58&1.54&+0.13 / -0.12\\
2350.33&0.11&+0.04 / -0.03&11.45&+5.13 / -3.54&1.41&+0.15 / -0.13\\
2430.75&0.12&+0.03 / -0.03&11.90&+3.98 / -2.98&1.48&+0.13 / -0.12\\
\hline
\hline
\end{tabular}
\end{table*}

\begin{table*}[!]
\centering
\caption{{\bf Mode heights, mode linewidths and mode amplitude with their associated uncertainties} for KIC 3424541}
\begin{tabular}{c c c c c c c} 
\hline
\hline
Frequency&Mode height & 1-$\sigma$ uncertainty & Linewidth & 1-$\sigma$ uncertainty&Amplitude&1-$\sigma$ uncertainty\\
(in $\mu$Hz)&(in ppm$^2 \mu$Hz$^{-1}$) & (in ppm$^2 \mu$Hz$^{-1}$)& (in $\mu$Hz) & (in $\mu$Hz)&(in ppm)&(in ppm)\\
\hline
\hline
475.22&1.55&+0.53 / -0.40&2.95&+1.17 / -0.84&2.68&+0.31 / -0.28\\
516.39&2.22&+0.58 / -0.46&3.18&+0.97 / -0.74&3.33&+0.27 / -0.25\\
555.52&2.57&+0.41 / -0.36&4.59&+0.83 / -0.70&4.31&+0.23 / -0.22\\
590.34&3.21&+0.44 / -0.38&5.00&+0.72 / -0.63&5.02&+0.20 / -0.20\\
636.66&2.96&+0.45 / -0.39&4.65&+0.83 / -0.70&4.65&+0.22 / -0.21\\
678.55&4.10&+0.44 / -0.40&5.66&+0.66 / -0.59&6.03&+0.19 / -0.18\\
719.77&4.42&+0.46 / -0.42&5.79&+0.65 / -0.58&6.34&+0.18 / -0.17\\
761.25&3.93&+0.41 / -0.37&6.29&+0.73 / -0.65&6.23&+0.18 / -0.18\\
801.48&3.40&+0.38 / -0.34&7.19&+0.93 / -0.82&6.19&+0.20 / -0.19\\
841.40&2.37&+0.27 / -0.25&9.27&+1.28 / -1.12&5.88&+0.20 / -0.20\\
884.73&1.82&+0.23 / -0.20&7.92&+1.15 / -1.01&4.76&+0.20 / -0.20\\
927.36&1.51&+0.23 / -0.20&6.67&+1.18 / -1.00&3.98&+0.20 / -0.19\\
965.79&0.95&+0.15 / -0.13&10.35&+2.00 / -1.68&3.94&+0.22 / -0.20\\
1012.61&0.84&+0.24 / -0.19&4.72&+1.81 / -1.31&2.49&+0.26 / -0.24\\
1054.05&0.48&+0.20 / -0.14&13.57&+8.57 / -5.25&3.20&+0.34 / -0.31\\
\hline
\hline
\end{tabular}
\end{table*}

\begin{table*}[!]
\centering
\caption{{\bf Mode heights, mode linewidths and mode amplitude with their associated uncertainties} for KIC 3733735}
\begin{tabular}{c c c c c c c} 
\hline
\hline
Frequency&Mode height & 1-$\sigma$ uncertainty & Linewidth & 1-$\sigma$ uncertainty&Amplitude&1-$\sigma$ uncertainty\\
(in $\mu$Hz)&(in ppm$^2 \mu$Hz$^{-1}$) & (in ppm$^2 \mu$Hz$^{-1}$)& (in $\mu$Hz) & (in $\mu$Hz)&(in ppm)&(in ppm)\\
\hline
\hline
1293.12&0.65&+0.23 / -0.17&1.45&+0.69 / -0.47&1.21&+0.13 / -0.12\\
1385.60&0.45&+0.10 / -0.08&4.01&+1.02 / -0.82&1.68&+0.11 / -0.10\\
1473.54&0.60&+0.10 / -0.08&4.83&+0.88 / -0.75&2.14&+0.09 / -0.09\\
1562.96&0.56&+0.07 / -0.07&6.91&+1.00 / -0.87&2.47&+0.09 / -0.08\\
1653.77&0.67&+0.08 / -0.07&6.85&+0.94 / -0.82&2.69&+0.08 / -0.08\\
1747.04&0.81&+0.08 / -0.07&6.22&+0.66 / -0.59&2.81&+0.08 / -0.08\\
1840.35&0.69&+0.07 / -0.07&7.79&+0.91 / -0.81&2.90&+0.08 / -0.08\\
1933.52&0.81&+0.07 / -0.06&7.63&+0.72 / -0.65&3.12&+0.07 / -0.07\\
2026.36&0.68&+0.06 / -0.06&10.18&+1.02 / -0.92&3.30&+0.07 / -0.07\\
2117.55&0.64&+0.05 / -0.05&11.53&+1.05 / -0.97&3.39&+0.07 / -0.07\\
2209.06&0.55&+0.04 / -0.04&11.94&+1.07 / -0.98&3.20&+0.08 / -0.07\\
2301.84&0.51&+0.05 / -0.05&11.65&+1.20 / -1.09&3.07&+0.07 / -0.07\\
2393.06&0.37&+0.04 / -0.03&13.67&+1.48 / -1.34&2.83&+0.08 / -0.08\\
2483.81&0.27&+0.03 / -0.03&14.13&+1.76 / -1.56&2.46&+0.08 / -0.08\\
2581.11&0.28&+0.04 / -0.04&9.82&+1.73 / -1.47&2.07&+0.09 / -0.08\\
2669.78&0.16&+0.03 / -0.03&14.81&+3.57 / -2.88&1.93&+0.10 / -0.10\\
2765.25&0.16&+0.02 / -0.02&14.61&+2.58 / -2.20&1.93&+0.10 / -0.09\\
2863.16&0.12&+0.02 / -0.02&21.47&+4.92 / -4.01&1.99&+0.11 / -0.11\\
\hline
\hline
\end{tabular}
\end{table*}

\begin{table*}[!]
\centering
\caption{{\bf Mode heights, mode linewidths and mode amplitude with their associated uncertainties} for KIC 6116048}
\begin{tabular}{c c c c c c c} 
\hline
\hline
Frequency&Mode height & 1-$\sigma$ uncertainty & Linewidth & 1-$\sigma$ uncertainty&Amplitude&1-$\sigma$ uncertainty\\
(in $\mu$Hz)&(in ppm$^2 \mu$Hz$^{-1}$) & (in ppm$^2 \mu$Hz$^{-1}$)& (in $\mu$Hz) & (in $\mu$Hz)&(in ppm)&(in ppm)\\
\hline
\hline
1349.35&0.92&+0.34 / -0.25&0.64&+0.26 / -0.18&0.96&+0.09 / -0.08\\
1450.06&0.85&+0.21 / -0.17&1.55&+0.46 / -0.35&1.44&+0.09 / -0.09\\
1550.22&1.28&+0.25 / -0.21&1.38&+0.30 / -0.24&1.66&+0.08 / -0.08\\
1649.77&2.03&+0.26 / -0.23&1.64&+0.20 / -0.18&2.29&+0.07 / -0.07\\
1748.30&3.02&+0.35 / -0.32&1.57&+0.17 / -0.15&2.73&+0.07 / -0.07\\
1847.86&5.83&+0.61 / -0.55&1.28&+0.11 / -0.10&3.42&+0.08 / -0.08\\
1948.40&7.58&+0.73 / -0.67&1.29&+0.10 / -0.09&3.92&+0.08 / -0.08\\
2049.42&8.48&+0.75 / -0.69&1.44&+0.10 / -0.09&4.38&+0.09 / -0.08\\
2149.99&6.78&+0.60 / -0.55&1.65&+0.11 / -0.11&4.20&+0.08 / -0.08\\
2250.50&4.46&+0.41 / -0.38&2.03&+0.15 / -0.14&3.77&+0.07 / -0.07\\
2351.72&1.93&+0.16 / -0.15&3.57&+0.27 / -0.25&3.28&+0.06 / -0.06\\
2453.13&0.92&+0.09 / -0.08&4.54&+0.42 / -0.39&2.55&+0.06 / -0.06\\
2554.95&0.56&+0.06 / -0.06&5.12&+0.58 / -0.52&2.13&+0.07 / -0.06\\
2656.57&0.31&+0.05 / -0.04&6.62&+1.19 / -1.01&1.80&+0.08 / -0.07\\
2760.44&0.20&+0.04 / -0.03&5.70&+1.25 / -1.03&1.34&+0.09 / -0.08\\
\hline
\hline
\end{tabular}
\end{table*}

\begin{table*}[!]
\centering
\caption{{\bf Mode heights, mode linewidths and mode amplitude with their associated uncertainties} for KIC 6508366}
\begin{tabular}{c c c c c c c} 
\hline
\hline
Frequency&Mode height & 1-$\sigma$ uncertainty & Linewidth & 1-$\sigma$ uncertainty&Amplitude&1-$\sigma$ uncertainty\\
(in $\mu$Hz)&(in ppm$^2 \mu$Hz$^{-1}$) & (in ppm$^2 \mu$Hz$^{-1}$)& (in $\mu$Hz) & (in $\mu$Hz)&(in ppm)&(in ppm)\\
\hline
\hline
574.64&1.76&+0.51 / -0.39&1.71&+0.57 / -0.43&2.17&+0.19 / -0.17\\
624.42&1.38&+0.26 / -0.22&4.75&+1.01 / -0.83&3.21&+0.16 / -0.15\\
671.33&1.62&+0.39 / -0.32&4.34&+1.30 / -1.00&3.33&+0.17 / -0.16\\
722.94&2.99&+0.36 / -0.32&3.39&+0.42 / -0.38&3.99&+0.12 / -0.12\\
775.18&3.25&+0.37 / -0.33&3.34&+0.38 / -0.34&4.13&+0.12 / -0.11\\
826.74&3.40&+0.32 / -0.29&4.61&+0.42 / -0.39&4.96&+0.11 / -0.11\\
878.24&3.84&+0.34 / -0.31&4.64&+0.39 / -0.36&5.29&+0.11 / -0.11\\
929.45&4.46&+0.36 / -0.33&4.77&+0.35 / -0.33&5.78&+0.11 / -0.10\\
979.09&3.54&+0.27 / -0.25&5.86&+0.42 / -0.39&5.71&+0.10 / -0.10\\
1031.43&3.67&+0.29 / -0.27&5.01&+0.36 / -0.34&5.38&+0.10 / -0.10\\
1083.52&2.67&+0.23 / -0.21&5.35&+0.44 / -0.41&4.74&+0.10 / -0.10\\
1136.29&1.88&+0.18 / -0.16&5.81&+0.56 / -0.51&4.14&+0.10 / -0.10\\
1189.27&1.40&+0.13 / -0.12&7.01&+0.69 / -0.63&3.93&+0.10 / -0.10\\
1240.74&1.04&+0.11 / -0.10&7.57&+0.88 / -0.78&3.52&+0.10 / -0.10\\
1292.89&0.64&+0.09 / -0.08&8.71&+1.38 / -1.19&2.96&+0.12 / -0.11\\
1345.14&0.36&+0.06 / -0.05&12.24&+2.53 / -2.10&2.63&+0.13 / -0.13\\
1400.84&0.33&+0.11 / -0.08&5.63&+2.73 / -1.84&1.70&+0.16 / -0.15\\
1451.29&0.26&+0.08 / -0.06&7.53&+3.34 / -2.31&1.74&+0.15 / -0.14\\
1501.05&0.25&+0.15 / -0.09&2.24&+1.56 / -0.92&0.94&+0.16 / -0.14\\
\hline
\hline
\end{tabular}
\end{table*}

\begin{table*}[!]
\centering
\caption{{\bf Mode heights, mode linewidths and mode amplitude with their associated uncertainties} for KIC 6679371}
\begin{tabular}{c c c c c c c} 
\hline
\hline
Frequency&Mode height & 1-$\sigma$ uncertainty & Linewidth & 1-$\sigma$ uncertainty&Amplitude&1-$\sigma$ uncertainty\\
(in $\mu$Hz)&(in ppm$^2 \mu$Hz$^{-1}$) & (in ppm$^2 \mu$Hz$^{-1}$)& (in $\mu$Hz) & (in $\mu$Hz)&(in ppm)&(in ppm)\\
\hline
\hline
555.30&1.75&+0.62 / -0.46&1.55&+0.58 / -0.42&2.06&+0.21 / -0.19\\
606.39&2.63&+0.54 / -0.45&2.50&+0.59 / -0.48&3.21&+0.16 / -0.15\\
656.99&3.10&+0.42 / -0.37&3.54&+0.50 / -0.44&4.16&+0.14 / -0.13\\
703.97&3.04&+0.34 / -0.30&4.42&+0.50 / -0.45&4.60&+0.13 / -0.13\\
752.07&3.51&+0.34 / -0.31&4.58&+0.44 / -0.40&5.02&+0.12 / -0.12\\
803.04&4.73&+0.46 / -0.42&3.74&+0.35 / -0.32&5.27&+0.11 / -0.11\\
854.17&5.15&+0.43 / -0.40&3.81&+0.29 / -0.27&5.55&+0.11 / -0.11\\
905.16&4.58&+0.37 / -0.34&4.74&+0.36 / -0.33&5.84&+0.11 / -0.11\\
956.54&4.42&+0.34 / -0.32&5.43&+0.40 / -0.37&6.14&+0.11 / -0.11\\
1007.18&3.74&+0.29 / -0.27&6.33&+0.47 / -0.44&6.10&+0.11 / -0.10\\
1056.59&3.56&+0.26 / -0.24&6.26&+0.43 / -0.40&5.92&+0.10 / -0.10\\
1107.14&2.37&+0.18 / -0.17&7.37&+0.57 / -0.52&5.24&+0.10 / -0.10\\
1159.10&1.89&+0.17 / -0.16&6.99&+0.67 / -0.61&4.55&+0.10 / -0.10\\
1209.91&1.43&+0.14 / -0.13&6.62&+0.68 / -0.62&3.86&+0.10 / -0.10\\
1262.32&0.95&+0.11 / -0.10&8.28&+1.08 / -0.95&3.51&+0.10 / -0.10\\
1313.96&0.70&+0.09 / -0.08&8.25&+1.23 / -1.07&3.02&+0.11 / -0.10\\
1363.49&0.40&+0.07 / -0.06&9.39&+2.00 / -1.65&2.44&+0.12 / -0.11\\
1415.10&0.37&+0.07 / -0.06&6.41&+1.45 / -1.18&1.92&+0.12 / -0.11\\
1469.36&0.25&+0.10 / -0.07&4.54&+1.55 / -1.25&1.34&+0.13 / -0.13\\
\hline
\hline
\end{tabular}
\end{table*}

\begin{table*}[!]
\centering
\caption{{\bf Mode heights, mode linewidths and mode amplitude with their associated uncertainties} for KIC 7103006}
\begin{tabular}{c c c c c c c} 
\hline
\hline
Frequency&Mode height & 1-$\sigma$ uncertainty & Linewidth & 1-$\sigma$ uncertainty&Amplitude&1-$\sigma$ uncertainty\\
(in $\mu$Hz)&(in ppm$^2 \mu$Hz$^{-1}$) & (in ppm$^2 \mu$Hz$^{-1}$)& (in $\mu$Hz) & (in $\mu$Hz)&(in ppm)&(in ppm)\\
\hline
\hline
722.07&0.87&+0.35 / -0.25&2.39&+1.19 / -0.80&1.81&+0.18 / -0.16\\
779.63&2.27&+0.51 / -0.42&1.73&+0.42 / -0.34&2.49&+0.12 / -0.12\\
836.17&1.89&+0.32 / -0.28&2.86&+0.53 / -0.45&2.91&+0.12 / -0.12\\
894.43&2.27&+0.31 / -0.28&2.83&+0.41 / -0.36&3.17&+0.11 / -0.11\\
954.16&2.31&+0.27 / -0.24&3.43&+0.40 / -0.36&3.53&+0.10 / -0.10\\
1014.29&2.68&+0.26 / -0.23&3.94&+0.36 / -0.33&4.07&+0.10 / -0.10\\
1074.11&3.53&+0.33 / -0.30&3.67&+0.32 / -0.30&4.51&+0.10 / -0.09\\
1132.88&3.74&+0.33 / -0.30&3.78&+0.31 / -0.28&4.71&+0.09 / -0.09\\
1191.43&2.91&+0.25 / -0.23&4.78&+0.38 / -0.35&4.68&+0.09 / -0.09\\
1251.44&2.64&+0.24 / -0.22&4.31&+0.37 / -0.34&4.23&+0.09 / -0.09\\
1310.94&2.12&+0.18 / -0.17&4.77&+0.38 / -0.36&3.98&+0.09 / -0.09\\
1372.91&1.49&+0.17 / -0.15&4.92&+0.55 / -0.50&3.40&+0.09 / -0.09\\
1432.52&0.97&+0.09 / -0.09&6.73&+0.65 / -0.59&3.20&+0.09 / -0.09\\
1494.41&0.79&+0.10 / -0.09&6.41&+0.86 / -0.76&2.81&+0.09 / -0.09\\
1554.29&0.54&+0.07 / -0.06&7.06&+1.01 / -0.91&2.44&+0.10 / -0.10\\
1616.01&0.34&+0.05 / -0.04&8.46&+1.29 / -1.12&2.13&+0.11 / -0.10\\
1677.47&0.22&+0.05 / -0.04&8.52&+2.47 / -1.92&1.72&+0.13 / -0.12\\
1733.10&0.21&+0.06 / -0.05&7.75&+3.04 / -2.18&1.59&+0.14 / -0.13\\
\hline
\hline
\end{tabular}
\end{table*}

\clearpage

\begin{table*}[!]
\centering
\caption{{\bf Mode heights, mode linewidths and mode amplitude with their associated uncertainties} for KIC 7206837}
\begin{tabular}{c c c c c c c} 
\hline
\hline
Frequency&Mode height & 1-$\sigma$ uncertainty & Linewidth & 1-$\sigma$ uncertainty&Amplitude&1-$\sigma$ uncertainty\\
(in $\mu$Hz)&(in ppm$^2 \mu$Hz$^{-1}$) & (in ppm$^2 \mu$Hz$^{-1}$)& (in $\mu$Hz) & (in $\mu$Hz)&(in ppm)&(in ppm)\\
\hline
\hline
1039.96&1.50&+0.40 / -0.38&0.85&+0.42 / -0.33&1.41&+0.17 / -0.16\\
1117.89&1.52&+0.55 / -0.40&1.25&+0.49 / -0.35&1.73&+0.17 / -0.15\\
1194.76&1.15&+0.37 / -0.28&2.62&+1.03 / -0.74&2.18&+0.18 / -0.16\\
1273.02&1.71&+0.48 / -0.37&2.22&+0.72 / -0.54&2.44&+0.15 / -0.14\\
1353.23&1.98&+0.13 / -0.12&2.93&+0.32 / -0.29&3.02&+0.13 / -0.13\\
1431.70&3.18&+0.48 / -0.42&2.42&+0.38 / -0.33&3.48&+0.12 / -0.12\\
1508.69&3.54&+0.44 / -0.39&3.08&+0.39 / -0.35&4.14&+0.12 / -0.12\\
1586.44&3.20&+0.38 / -0.34&3.79&+0.47 / -0.42&4.37&+0.12 / -0.12\\
1665.21&3.36&+0.39 / -0.35&3.30&+0.39 / -0.35&4.17&+0.12 / -0.11\\
1745.08&2.79&+0.31 / -0.28&3.81&+0.43 / -0.39&4.09&+0.12 / -0.12\\
1825.90&2.51&+0.32 / -0.29&3.97&+0.55 / -0.48&3.95&+0.12 / -0.12\\
1905.22&1.31&+0.17 / -0.15&6.08&+0.85 / -0.75&3.54&+0.13 / -0.13\\
1984.16&1.10&+0.16 / -0.14&5.35&+0.91 / -0.78&3.04&+0.14 / -0.13\\
2065.74&0.82&+0.16 / -0.14&5.83&+1.40 / -1.13&2.74&+0.15 / -0.15\\
2143.49&0.40&+0.09 / -0.07&9.30&+2.57 / -2.01&2.41&+0.18 / -0.17\\
\hline
\hline
\end{tabular}
\end{table*}

\begin{table*}[!]
\centering
\caption{{\bf Mode heights, mode linewidths and mode amplitude with their associated uncertainties} for KIC 8379927}
\begin{tabular}{c c c c c c c} 
\hline
\hline
Frequency&Mode height & 1-$\sigma$ uncertainty & Linewidth & 1-$\sigma$ uncertainty&Amplitude&1-$\sigma$ uncertainty\\
(in $\mu$Hz)&(in ppm$^2 \mu$Hz$^{-1}$) & (in ppm$^2 \mu$Hz$^{-1}$)& (in $\mu$Hz) & (in $\mu$Hz)&(in ppm)&(in ppm)\\
\hline
\hline
1847.13&0.25&+0.10 / -0.07&0.86&+0.41 / -0.28&0.59&+0.07 / -0.06\\
1967.81&0.28&+0.06 / -0.05&1.85&+0.48 / -0.38&0.90&+0.05 / -0.05\\
2087.99&0.34&+0.05 / -0.04&2.41&+0.39 / -0.34&1.14&+0.05 / -0.05\\
2206.63&0.70&+0.09 / -0.08&1.92&+0.25 / -0.22&1.46&+0.04 / -0.04\\
2324.50&0.97&+0.10 / -0.09&1.94&+0.18 / -0.17&1.71&+0.04 / -0.04\\
2443.14&1.37&+0.13 / -0.12&1.90&+0.16 / -0.15&2.02&+0.04 / -0.04\\
2563.59&1.84&+0.16 / -0.14&1.84&+0.13 / -0.12&2.31&+0.04 / -0.04\\
2683.93&1.91&+0.15 / -0.14&2.14&+0.14 / -0.13&2.53&+0.04 / -0.04\\
2804.39&1.76&+0.13 / -0.12&2.41&+0.15 / -0.14&2.58&+0.04 / -0.04\\
2924.44&1.34&+0.10 / -0.10&2.63&+0.16 / -0.15&2.35&+0.04 / -0.04\\
3044.83&0.73&+0.06 / -0.05&3.86&+0.27 / -0.25&2.10&+0.04 / -0.04\\
3165.64&0.39&+0.03 / -0.03&4.84&+0.40 / -0.37&1.72&+0.04 / -0.04\\
3286.38&0.20&+0.02 / -0.02&7.32&+0.73 / -0.66&1.52&+0.04 / -0.04\\
3408.71&0.15&+0.02 / -0.02&6.50&+0.78 / -0.69&1.25&+0.04 / -0.04\\
3529.21&0.06&+0.01 / -0.01&10.44&+2.05 / -1.71&1.01&+0.05 / -0.05\\
3652.33&0.05&+0.01 / -0.01&7.90&+2.00 / -1.59&0.82&+0.06 / -0.05\\
\hline
\hline
\end{tabular}
\end{table*}

\begin{table*}[!]
\centering
\caption{{\bf Mode heights, mode linewidths and mode amplitude with their associated uncertainties} for KIC 8694723}
\begin{tabular}{c c c c c c c} 
\hline
\hline
Frequency&Mode height & 1-$\sigma$ uncertainty & Linewidth & 1-$\sigma$ uncertainty&Amplitude&1-$\sigma$ uncertainty\\
(in $\mu$Hz)&(in ppm$^2 \mu$Hz$^{-1}$) & (in ppm$^2 \mu$Hz$^{-1}$)& (in $\mu$Hz) & (in $\mu$Hz)&(in ppm)&(in ppm)\\
\hline
\hline
772.29&0.94&+0.51 / -0.33&0.66&+0.45 / -0.27&0.98&+0.16 / -0.14\\
846.11&0.70&+0.21 / -0.16&1.79&+0.64 / -0.47&1.40&+0.15 / -0.14\\
917.99&1.02&+0.13 / -0.12&2.13&+0.36 / -0.31&1.85&+0.12 / -0.11\\
990.44&1.41&+0.21 / -0.18&2.59&+0.44 / -0.37&2.40&+0.11 / -0.11\\
1064.45&1.93&+0.22 / -0.20&2.83&+0.30 / -0.27&2.93&+0.10 / -0.09\\
1139.38&2.57&+0.22 / -0.20&3.58&+0.28 / -0.26&3.80&+0.09 / -0.09\\
1212.46&3.66&+0.29 / -0.27&3.27&+0.22 / -0.21&4.34&+0.09 / -0.09\\
1285.68&5.73&+0.41 / -0.38&3.15&+0.18 / -0.17&5.33&+0.09 / -0.09\\
1359.86&5.69&+0.41 / -0.38&3.17&+0.18 / -0.17&5.32&+0.09 / -0.09\\
1435.50&6.42&+0.45 / -0.42&3.12&+0.17 / -0.16&5.61&+0.09 / -0.09\\
1510.92&5.90&+0.41 / -0.39&3.09&+0.17 / -0.16&5.35&+0.09 / -0.09\\
1586.77&3.75&+0.25 / -0.24&4.34&+0.23 / -0.22&5.06&+0.08 / -0.08\\
1661.79&2.72&+0.20 / -0.18&4.48&+0.28 / -0.26&4.37&+0.08 / -0.08\\
1737.97&1.60&+0.12 / -0.11&5.49&+0.37 / -0.35&3.71&+0.08 / -0.08\\
1813.23&1.04&+0.02 / -0.02&6.68&+0.34 / -0.32&3.31&+0.08 / -0.07\\
1890.22&0.68&+0.08 / -0.07&6.28&+0.78 / -0.70&2.59&+0.09 / -0.08\\
1965.46&0.38&+0.04 / -0.04&9.25&+1.07 / -0.96&2.35&+0.09 / -0.09\\
2042.16&0.26&+0.04 / -0.04&7.76&+1.42 / -1.20&1.78&+0.10 / -0.10\\
2117.60&0.18&+0.16 / -0.09&5.23&+8.47 / -3.23&1.23&+0.28 / -0.22\\
2193.59&0.23&+0.11 / -0.08&1.96&+1.34 / -0.80&0.84&+0.13 / -0.12\\
\hline
\hline
\end{tabular}
\end{table*}

\begin{table*}[!]
\centering
\caption{{\bf Mode heights, mode linewidths and mode amplitude with their associated uncertainties} for KIC 9139151}
\begin{tabular}{c c c c c c c} 
\hline
\hline
Frequency&Mode height & 1-$\sigma$ uncertainty & Linewidth & 1-$\sigma$ uncertainty&Amplitude&1-$\sigma$ uncertainty\\
(in $\mu$Hz)&(in ppm$^2 \mu$Hz$^{-1}$) & (in ppm$^2 \mu$Hz$^{-1}$)& (in $\mu$Hz) & (in $\mu$Hz)&(in ppm)&(in ppm)\\
\hline
\hline
1921.75&0.62&+0.28 / -0.19&1.63&+0.96 / -0.61&1.26&+0.14 / -0.13\\
2038.67&1.01&+0.29 / -0.22&1.27&+0.42 / -0.32&1.42&+0.11 / -0.10\\
2154.39&1.37&+0.26 / -0.22&1.45&+0.29 / -0.24&1.77&+0.10 / -0.09\\
2269.43&1.31&+0.22 / -0.19&2.07&+0.36 / -0.31&2.06&+0.09 / -0.09\\
2385.90&2.13&+0.28 / -0.25&1.98&+0.26 / -0.23&2.58&+0.09 / -0.09\\
2502.76&2.93&+0.34 / -0.31&1.83&+0.20 / -0.18&2.90&+0.09 / -0.08\\
2620.20&3.66&+0.39 / -0.35&1.73&+0.17 / -0.15&3.15&+0.09 / -0.08\\
2737.42&2.74&+0.31 / -0.28&2.12&+0.23 / -0.21&3.02&+0.09 / -0.08\\
2854.93&1.86&+0.22 / -0.20&2.81&+0.33 / -0.29&2.86&+0.09 / -0.08\\
2972.46&1.22&+0.19 / -0.16&2.91&+0.49 / -0.42&2.36&+0.09 / -0.09\\
3089.73&0.52&+0.10 / -0.08&5.18&+1.28 / -1.02&2.05&+0.12 / -0.11\\
3208.74&0.28&+0.07 / -0.05&6.60&+1.89 / -1.47&1.71&+0.13 / -0.12\\
\hline
\hline
\end{tabular}
\end{table*}

\begin{table*}[!]
\centering
\caption{{\bf Mode heights, mode linewidths and mode amplitude with their associated uncertainties} for KIC 9139163}
\begin{tabular}{c c c c c c c} 
\hline
\hline
Frequency&Mode height & 1-$\sigma$ uncertainty & Linewidth & 1-$\sigma$ uncertainty&Amplitude&1-$\sigma$ uncertainty\\
(in $\mu$Hz)&(in ppm$^2 \mu$Hz$^{-1}$) & (in ppm$^2 \mu$Hz$^{-1}$)& (in $\mu$Hz) & (in $\mu$Hz)&(in ppm)&(in ppm)\\
\hline
\hline
984.90&0.45&+0.16 / -0.12&1.60&+0.73 / -0.50&1.07&+0.14 / -0.12\\
1065.30&0.78&+0.16 / -0.13&1.78&+0.45 / -0.36&1.48&+0.11 / -0.10\\
1142.93&1.06&+0.19 / -0.16&1.76&+0.33 / -0.28&1.72&+0.09 / -0.09\\
1221.23&0.87&+0.11 / -0.10&3.81&+0.53 / -0.46&2.28&+0.09 / -0.09\\
1301.73&1.11&+0.13 / -0.12&3.37&+0.40 / -0.36&2.42&+0.08 / -0.08\\
1383.10&1.43&+0.13 / -0.12&4.07&+0.35 / -0.33&3.02&+0.08 / -0.08\\
1464.51&1.67&+0.13 / -0.12&4.80&+0.34 / -0.32&3.55&+0.07 / -0.07\\
1544.64&1.86&+0.13 / -0.12&5.03&+0.33 / -0.31&3.83&+0.07 / -0.07\\
1624.19&1.83&+0.13 / -0.12&5.99&+0.38 / -0.36&4.15&+0.07 / -0.07\\
1704.20&2.05&+0.14 / -0.13&5.16&+0.31 / -0.29&4.07&+0.07 / -0.07\\
1785.97&1.78&+0.12 / -0.12&5.15&+0.32 / -0.30&3.80&+0.07 / -0.07\\
1867.27&1.26&+0.09 / -0.09&6.14&+0.42 / -0.39&3.49&+0.07 / -0.07\\
1949.31&1.13&+0.09 / -0.08&6.01&+0.43 / -0.40&3.26&+0.07 / -0.07\\
2031.93&0.75&+0.06 / -0.06&7.59&+0.61 / -0.57&2.99&+0.07 / -0.07\\
2114.63&0.63&+0.05 / -0.05&8.22&+0.70 / -0.65&2.84&+0.07 / -0.07\\
2196.19&0.39&+0.04 / -0.04&9.09&+1.06 / -0.95&2.37&+0.07 / -0.07\\
2277.68&0.28&+0.04 / -0.03&10.00&+1.58 / -1.37&2.08&+0.08 / -0.08\\
2359.55&0.20&+0.03 / -0.03&10.54&+1.96 / -1.65&1.80&+0.09 / -0.08\\
2444.57&0.16&+0.04 / -0.03&5.97&+2.05 / -1.53&1.21&+0.11 / -0.10\\
2522.49&0.16&+0.04 / -0.03&1.78&+0.39 / -0.32&0.66&+0.10 / -0.08\\
\hline
\hline
\end{tabular}
\end{table*}

\begin{table*}[!]
\centering
\caption{{\bf Mode heights, mode linewidths and mode amplitude with their associated uncertainties} for KIC 9206432}
\begin{tabular}{c c c c c c c} 
\hline
\hline
Frequency&Mode height & 1-$\sigma$ uncertainty & Linewidth & 1-$\sigma$ uncertainty&Amplitude&1-$\sigma$ uncertainty\\
(in $\mu$Hz)&(in ppm$^2 \mu$Hz$^{-1}$) & (in ppm$^2 \mu$Hz$^{-1}$)& (in $\mu$Hz) & (in $\mu$Hz)&(in ppm)&(in ppm)\\
\hline
\hline
1108.68&0.48&+0.17 / -0.13&2.15&+0.98 / -0.67&1.27&+0.16 / -0.14\\
1194.96&0.63&+0.19 / -0.15&2.21&+0.85 / -0.61&1.48&+0.14 / -0.13\\
1274.82&0.71&+0.12 / -0.11&3.60&+0.68 / -0.57&2.01&+0.12 / -0.11\\
1356.25&0.75&+0.11 / -0.10&4.81&+0.83 / -0.71&2.38&+0.12 / -0.11\\
1440.64&1.28&+0.16 / -0.14&3.49&+0.45 / -0.40&2.65&+0.10 / -0.10\\
1525.14&0.97&+0.12 / -0.11&4.47&+0.63 / -0.55&2.61&+0.10 / -0.10\\
1611.17&1.23&+0.13 / -0.11&4.53&+0.47 / -0.43&2.96&+0.09 / -0.09\\
1696.45&1.38&+0.12 / -0.11&5.57&+0.46 / -0.42&3.48&+0.09 / -0.09\\
1781.27&1.36&+0.11 / -0.10&6.81&+0.55 / -0.51&3.81&+0.09 / -0.09\\
1864.21&1.39&+0.12 / -0.11&7.00&+0.60 / -0.56&3.91&+0.09 / -0.09\\
1948.60&1.38&+0.11 / -0.10&6.46&+0.53 / -0.49&3.74&+0.09 / -0.08\\
2032.93&0.93&+0.08 / -0.08&7.66&+0.71 / -0.65&3.35&+0.09 / -0.09\\
2120.08&0.95&+0.09 / -0.09&6.05&+0.61 / -0.55&3.00&+0.09 / -0.09\\
2204.60&0.67&+0.07 / -0.06&8.09&+0.82 / -0.75&2.92&+0.09 / -0.09\\
2289.92&0.49&+0.06 / -0.05&8.77&+1.14 / -1.01&2.59&+0.10 / -0.09\\
2374.11&0.34&+0.06 / -0.05&6.76&+1.43 / -1.18&1.91&+0.12 / -0.11\\
2461.30&0.18&+0.04 / -0.03&12.29&+3.26 / -2.58&1.89&+0.13 / -0.12\\
2549.94&0.19&+0.31 / -0.12&9.88&+21.97 / -6.81&1.72&+0.23 / -0.21\\
\hline
\hline
\end{tabular}
\end{table*}

\begin{table*}[!]
\centering
\caption{{\bf Mode heights, mode linewidths and mode amplitude with their associated uncertainties} for KIC 9812850}
\begin{tabular}{c c c c c c c} 
\hline
\hline
Frequency&Mode height & 1-$\sigma$ uncertainty & Linewidth & 1-$\sigma$ uncertainty&Amplitude&1-$\sigma$ uncertainty\\
(in $\mu$Hz)&(in ppm$^2 \mu$Hz$^{-1}$) & (in ppm$^2 \mu$Hz$^{-1}$)& (in $\mu$Hz) & (in $\mu$Hz)&(in ppm)&(in ppm)\\
\hline
\hline
788.78&1.32&+0.53 / -0.38&1.51&+0.70 / -0.48&1.77&+0.18 / -0.16\\
850.27&1.53&+0.49 / -0.37&1.61&+0.59 / -0.43&1.96&+0.16 / -0.15\\
912.48&1.68&+0.47 / -0.28&2.79&+0.71 / -0.54&2.71&+0.16 / -0.14\\
977.35&1.98&+0.46 / -0.38&2.90&+0.77 / -0.61&3.00&+0.14 / -0.13\\
1042.18&2.48&+0.31 / -0.27&3.74&+0.47 / -0.42&3.82&+0.12 / -0.12\\
1107.52&3.29&+0.39 / -0.35&3.45&+0.41 / -0.37&4.22&+0.11 / -0.11\\
1170.37&3.34&+0.38 / -0.34&3.84&+0.43 / -0.39&4.49&+0.11 / -0.11\\
1234.04&3.15&+0.32 / -0.29&4.36&+0.44 / -0.40&4.65&+0.11 / -0.11\\
1298.25&2.55&+0.27 / -0.24&4.88&+0.51 / -0.47&4.43&+0.11 / -0.11\\
1364.38&2.74&+0.30 / -0.27&4.03&+0.43 / -0.39&4.16&+0.11 / -0.10\\
1430.42&1.78&+0.20 / -0.18&5.55&+0.64 / -0.57&3.94&+0.11 / -0.11\\
1495.50&1.04&+0.13 / -0.12&7.10&+0.94 / -0.83&3.41&+0.12 / -0.12\\
1559.44&0.76&+0.11 / -0.09&7.94&+1.22 / -1.06&3.09&+0.13 / -0.12\\
1623.25&0.71&+0.14 / -0.12&5.63&+1.38 / -1.11&2.50&+0.14 / -0.13\\
1694.02&0.31&+0.09 / -0.07&11.36&+3.77 / -2.83&2.36&+0.17 / -0.16\\
1762.75&0.62&+0.31 / -0.21&2.95&+1.97 / -1.18&1.70&+0.18 / -0.16\\
1822.95&0.21&+0.14 / -0.08&5.68&+5.39 / -2.77&1.36&+0.25 / -0.21\\
\hline
\hline
\end{tabular}
\end{table*}

\begin{table*}[!]
\centering
\caption{{\bf Mode heights, mode linewidths and mode amplitude with their associated uncertainties} for KIC 10162436}
\begin{tabular}{c c c c c c c} 
\hline
\hline
Frequency&Mode height & 1-$\sigma$ uncertainty & Linewidth & 1-$\sigma$ uncertainty&Amplitude&1-$\sigma$ uncertainty\\
(in $\mu$Hz)&(in ppm$^2 \mu$Hz$^{-1}$) & (in ppm$^2 \mu$Hz$^{-1}$)& (in $\mu$Hz) & (in $\mu$Hz)&(in ppm)&(in ppm)\\
\hline
\hline
626.74&2.05&+0.55 / -0.44&1.24&+0.42 / -0.31&2.00&+0.16 / -0.15\\
678.93&2.13&+0.34 / -0.29&2.28&+0.38 / -0.32&2.76&+0.13 / -0.12\\
733.04&2.95&+0.40 / -0.35&2.10&+0.29 / -0.25&3.12&+0.12 / -0.11\\
788.79&2.99&+0.33 / -0.29&2.98&+0.32 / -0.29&3.74&+0.11 / -0.11\\
843.98&5.03&+0.45 / -0.42&2.76&+0.22 / -0.20&4.67&+0.11 / -0.10\\
898.99&6.76&+0.56 / -0.52&2.92&+0.20 / -0.19&5.57&+0.11 / -0.11\\
953.14&8.01&+0.59 / -0.55&2.81&+0.16 / -0.15&5.95&+0.11 / -0.11\\
1008.10&9.16&+0.70 / -0.65&2.64&+0.15 / -0.14&6.17&+0.11 / -0.11\\
1064.50&8.24&+0.66 / -0.61&2.67&+0.16 / -0.15&5.87&+0.11 / -0.11\\
1120.60&6.25&+0.46 / -0.43&3.20&+0.18 / -0.17&5.61&+0.10 / -0.10\\
1177.38&4.21&+0.34 / -0.31&3.60&+0.24 / -0.22&4.88&+0.09 / -0.09\\
1233.31&2.83&+0.23 / -0.21&4.35&+0.32 / -0.29&4.40&+0.09 / -0.08\\
1289.95&1.96&+0.18 / -0.16&4.63&+0.40 / -0.37&3.78&+0.09 / -0.08\\
1346.08&1.02&+0.12 / -0.10&5.95&+0.68 / -0.61&3.09&+0.09 / -0.08\\
1402.02&0.62&+0.08 / -0.07&5.40&+0.76 / -0.67&2.29&+0.09 / -0.09\\
1457.95&0.44&+0.06 / -0.05&6.05&+0.88 / -0.77&2.05&+0.10 / -0.09\\
1513.55&0.17&+0.04 / -0.03&7.32&+2.27 / -1.73&1.41&+0.13 / -0.12\\
1573.95&0.08&+0.05 / -0.03&7.86&+7.51 / -3.84&0.99&+0.20 / -0.17\\
\hline
\hline
\end{tabular}
\end{table*}

\begin{table*}[!]
\centering
\caption{{\bf Mode heights, mode linewidths and mode amplitude with their associated uncertainties} for KIC 10355856}
\begin{tabular}{c c c c c c c} 
\hline
\hline
Frequency&Mode height & 1-$\sigma$ uncertainty & Linewidth & 1-$\sigma$ uncertainty&Amplitude&1-$\sigma$ uncertainty\\
(in $\mu$Hz)&(in ppm$^2 \mu$Hz$^{-1}$) & (in ppm$^2 \mu$Hz$^{-1}$)& (in $\mu$Hz) & (in $\mu$Hz)&(in ppm)&(in ppm)\\
\hline
\hline
880.90&1.30&+0.20 / -0.17&3.84&+0.73 / -0.61&2.80&+0.17 / -0.16\\
945.94&1.60&+0.34 / -0.28&2.65&+0.64 / -0.52&2.57&+0.16 / -0.15\\
1011.13&1.43&+0.25 / -0.21&4.16&+0.79 / -0.67&3.06&+0.15 / -0.15\\
1078.80&2.17&+0.33 / -0.29&2.99&+0.45 / -0.39&3.19&+0.14 / -0.13\\
1146.43&2.98&+0.36 / -0.32&3.49&+0.39 / -0.35&4.04&+0.13 / -0.13\\
1214.81&2.43&+0.26 / -0.24&4.79&+0.48 / -0.43&4.28&+0.13 / -0.12\\
1280.09&2.36&+0.27 / -0.24&6.20&+0.68 / -0.61&4.79&+0.13 / -0.12\\
1345.84&2.31&+0.22 / -0.20&6.02&+0.53 / -0.49&4.68&+0.13 / -0.12\\
1414.42&2.18&+0.25 / -0.22&5.16&+0.58 / -0.52&4.20&+0.12 / -0.12\\
1482.32&1.86&+0.21 / -0.19&5.35&+0.60 / -0.54&3.96&+0.12 / -0.12\\
1549.84&1.09&+0.13 / -0.12&7.06&+0.86 / -0.76&3.48&+0.13 / -0.12\\
1620.84&0.86&+0.12 / -0.11&8.05&+1.22 / -1.06&3.29&+0.13 / -0.13\\
1688.45&0.98&+0.15 / -0.13&5.39&+0.93 / -0.80&2.88&+0.13 / -0.13\\
1758.72&0.74&+0.11 / -0.10&7.19&+1.21 / -1.04&2.89&+0.13 / -0.13\\
1822.35&0.47&+0.14 / -0.11&4.05&+1.56 / -1.13&1.74&+0.18 / -0.16\\
1890.94&0.26&+0.08 / -0.06&8.48&+3.11 / -2.28&1.85&+0.19 / -0.17\\
2027.09&0.23&+0.17 / -0.10&2.64&+2.91 / -1.39&0.97&+0.24 / -0.19\\
\hline
\hline
\end{tabular}
\end{table*}

\clearpage

\begin{table*}[!]
\centering
\caption{{\bf Mode heights, mode linewidths and mode amplitude with their associated uncertainties} for KIC 10454113}
\begin{tabular}{c c c c c c c} 
\hline
\hline
Frequency&Mode height & 1-$\sigma$ uncertainty & Linewidth & 1-$\sigma$ uncertainty&Amplitude&1-$\sigma$ uncertainty\\
(in $\mu$Hz)&(in ppm$^2 \mu$Hz$^{-1}$) & (in ppm$^2 \mu$Hz$^{-1}$)& (in $\mu$Hz) & (in $\mu$Hz)&(in ppm)&(in ppm)\\
\hline
\hline
1602.80&0.49&+0.13 / -0.10&1.90&+0.60 / -0.45&1.21&+0.10 / -0.09\\
1706.89&0.62&+0.12 / -0.10&2.24&+0.48 / -0.39&1.48&+0.09 / -0.08\\
1812.46&0.71&+0.11 / -0.09&2.77&+0.46 / -0.40&1.75&+0.08 / -0.08\\
1916.12&1.32&+0.16 / -0.14&2.44&+0.28 / -0.25&2.25&+0.07 / -0.07\\
2018.81&1.32&+0.13 / -0.12&3.42&+0.32 / -0.29&2.66&+0.07 / -0.07\\
2122.76&1.79&+0.15 / -0.14&3.36&+0.26 / -0.24&3.08&+0.07 / -0.07\\
2227.55&1.73&+0.14 / -0.13&3.64&+0.28 / -0.26&3.15&+0.07 / -0.07\\
2333.08&1.79&+0.15 / -0.14&3.52&+0.26 / -0.25&3.15&+0.07 / -0.07\\
2438.94&1.79&+0.16 / -0.14&3.40&+0.27 / -0.25&3.09&+0.07 / -0.07\\
2544.36&1.23&+0.11 / -0.10&4.50&+0.36 / -0.34&2.95&+0.07 / -0.07\\
2649.44&0.76&+0.07 / -0.06&6.11&+0.56 / -0.51&2.70&+0.07 / -0.07\\
2754.89&0.49&+0.06 / -0.05&6.86&+0.88 / -0.78&2.31&+0.08 / -0.08\\
2859.98&0.41&+0.05 / -0.05&6.21&+0.92 / -0.80&2.00&+0.08 / -0.08\\
2967.58&0.22&+0.04 / -0.04&7.11&+1.70 / -1.37&1.55&+0.10 / -0.09\\
3071.12&0.13&+0.03 / -0.03&9.69&+3.18 / -2.39&1.40&+0.12 / -0.11\\
3174.22&0.09&+0.02 / -0.02&12.95&+3.62 / -2.83&1.34&+0.13 / -0.11\\
\hline
\hline
\end{tabular}
\end{table*}

\begin{table*}[!]
\centering
\caption{{\bf Mode heights, mode linewidths and mode amplitude with their associated uncertainties} for KIC 10909629}
\begin{tabular}{c c c c c c c} 
\hline
\hline
Frequency&Mode height & 1-$\sigma$ uncertainty & Linewidth & 1-$\sigma$ uncertainty&Amplitude&1-$\sigma$ uncertainty\\
(in $\mu$Hz)&(in ppm$^2 \mu$Hz$^{-1}$) & (in ppm$^2 \mu$Hz$^{-1}$)& (in $\mu$Hz) & (in $\mu$Hz)&(in ppm)&(in ppm)\\
\hline
\hline
549.55&2.71&+0.85 / -0.65&2.65&+1.09 / -0.77&3.36&+0.34 / -0.31\\
601.13&2.35&+1.09 / -0.74&1.96&+1.32 / -0.79&2.69&+0.40 / -0.35\\
647.42&2.21&+0.54 / -0.44&4.79&+1.64 / -1.22&4.08&+0.34 / -0.32\\
696.49&4.73&+0.16 / -0.15&2.53&+0.32 / -0.28&4.34&+0.25 / -0.24\\
746.71&5.85&+0.91 / -0.79&2.73&+0.44 / -0.38&5.00&+0.22 / -0.21\\
795.86&6.88&+0.90 / -0.79&3.00&+0.38 / -0.34&5.69&+0.21 / -0.20\\
844.24&9.70&+1.16 / -1.03&2.86&+0.33 / -0.30&6.60&+0.21 / -0.20\\
892.62&7.90&+0.96 / -0.86&2.74&+0.31 / -0.28&5.83&+0.20 / -0.19\\
941.30&5.98&+0.78 / -0.69&3.71&+0.50 / -0.44&5.90&+0.21 / -0.20\\
992.82&4.81&+0.52 / -0.47&3.72&+0.42 / -0.37&5.30&+0.20 / -0.20\\
1042.14&3.08&+0.49 / -0.43&5.27&+0.93 / -0.79&5.05&+0.22 / -0.21\\
1095.05&2.63&+0.48 / -0.40&4.41&+0.87 / -0.73&4.27&+0.24 / -0.22\\
1144.31&2.44&+0.50 / -0.42&3.57&+0.80 / -0.66&3.70&+0.24 / -0.22\\
1194.62&1.82&+0.60 / -0.45&3.53&+1.58 / -1.09&3.18&+0.31 / -0.29\\
\hline
\hline
\end{tabular}
\end{table*}

\begin{table*}[!]
\centering
\caption{{\bf Mode heights, mode linewidths and mode amplitude with their associated uncertainties} for KIC 11081729}
\begin{tabular}{c c c c c c c} 
\hline
\hline
Frequency&Mode height & 1-$\sigma$ uncertainty & Linewidth & 1-$\sigma$ uncertainty&Amplitude&1-$\sigma$ uncertainty\\
(in $\mu$Hz)&(in ppm$^2 \mu$Hz$^{-1}$) & (in ppm$^2 \mu$Hz$^{-1}$)& (in $\mu$Hz) & (in $\mu$Hz)&(in ppm)&(in ppm)\\
\hline
\hline
1270.73&0.40&+0.15 / -0.11&2.60&+1.15 / -0.80&1.27&+0.15 / -0.13\\
1357.01&0.71&+0.16 / -0.13&2.57&+0.64 / -0.51&1.69&+0.11 / -0.11\\
1446.98&1.03&+1.01 / -0.51&2.55&+2.37 / -1.23&2.04&+0.11 / -0.10\\
1535.66&0.83&+0.13 / -0.11&4.35&+0.77 / -0.65&2.38&+0.10 / -0.10\\
1626.06&0.98&+0.14 / -0.12&4.89&+0.78 / -0.67&2.75&+0.10 / -0.09\\
1715.69&1.23&+0.13 / -0.11&5.44&+0.58 / -0.52&3.24&+0.09 / -0.08\\
1802.58&1.27&+0.13 / -0.11&5.51&+0.56 / -0.51&3.32&+0.08 / -0.08\\
1893.60&1.02&+0.10 / -0.09&8.07&+0.83 / -0.75&3.60&+0.09 / -0.08\\
1982.21&1.01&+0.10 / -0.09&7.90&+0.80 / -0.73&3.55&+0.09 / -0.08\\
2073.47&0.94&+0.10 / -0.09&7.45&+0.84 / -0.76&3.31&+0.09 / -0.08\\
2165.12&0.72&+0.09 / -0.08&8.02&+1.04 / -0.92&3.01&+0.09 / -0.09\\
2257.06&0.58&+0.08 / -0.07&7.68&+1.12 / -0.97&2.65&+0.09 / -0.09\\
2346.77&0.41&+0.05 / -0.05&10.94&+1.77 / -1.53&2.65&+0.11 / -0.10\\
2433.23&0.30&+0.04 / -0.04&13.68&+2.48 / -2.10&2.52&+0.11 / -0.11\\
2529.25&0.33&+0.07 / -0.06&8.07&+2.02 / -1.61&2.04&+0.12 / -0.11\\
2626.16&0.16&+0.05 / -0.04&12.00&+4.57 / -3.31&1.75&+0.14 / -0.13\\
2724.45&0.12&+0.05 / -0.04&12.24&+6.59 / -4.28&1.53&+0.16 / -0.15\\
2798.25&0.14&+0.09 / -0.06&5.87&+5.52 / -2.84&1.15&+0.18 / -0.15\\
\hline
\hline
\end{tabular}
\end{table*}

\begin{table*}[!]
\centering
\caption{{\bf Mode heights, mode linewidths and mode amplitude with their associated uncertainties} for KIC 12009504}
\begin{tabular}{c c c c c c c} 
\hline
\hline
Frequency&Mode height & 1-$\sigma$ uncertainty & Linewidth & 1-$\sigma$ uncertainty&Amplitude&1-$\sigma$ uncertainty\\
(in $\mu$Hz)&(in ppm$^2 \mu$Hz$^{-1}$) & (in ppm$^2 \mu$Hz$^{-1}$)& (in $\mu$Hz) & (in $\mu$Hz)&(in ppm)&(in ppm)\\
\hline
\hline
1172.10&0.71&+0.58 / -0.32&1.12&+1.21 / -0.58&1.12&+0.20 / -0.17\\
1258.79&1.11&+0.34 / -0.26&1.52&+0.53 / -0.39&1.63&+0.14 / -0.13\\
1345.99&2.43&+0.53 / -0.44&1.07&+0.24 / -0.20&2.02&+0.11 / -0.11\\
1433.86&1.77&+0.35 / -0.29&2.09&+0.47 / -0.38&2.42&+0.11 / -0.11\\
1520.57&2.22&+0.33 / -0.28&2.57&+0.40 / -0.35&3.00&+0.10 / -0.10\\
1606.48&3.58&+0.44 / -0.39&2.19&+0.26 / -0.23&3.51&+0.10 / -0.10\\
1693.79&4.91&+0.56 / -0.50&2.00&+0.22 / -0.20&3.93&+0.10 / -0.10\\
1781.94&5.49&+0.60 / -0.54&2.12&+0.21 / -0.19&4.27&+0.10 / -0.10\\
1870.58&5.11&+0.51 / -0.47&2.26&+0.21 / -0.19&4.25&+0.10 / -0.09\\
1958.69&3.83&+0.40 / -0.37&3.04&+0.31 / -0.28&4.28&+0.10 / -0.09\\
2047.07&1.99&+0.22 / -0.20&4.40&+0.51 / -0.45&3.71&+0.10 / -0.10\\
2135.30&1.68&+0.23 / -0.20&3.33&+0.48 / -0.42&2.97&+0.10 / -0.10\\
2224.00&0.60&+0.09 / -0.08&7.58&+1.30 / -1.11&2.68&+0.12 / -0.11\\
2312.47&0.34&+0.07 / -0.06&9.56&+2.44 / -1.95&2.24&+0.14 / -0.13\\
2405.50&0.20&+0.12 / -0.08&8.25&+6.22 / -3.54&1.63&+0.19 / -0.17\\
2491.33&0.29&+0.11 / -0.08&5.03&+2.66 / -1.74&1.52&+0.17 / -0.16\\
\hline
\hline
\end{tabular}
\end{table*}

\begin{table*}[!]
\centering
\caption{{\bf Mode heights, mode linewidths and mode amplitude with their associated uncertainties} for KIC 12258514}
\begin{tabular}{c c c c c c c} 
\hline
\hline
Frequency&Mode height & 1-$\sigma$ uncertainty & Linewidth & 1-$\sigma$ uncertainty&Amplitude&1-$\sigma$ uncertainty\\
(in $\mu$Hz)&(in ppm$^2 \mu$Hz$^{-1}$) & (in ppm$^2 \mu$Hz$^{-1}$)& (in $\mu$Hz) & (in $\mu$Hz)&(in ppm)&(in ppm)\\
\hline
\hline
997.08&1.48&+0.39 / -0.31&1.18&+0.37 / -0.28&1.65&+0.11 / -0.10\\
1071.81&1.80&+0.28 / -0.24&1.60&+0.26 / -0.22&2.13&+0.09 / -0.09\\
1146.47&2.94&+0.38 / -0.34&1.58&+0.19 / -0.17&2.70&+0.09 / -0.08\\
1219.91&3.73&+0.37 / -0.34&1.99&+0.18 / -0.16&3.42&+0.08 / -0.08\\
1293.08&5.45&+0.54 / -0.49&1.70&+0.14 / -0.13&3.82&+0.09 / -0.08\\
1367.10&8.64&+0.78 / -0.71&1.63&+0.11 / -0.10&4.70&+0.10 / -0.09\\
1442.14&10.23&+0.94 / -0.86&1.54&+0.10 / -0.10&4.98&+0.10 / -0.10\\
1517.28&13.50&+1.21 / -1.11&1.43&+0.09 / -0.08&5.51&+0.11 / -0.11\\
1592.10&7.05&+0.57 / -0.53&2.17&+0.13 / -0.12&4.90&+0.09 / -0.09\\
1666.78&3.97&+0.32 / -0.30&2.69&+0.17 / -0.16&4.09&+0.08 / -0.08\\
1741.74&1.81&+0.15 / -0.14&3.98&+0.28 / -0.26&3.36&+0.07 / -0.07\\
1816.86&0.92&+0.08 / -0.08&5.43&+0.49 / -0.45&2.81&+0.07 / -0.07\\
1893.40&0.67&+0.07 / -0.06&5.13&+0.54 / -0.49&2.32&+0.07 / -0.06\\
1967.82&0.26&+0.03 / -0.03&6.91&+0.96 / -0.84&1.68&+0.08 / -0.08\\
2044.91&0.25&+0.08 / -0.06&3.27&+1.47 / -1.01&1.13&+0.11 / -0.10\\
2117.69&0.09&+0.05 / -0.03&4.29&+2.98 / -1.76&0.78&+0.13 / -0.11\\
\hline
\hline
\end{tabular}
\end{table*}

\begin{table*}[!]
\centering
\caption{{\bf Mode heights, mode linewidths and mode amplitude with their associated uncertainties} for KIC 12317678}
\begin{tabular}{c c c c c c c} 
\hline
\hline
Frequency&Mode height & 1-$\sigma$ uncertainty & Linewidth & 1-$\sigma$ uncertainty&Amplitude&1-$\sigma$ uncertainty\\
(in $\mu$Hz)&(in ppm$^2 \mu$Hz$^{-1}$) & (in ppm$^2 \mu$Hz$^{-1}$)& (in $\mu$Hz) & (in $\mu$Hz)&(in ppm)&(in ppm)\\
\hline
\hline
790.85&2.39&+0.44 / -0.37&1.73&+0.31 / -0.27&2.55&+0.12 / -0.12\\
853.18&2.38&+0.34 / -0.30&2.97&+0.44 / -0.38&3.34&+0.12 / -0.11\\
913.18&2.48&+0.29 / -0.26&3.87&+0.47 / -0.42&3.89&+0.11 / -0.11\\
975.01&3.64&+0.41 / -0.37&3.04&+0.36 / -0.32&4.17&+0.11 / -0.10\\
1038.34&3.21&+0.29 / -0.27&4.06&+0.36 / -0.33&4.52&+0.10 / -0.10\\
1101.93&3.99&+0.37 / -0.34&3.56&+0.32 / -0.29&4.73&+0.10 / -0.10\\
1166.74&4.01&+0.31 / -0.28&4.59&+0.33 / -0.30&5.38&+0.10 / -0.10\\
1230.49&3.65&+0.27 / -0.25&5.25&+0.36 / -0.34&5.49&+0.10 / -0.09\\
1293.75&3.12&+0.23 / -0.21&5.63&+0.39 / -0.36&5.25&+0.09 / -0.09\\
1356.62&2.60&+0.20 / -0.19&6.08&+0.46 / -0.43&4.99&+0.09 / -0.09\\
1420.82&1.64&+0.14 / -0.13&7.07&+0.63 / -0.58&4.27&+0.09 / -0.09\\
1484.68&1.40&+0.13 / -0.12&6.35&+0.65 / -0.59&3.74&+0.09 / -0.09\\
1550.06&1.07&+0.12 / -0.11&6.30&+0.78 / -0.70&3.26&+0.10 / -0.09\\
1613.83&0.62&+0.07 / -0.06&9.35&+1.20 / -1.06&3.01&+0.10 / -0.10\\
1682.67&0.54&+0.12 / -0.10&6.34&+1.66 / -1.31&2.31&+0.11 / -0.11\\
1747.52&0.41&+0.06 / -0.05&8.64&+2.45 / -1.65&2.35&+0.12 / -0.12\\
1814.95&0.25&+0.06 / -0.05&7.61&+2.68 / -1.98&1.73&+0.15 / -0.14\\
1872.55&0.31&+0.22 / -0.13&3.49&+3.64 / -1.78&1.31&+0.19 / -0.16\\
1935.70&0.24&+0.25 / -0.12&3.15&+4.66 / -1.88&1.09&+0.19 / -0.16\\
\hline
\hline
\end{tabular}
\end{table*}

\appendix

\section{Source of systematic errors and their correction}
In this appendix, we mainly focus on internal differences resulting from different fit assumptions, which led to different fitted results.  The observation duration, being common to all fitters, was excluded from the possible source of systematic errors; {in addition since} the observation duration was larger than 30 times the longest mode lifetime \citep[See][]{Chaplin2008a} this source of external systematic error was negligible.   We also note, that the values of mode linewidths and mode heights obtained with MLE are derived from the mean value of the natural logarithm of these parameters, while the Bayesian values are obtained from the median of these parameters \citep{Benomar2009}, resulting in differences between the two statistical values of less than a few percent, thanks to the long observation duration.  

The large deviations between the results obtained leads to an investigation of the possible sources of systematic errors, which are:
\begin{itemize}
\item Bias on the estimated stellar background
\item Bias on the estimated splitting and inclination angle
\item Bias on the assumed mode height ratio
\item Different definition of the mean frequency
\end{itemize}
Hereafter we detail each source of systematic error from a theoretical and practical point of view.  The theoretical understanding of the systematic errors is based on the use of MLE for mode extraction.  Other extracting methods used in this paper derive the mode parameters in a Bayesian framework \citep{Benomar2009,Handberg2011}.  The two methods are obviously not identical since the Bayesian approach used a priori information on the parameters to be fitted which will slant the outcome in a different manner.  We deliberately chose not to take account of these additional systematic errors, which are far more difficult to model than those resulting from using the MLE approach. The correction approach based upon an MLE fit proved its efficacy, showing that there was no need to take into account the Bayesian parameter derivation in the correction scheme.

Table A.1 lists the model characteristics of each fitter whose differences lead to systematic errors in the mode linewidth and height.

\begin{table*}[htbp]
\caption{Model parameters affecting directly the stellar background, mode linewidth and mode height.  The first column provides the fitter name.  The second column provides the mode height ratio between the $l$=0,1,2,3 mode, respectively.  The third column provides the model of the stellar background with the mention of the number of components of the modified Harvey model.  The fourth column provides how the linewidth is assigned to the modes per order.  The last column provides the location of the $l=1$ mode with respect to the $l=0$ mode: Configuration A: $l=0$ mode at the left hand side of the $l=1$ mode;  Configuration B: $l=0$ mode at the right hand side of the $l=1$ mode.  Configuration {\it a} follows the formal order definition.}             
\label{tab_summary2}      
\centering                          
\begin{tabular}{c c c c c c}        
\hline                 
\hline     
Fitter&Mode&Background& Linewidth&Mode\\
&ratio&&configuration&configuration\\
\hline  
\hline                           
Appourchaux, IAS&1.0 / 1.5 /0.5 / 0.0&One Harvey+white noise&Single per $\Delta \nu$& A\\
Howe, BIR&1.0 /1.5 / 0.5 / 0.2&Two Harvey+white noise&Single per true order& {\it a}\\
Davies, BIR&Free&One Harvey+white noise&Single per $\Delta \nu$&A\\
Antia, TAT&Free&One Harvey+white noise&Single per $\Delta \nu$&A\\
R\'egulo, IAC&Free&One Harvey + power law+white noise&Single per $\Delta \nu$&A\\
Campante, BIR&1.0 / 1.49 / 0.5 / 0.0&One Harvey+white noise&Single per $\Delta \nu$&B\\
Benomar, SYD&Free&One Harvey+white noise&Single per $\Delta \nu$&A, B\\
Handberg, AAU&1.0 /Free/ 0.5 / 0.0&One Harvey+white noise&Single per $\Delta \nu$&A\\
\hline  
\hline  
\end{tabular}
\label{methods2}
\end{table*}

\begin{figure*}[htbp]
\centering
\hspace{1.5cm}
\hbox{
\includegraphics[width=5.5 cm,angle=90]{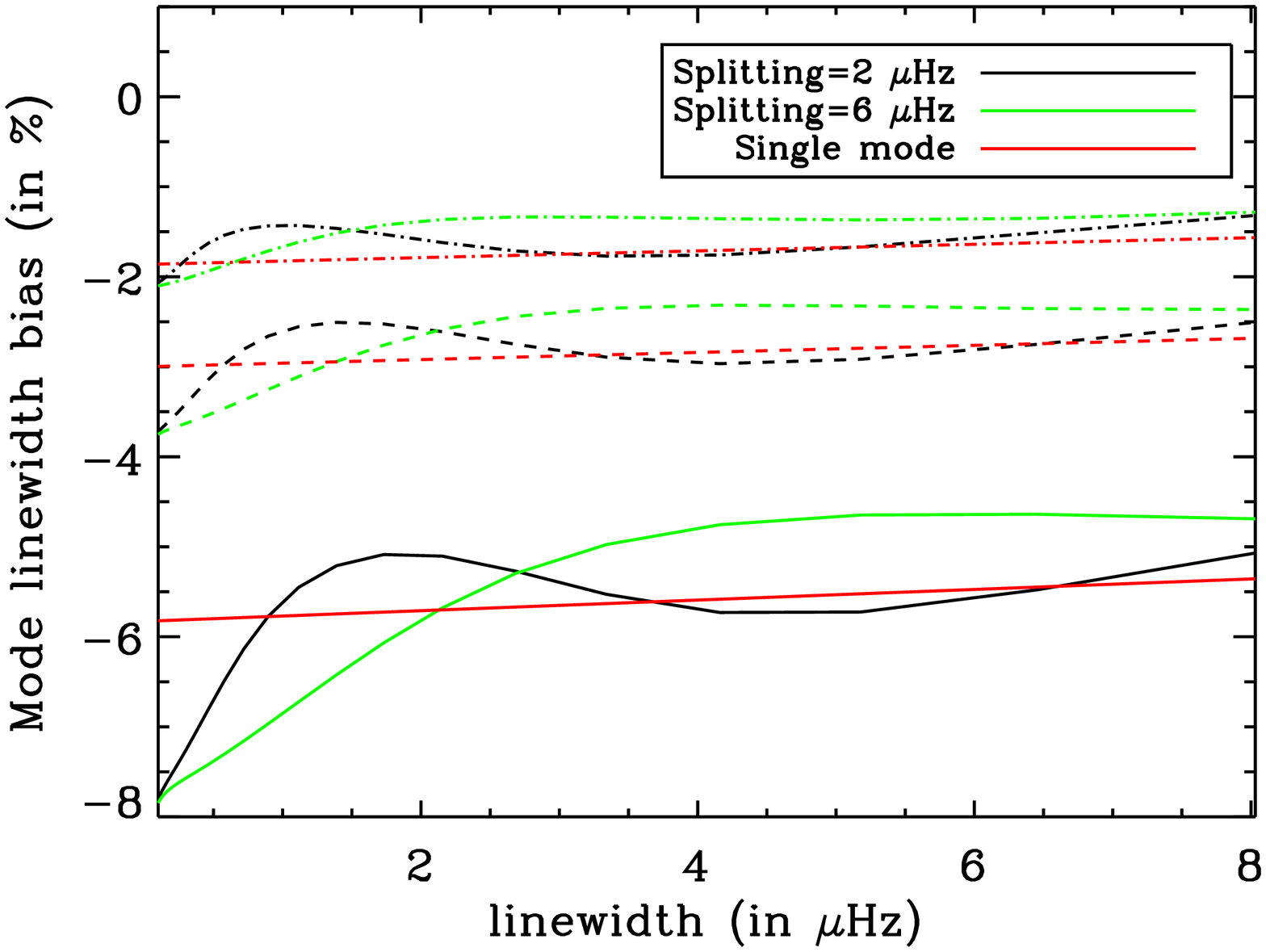}
\hspace{2.0cm}\includegraphics[width=5.5 cm,angle=90]{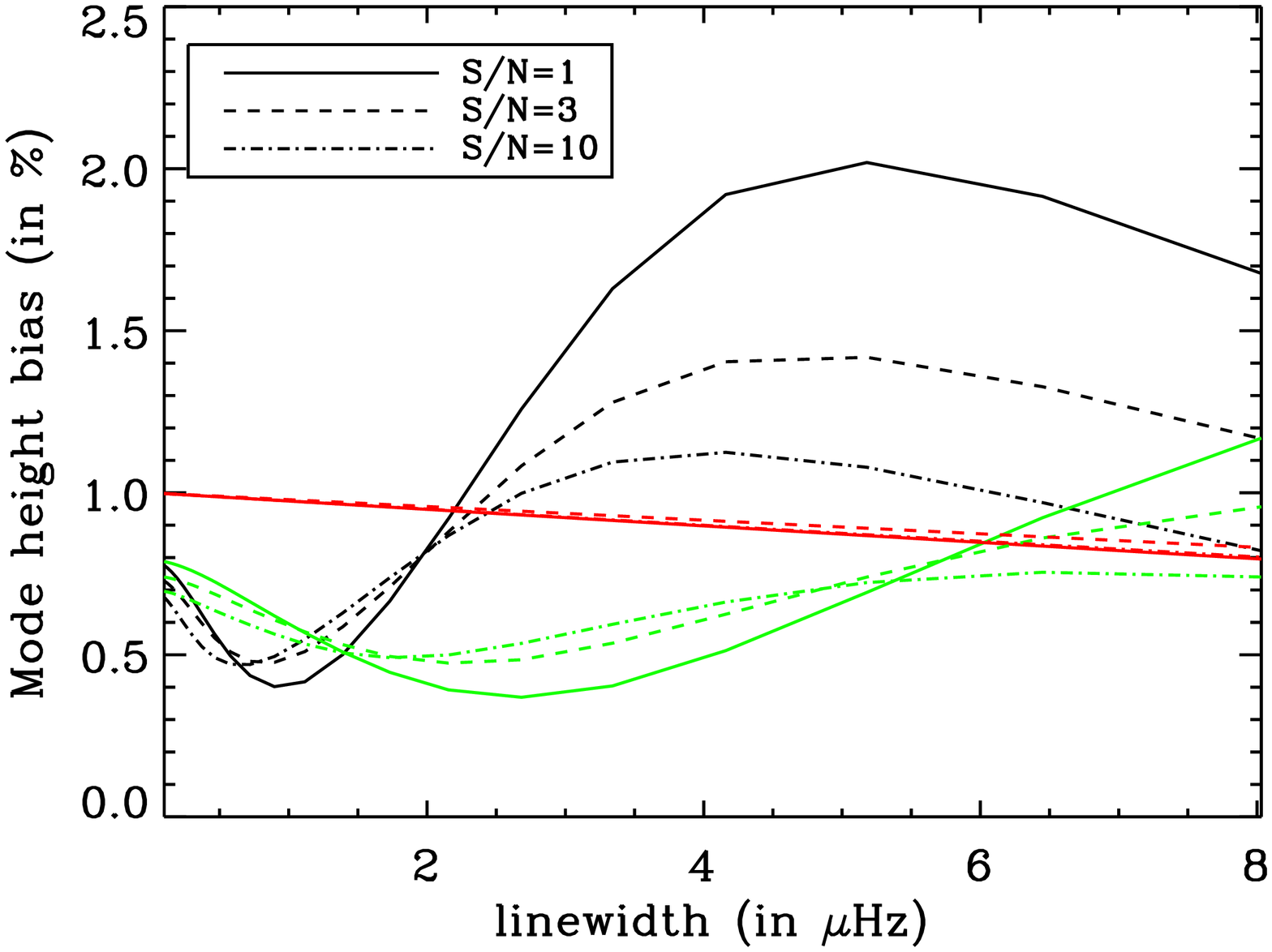}
}
\hbox{
\includegraphics[width=5.5 cm,angle=90]{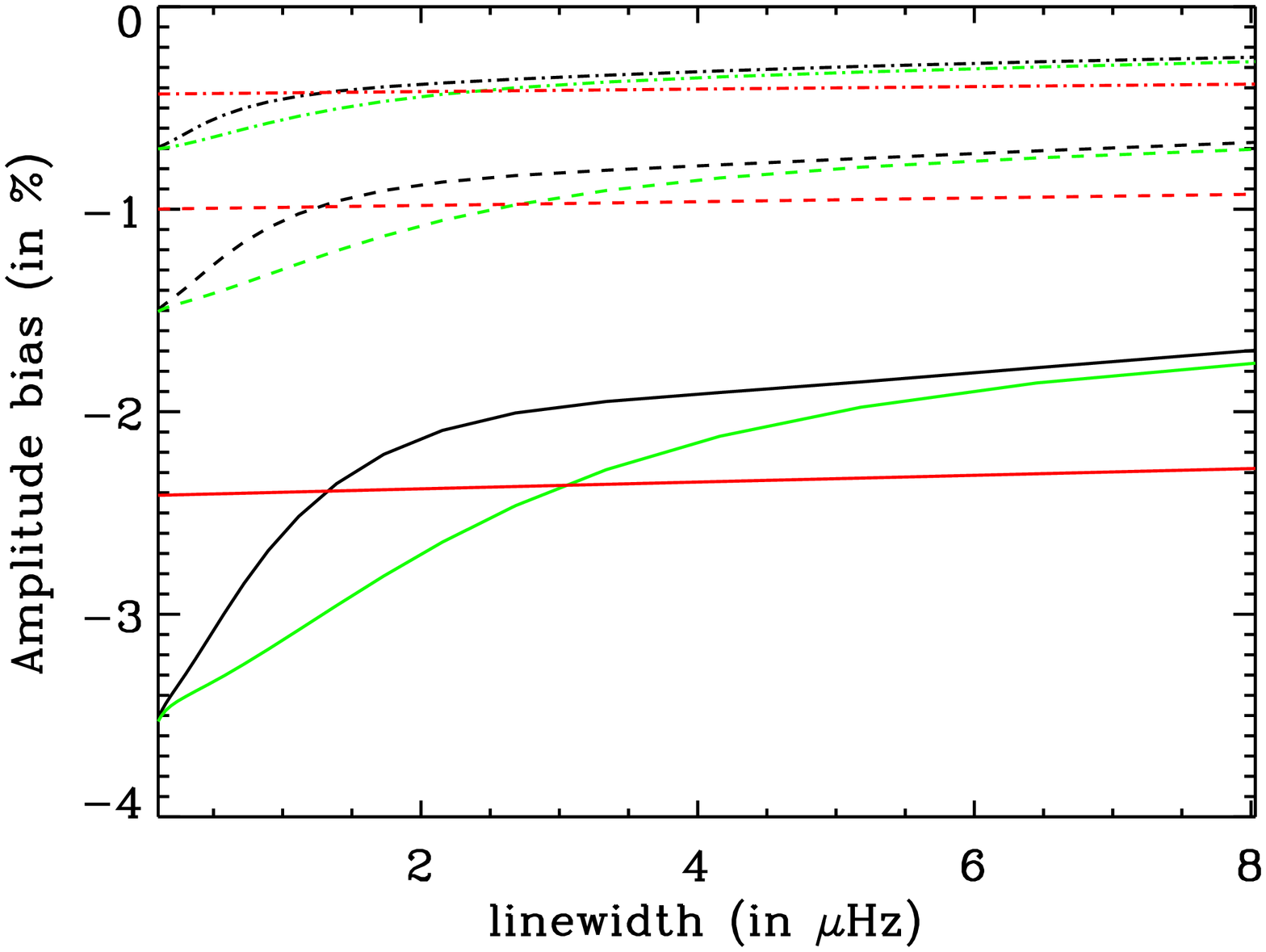}
\hspace{2.4cm}
\parbox[b]{70mm}{
\caption{Mode parameter biases as a function of linewidth for a stellar background bias of +1\%, for several signal-to-noise ratios and for three different kinds of profiles which are: a triplet $l=0,1,2,3$ with a splitting of 2 $\mu$Hz with the mode pairs $l=0-2$ and $l=1-3$ separated by 120 $\mu$Hz (black line); a triplet with a splitting of 6 $\mu$Hz with the mode pair $l=0-2$ and $l=1-3$ separated by 120 $\mu$Hz (green line); and a single mode profile (red line). (Top, left) For mode linewidth, (Top, right) For mode height ratio, (Bottom, left) For mode amplitude \vspace{8.75mm}}
}}
\label{ratio1}
\end{figure*}

\subsection{Effect of an incorrect estimation of the stellar background}
One might naively assume that an underestimated stellar background would lead to an overestimated mode height compensating the reduced stellar background.  This naive approach would provide an underestimated linewidth.  Fortunately, the effect of an incorrect estimation of the stellar background on the mode can be analytically calculated which provides a counter intuitive result.  The relations between the mode-linewidth, mode-height and mode-amplitude systematic errors which lead to an incorrect estimated stellar background are given in Appendix B.2. In short, an underestimation of the stellar background will lead to an underestimation of the mode height and an overestimation of the linewidth.  In addition, the lower the signal-to-noise ratio the higher the systematic error on mode linewidth and height (the signal-to-noise ratio is the ratio of mode height to the stellar background).  The systematic errors were also numerically derived using the technique of the fit of a single (not many) profile of \citet{TT2005b}, i.e. by simulation of the MLE for a single profile fitted by a profile with a different stellar background. Figure~A.1 shows the results of the numerical calculation of the mode linewidth, mode height and mode amplitudes biases as a function of mode linewidth, for several signal-to-noise ratio.  Assuming that the background is overestimated by 1\%, then when the signal-to-noise ratio is about 10, the mode linewidth is typically underestimated by about 1.75\%, while the mode height is overestimated by about 0.75 \%.  The sign of the mode-linewidth systematic error  is opposite to that of the mode-height systematic error, such that they almost compensate each other when computing the mode-amplitude systematic errors. It means that an overestimation of the stellar background of 1\% will provide a mode amplitude underestimated by 0.5\%.  




When the signal-to-noise ratio is close to 1, the mode-linewidth systematic errors increases to about -8\% while the mode-height systematic errors are only 2\%.  This is what was observed in some stars for which we could not understand why the systematic error on the linewidth was so much larger than for the mode height.  As a consequence, care should be taken when comparing amplitudes of different fitters when the signal-to-noise ratio is low.  

We also note that when the signal-to-noise ratio increases, the mode-amplitude systematic error gets smaller and smaller, i.e.  a 1\% stellar background overestimation would provide a 0.1\% mode amplitude underestimation.  This explains why the mode amplitude differences between different fits may be very small but not necessarily absent when the signal-to-noise ratio is high.  

The differences between the different fitted stellar backgrounds can vary between 30\% in the worst case to about 10\% in the best cases.  The differences are mainly due to the model of the stellar background used having one or several components as shown in Table A.1.  It must be pointed out that since the background bias varies with frequency, this will also introduce a varying effect on the relative variation of the mode linewidth and mode height.  


\begin{figure*}[htbp]
\centering
\hspace{1.5cm}
\hbox{
\includegraphics[width=5.5 cm,angle=90]{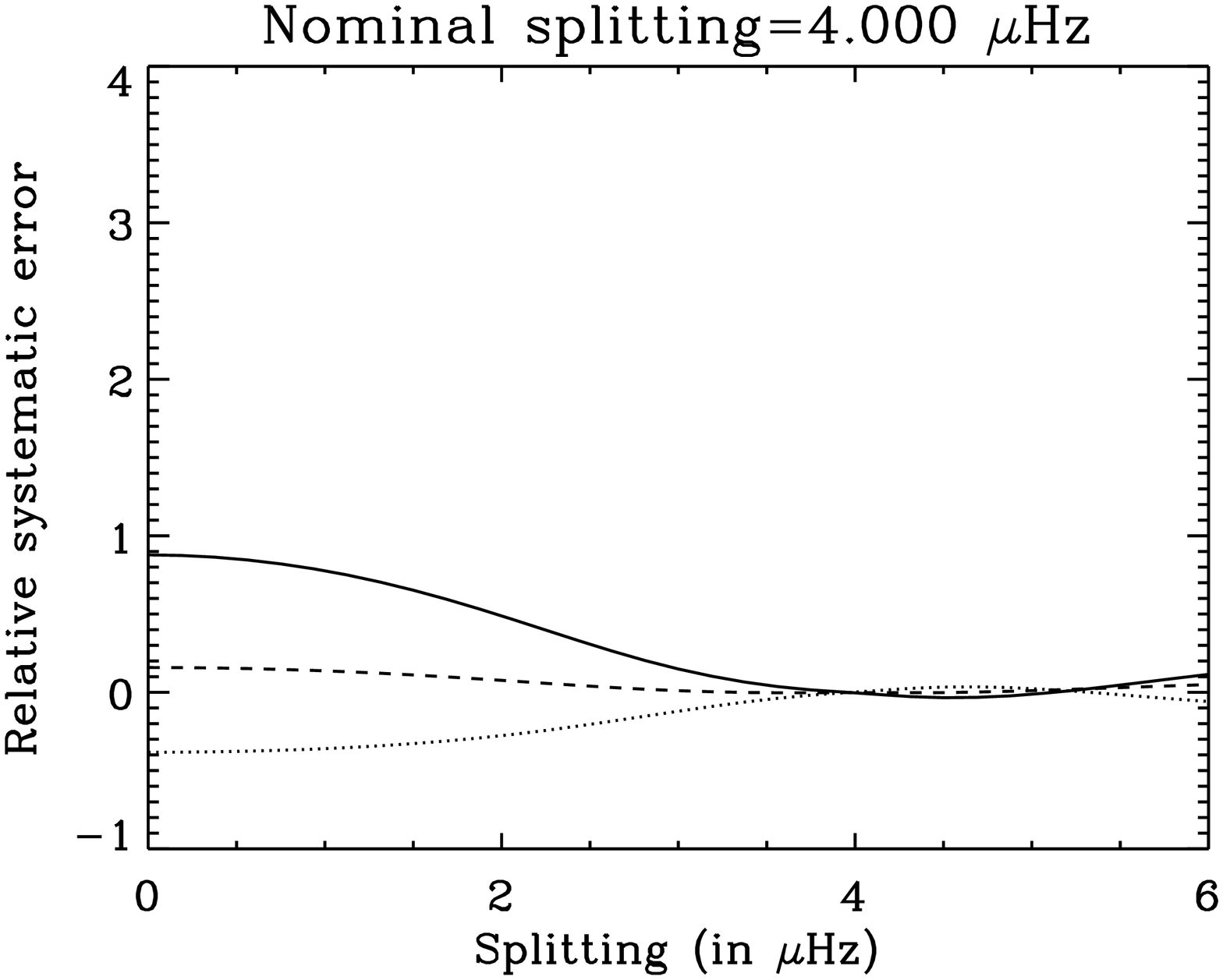}
\hspace{2.0cm}\includegraphics[width=5.5 cm,angle=90]{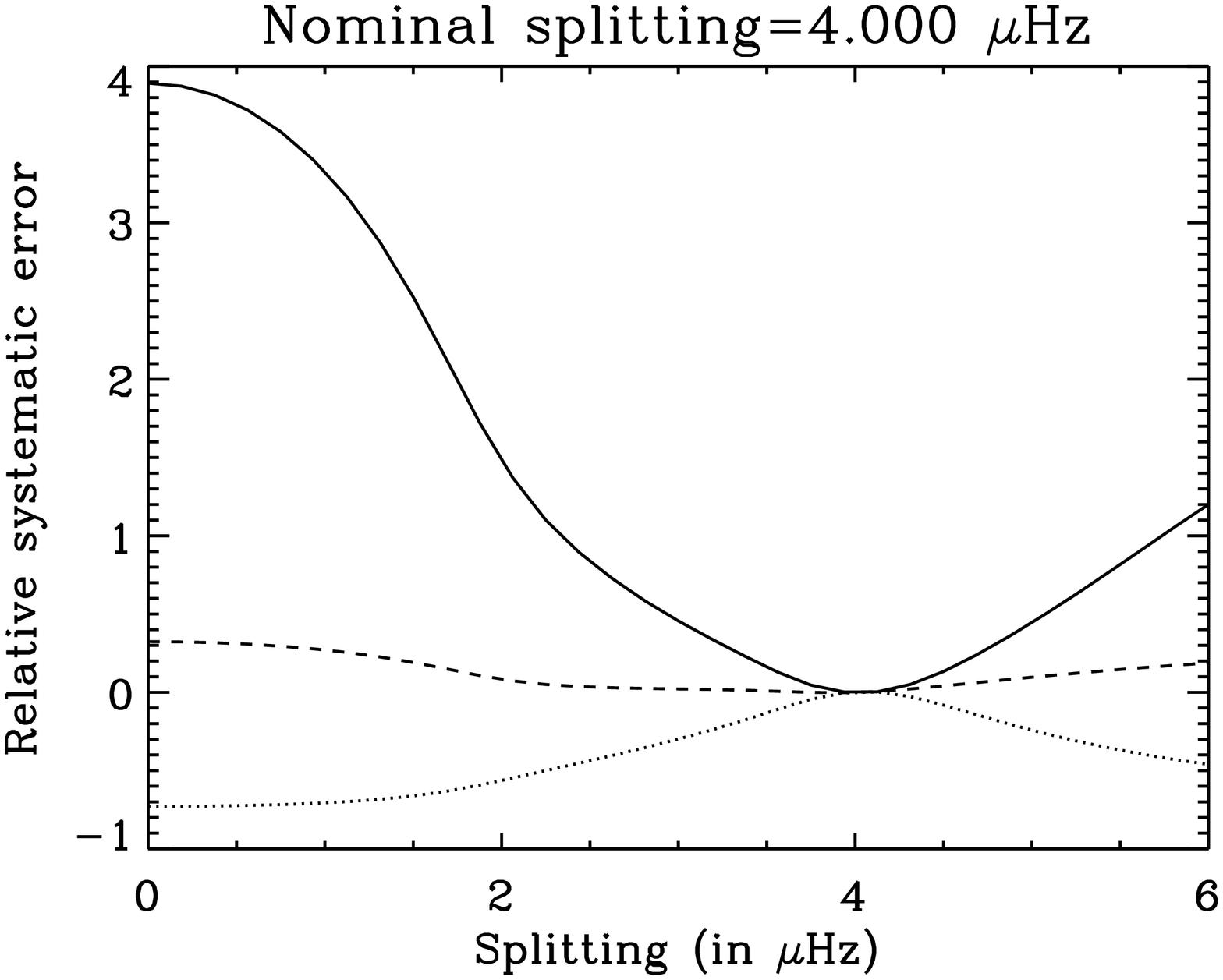}
}
\caption{Relative systematic error of three parameters of the $l=0-2$ and $l=1-3$ modes as a function of the fitted splitting: mode linewidth (solid line),  mode height (dotted line), mode amplitude (dashed line) for a signal-to-noise ratio of 1,  a nominal splitting of 4 $\mu$Hz and an inclination angle of 90 degrees.  (Left) The multiplet has a nominal linewidth of 6 $\mu$Hz.  (Right) The multiplet has a nominal linewidth of 3 $\mu$Hz.}
\label{splitting_1}
\end{figure*}


\begin{figure*}[htbp]
\centering
\hspace{1.5cm}
\hbox{
\includegraphics[width=5.5 cm,angle=90]{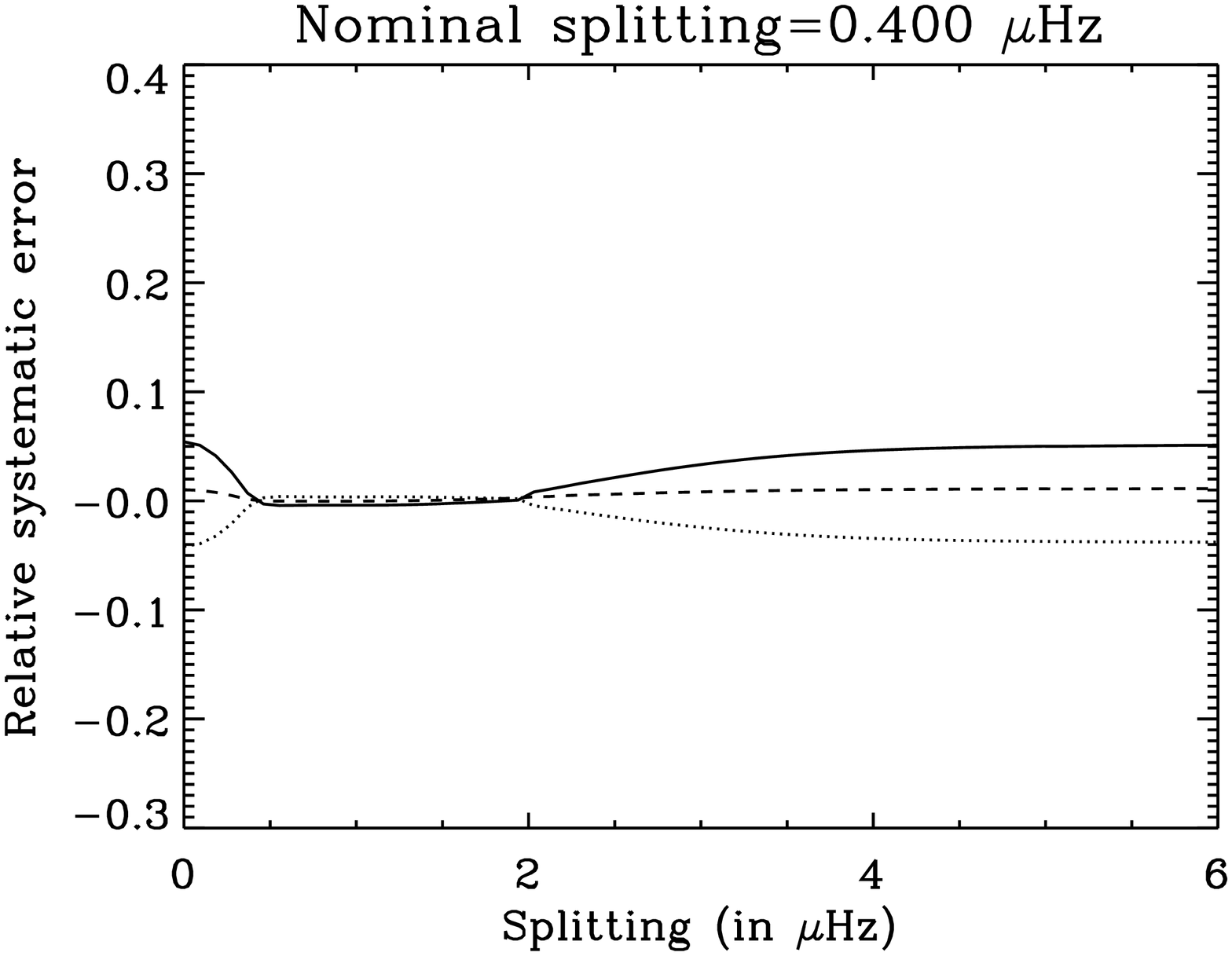}
\hspace{2.0cm}\includegraphics[width=5.5 cm,angle=90]{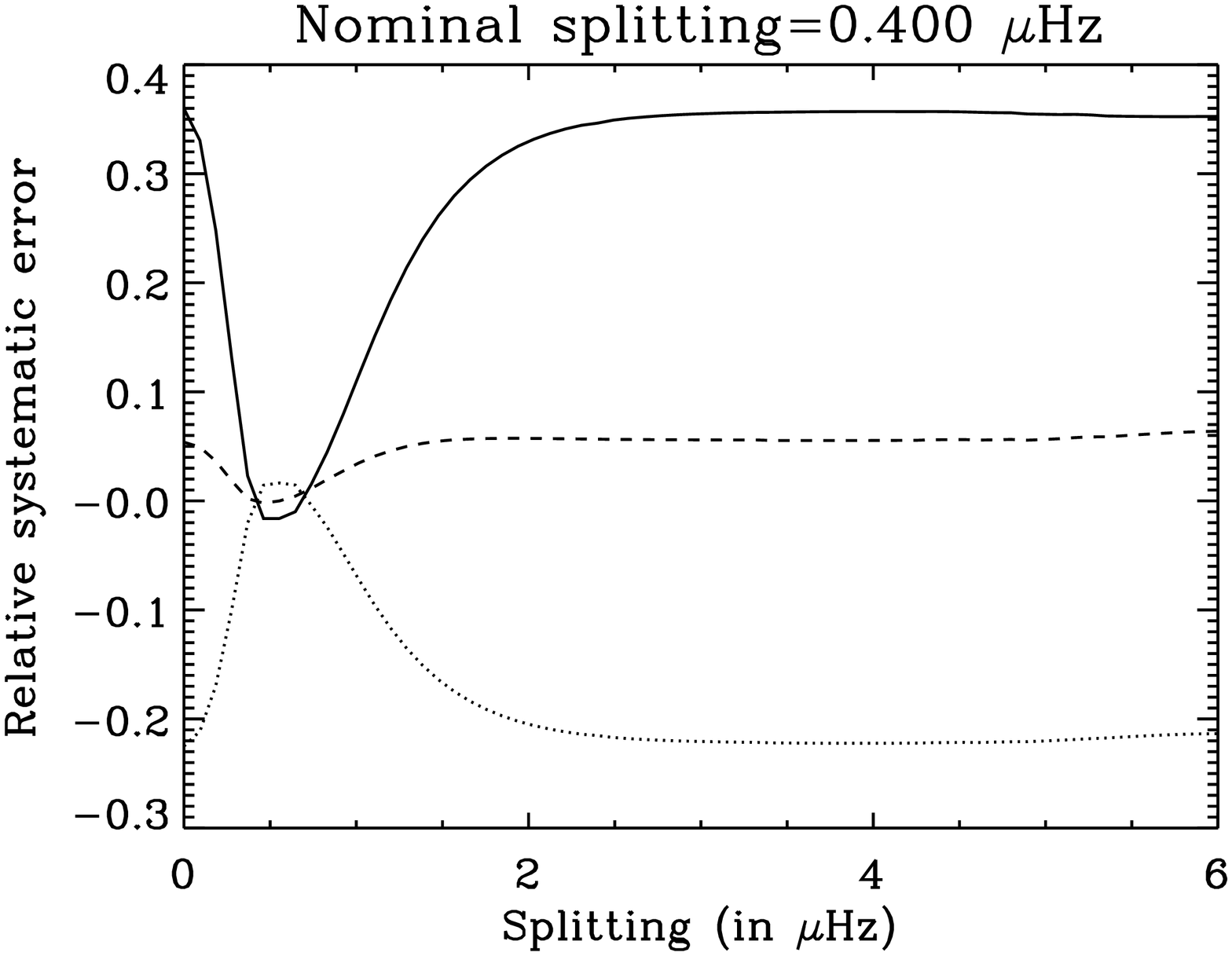}
}
\caption{Relative systematic error of three parameters of the $l=0-2$ and $l=1-3$ modes as a function of the fitted splitting: mode linewidth (solid line),  mode height (dotted line), mode amplitude (dashed line) for a signal-to-noise ratio of 1, a nominal splitting of 0.4 $\mu$Hz and an inclination angle of 90 degrees.  (Left) The multiplet has a nominal linewidth of 3 $\mu$Hz.  (Right) The multiplet has a nominal linewidth of 1 $\mu$Hz.}
\label{splitting_2}
\end{figure*}


\begin{figure*}[htbp]
\centering
\hspace{1.5cm}
\hbox{
\includegraphics[width=5.5 cm,angle=90]{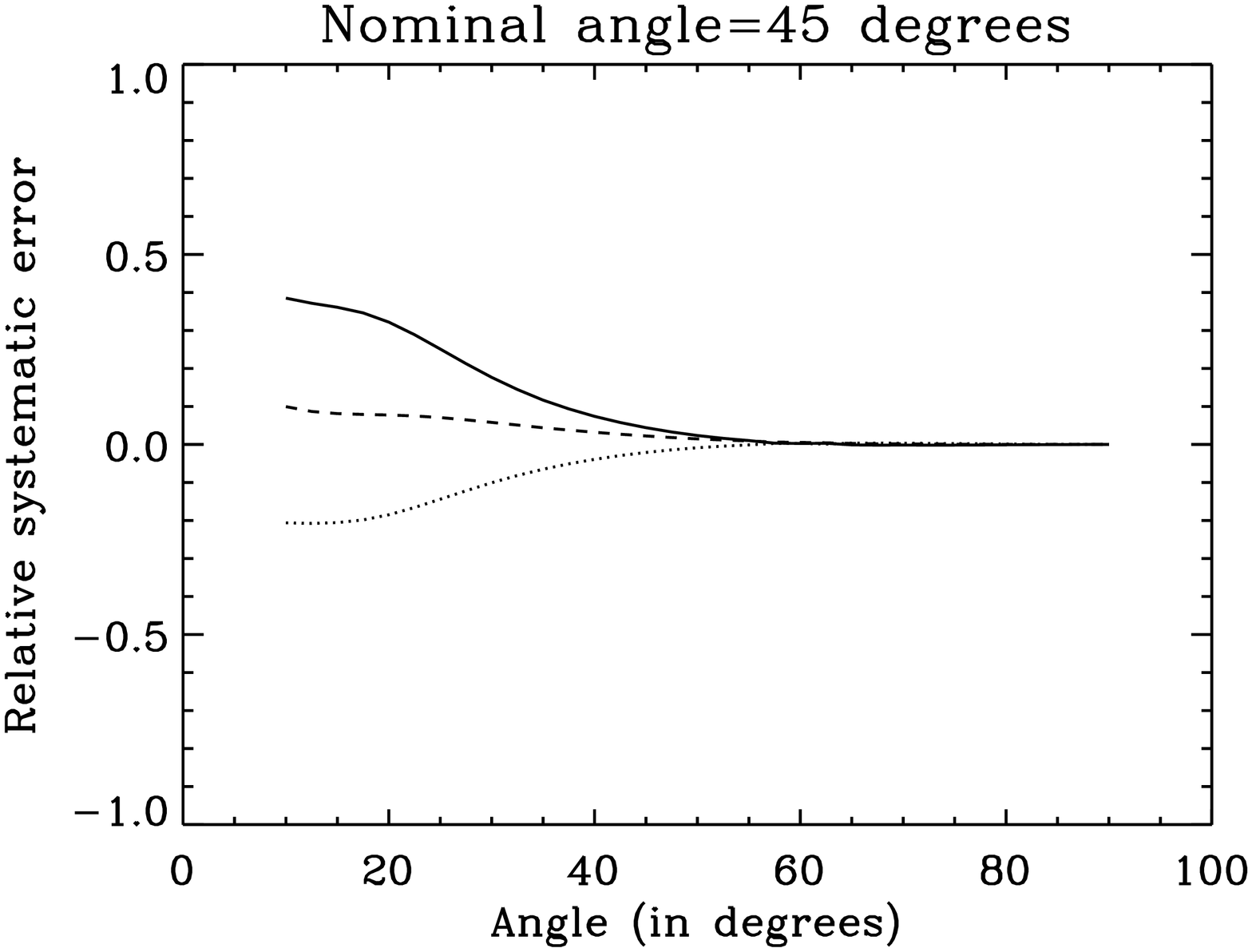}
\hspace{2.0cm}
\includegraphics[width=5.5 cm,angle=90]{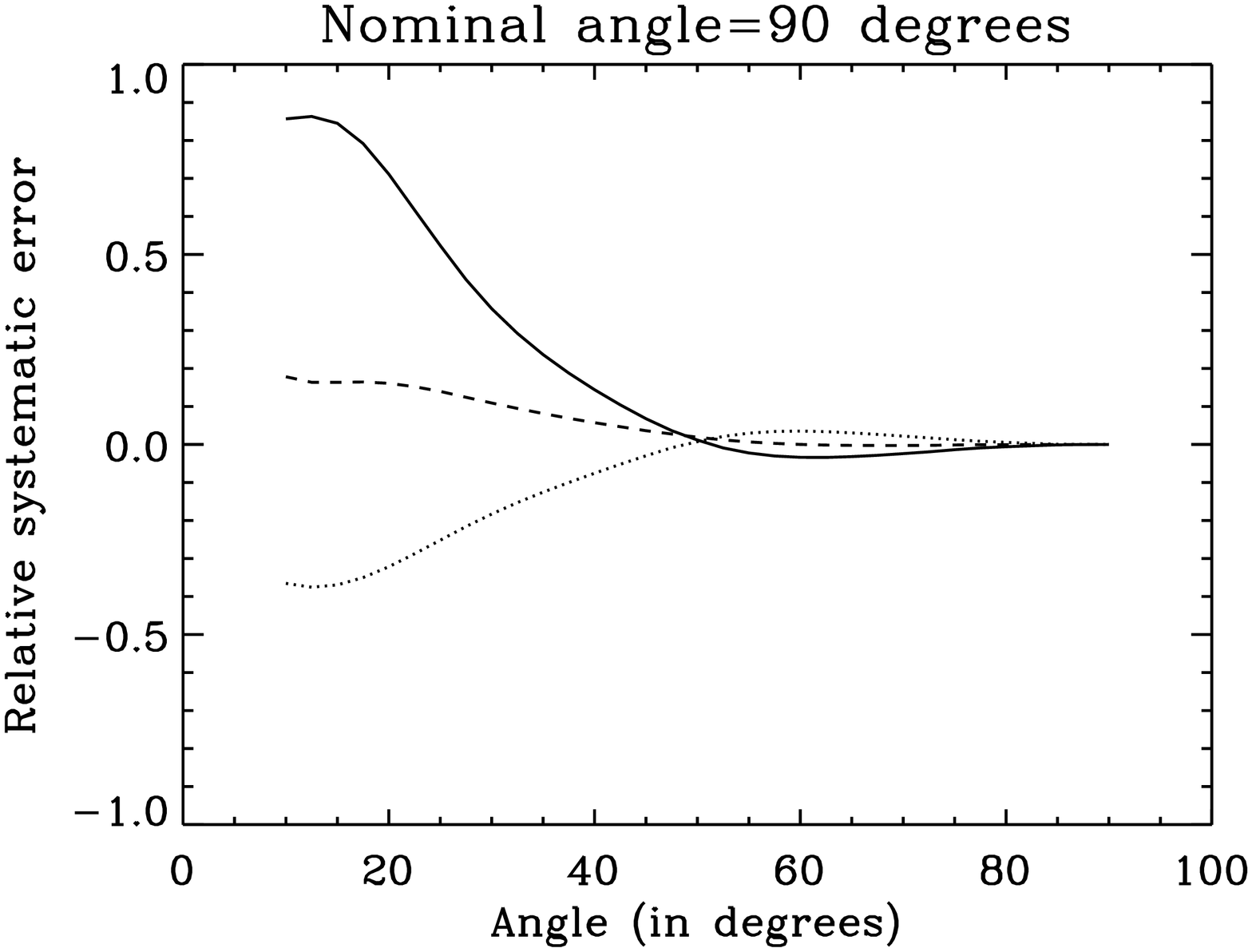}
}
\caption{Relative systematic error on three parameters of the $l=0-2$ and $l=1-3$ multiplet as function of the fitted inclination angle: mode linewidth (solid line),  mode height (dotted line), mode amplitude (dashed line) for a signal-to-noise ratio of 1. The multiplet has a nominal linewidth of 6 $\mu$Hz and a nominal splitting of 4 $\mu$Hz, with a nominal inclination angle of 45 degrees (Left), and a nominal inclination angle of 90 degrees (Right).}
\label{splitting_3}
\end{figure*}

\subsection{Effect of a different estimated splitting and inclination angle}
The effect of a different estimated splitting can also be simply understood when, for example, a null splitting is fitted when the modes are indeed genuinely split: this will lead to an overestimation of the linewidth.  The analytical calculations of the effect are detailed in Appendix B.3 but do not lead to simple useful formula; the only result being that second order effects dominate.  This is the reason why we simulated an $l=0-2$ pair and an $l=1-3$ pair with a symmetrical rotational splitting and with mode height ratio of 1 / 1.5 / 0.5 / 0.05 for the $l=0,1,2,3$, respectively.  We then applied the one-fit method of \citet{TT2005b} for computing the systematic errors. Figures \ref{splitting_1} to \ref{splitting_3} show results for three different cases.

In Fig.~\ref{splitting_1}, for a 90-degree inclination (perpendicular to the line of sight), when the nominal splitting is of the order of the linewidth, the fitted linewidth with a null fitted splitting will be typically overestimated by the value in excess of a few times the nominal splitting.  In Fig.~\ref{splitting_2}, for a 90-degree inclination (perpendicular to the line of sight), when the nominal splitting is about 10\% to 20\% of the linewidth, the fitted linewidth with a null fitted splitting will be typically overestimated by the value smaller than the nominal splitting.  We also note that the dependence of the linewidth bias on to the splitting is quadratic close to the nominal splitting but {\it not} exactly at the nominal splitting.  At the local minimum, the systematic error  is slightly negative, although the systematic error at the nominal splitting is null.

In Fig.~\ref{splitting_3}, we also simulated a varying inclination,  assuming that the projected splitting was kept constant ($\nu_s^{\rm fit} \sin \alpha^{\rm fit}=\nu_s \sin \alpha$).  The reason for using such a formulation is that the fit preferentially converges towards a value for which $\nu_s^{\rm fit} \sin \alpha^{\rm fit}$ is constant as shown by \citet{Ballot2006}. The systematic errors remain small when the fitted inclination angle is greater than 45 degrees.  In addition the systematic errors decrease with the inclination angle.  Therefore a correction is clearly necessary when the fitted inclination angles are less than 45 degrees while the fitted reference inclination angle is above 45 degrees, i.e. a difference of about 45 degrees will produce large systematic errors in the mode parameters.

If one assumes the same splitting and inclination angle for all modes, the splitting and angle systematic errors introduce a frequency dependence on the relative variation of the mode linewidth and mode height when compared to the reference fit values.  The frequency dependences on the relative errors are larger at smaller linewidths, emphasising the need for a proper estimation of the splitting at the low-frequency part of the spectrum.

\subsection{Effect of different assumptions on mode height ratio}
\begin{figure*}[htbp]
\centering
\vbox{
\hspace{1.5cm}
\hbox{\includegraphics[width=5.5 cm,angle=90]{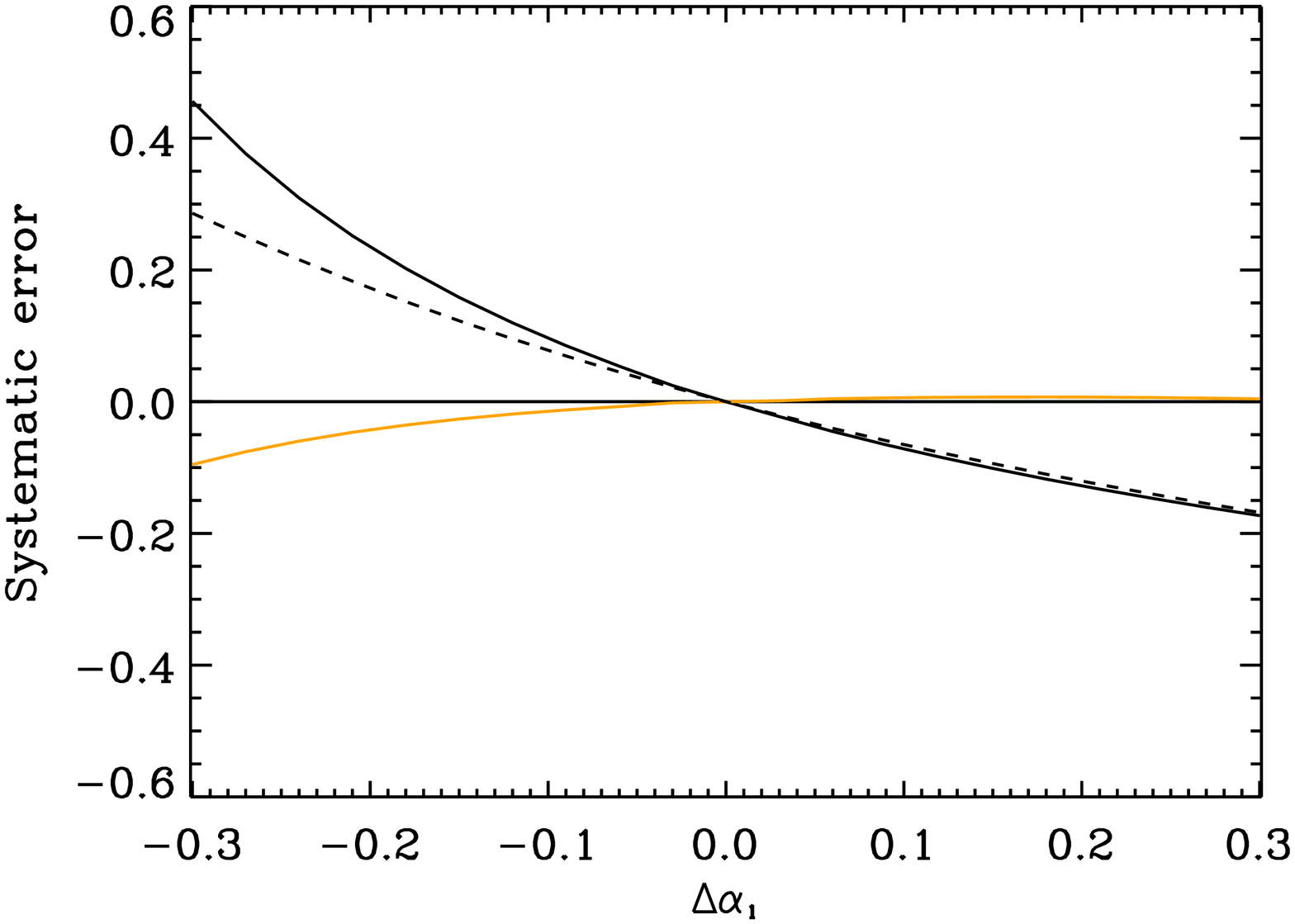}
\hspace{2.0cm}
\includegraphics[width=5.5 cm,angle=90]{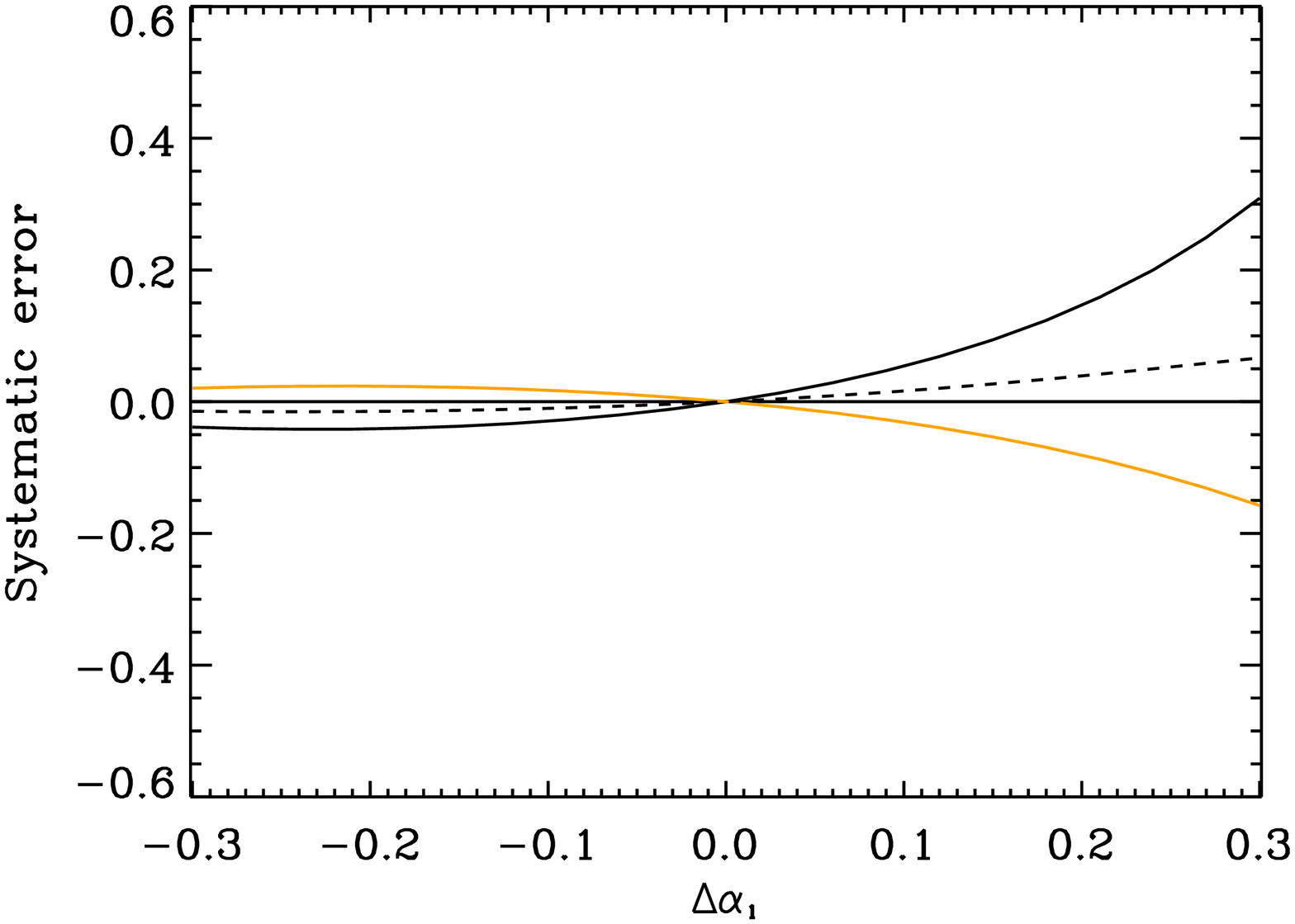}
}
\hbox{
\includegraphics[width=5.5 cm,angle=90]{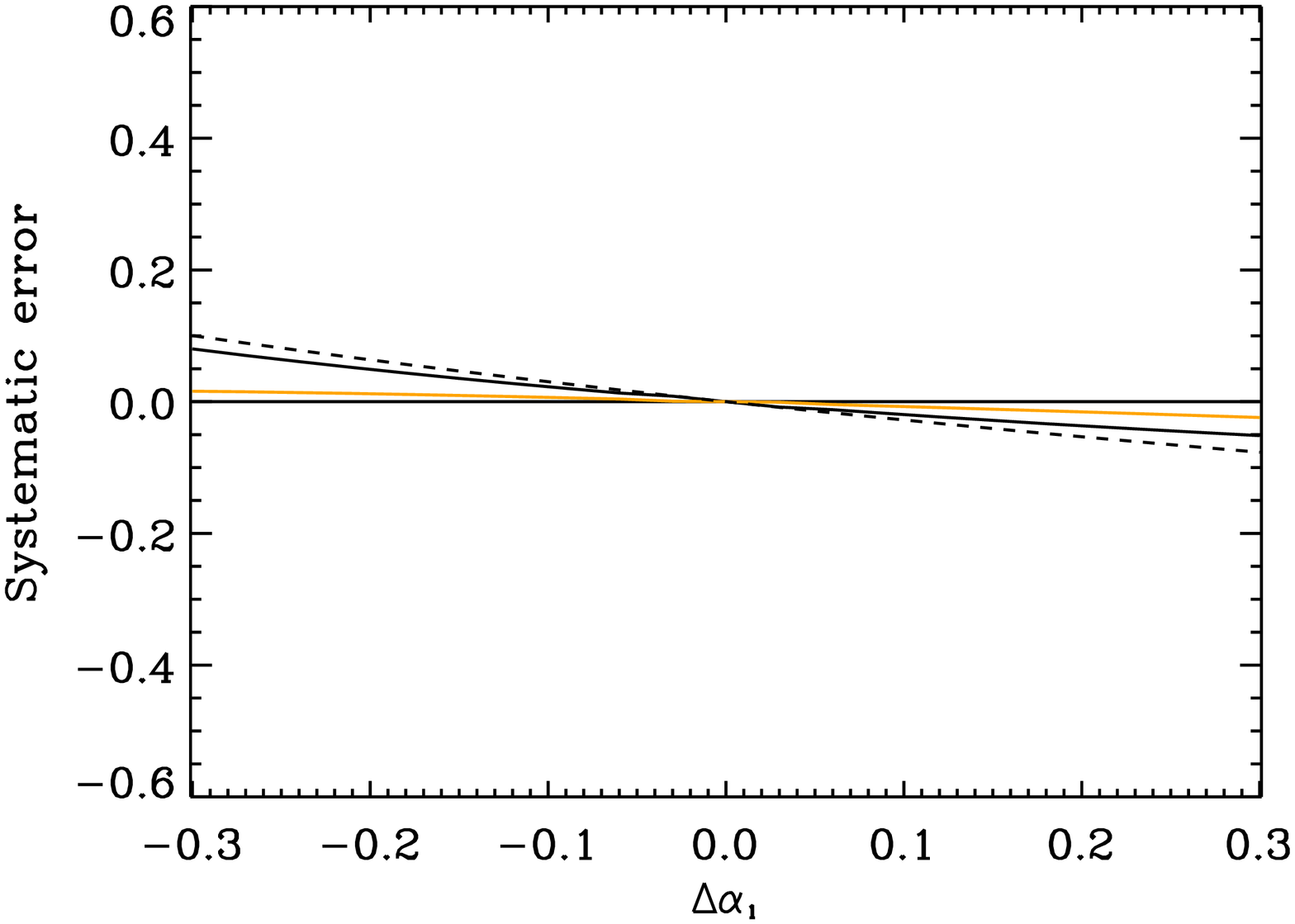}
\hspace{2.cm}
\includegraphics[width=5.5 cm,angle=90]{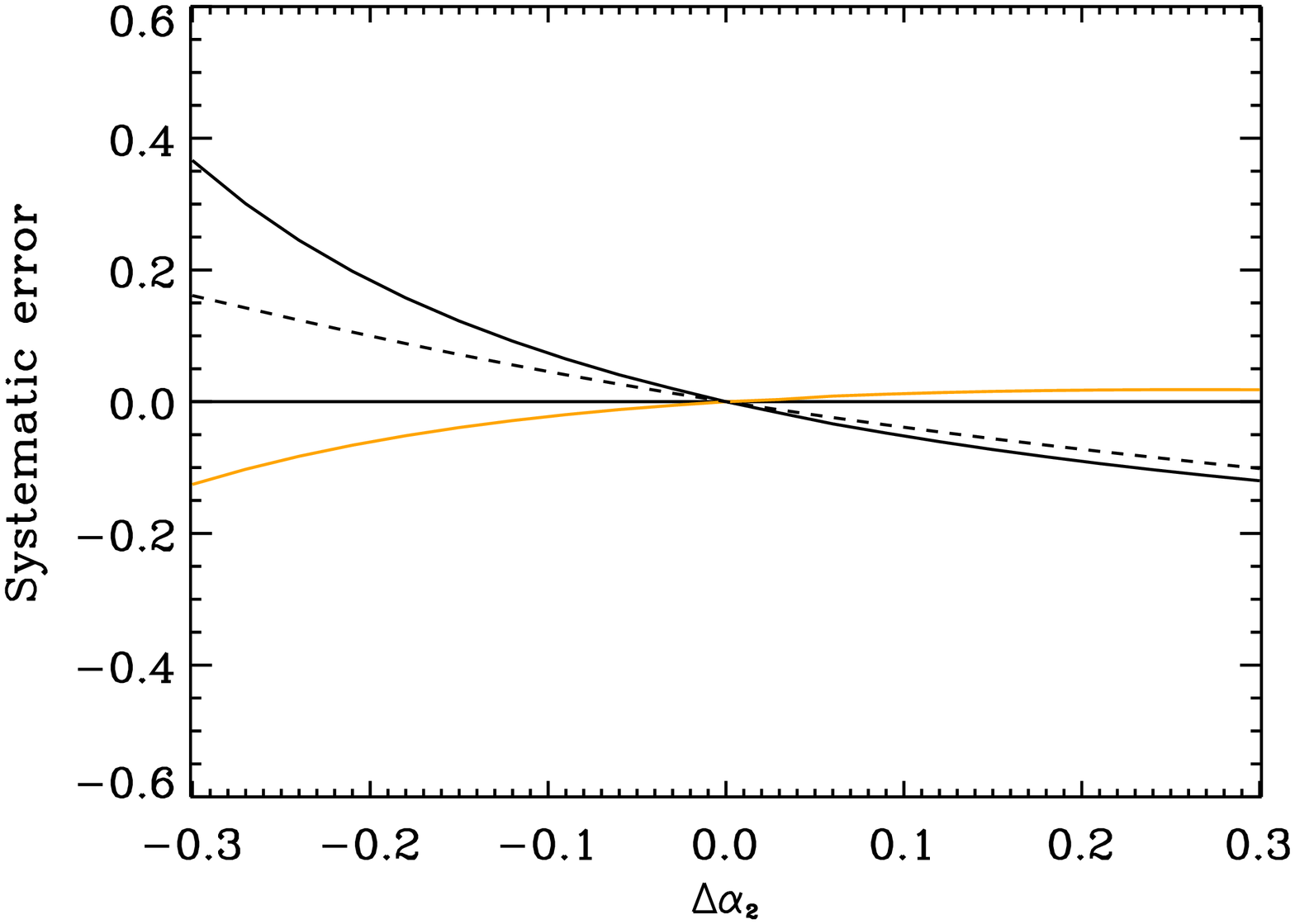}
}}
\caption{Relative systematic error as a function of the deviation to the reference mode visibility for the mode height (black line) and linewidth (orange line) for a set of $l=0,1,2$ modes with a 2-$\mu$Hz linewidth, and a signal-to-noise ratio 10 for the $l=0$ mode.  The continuous lines are the systematic error computed using the method of \citet{TT2005b}.  The dashed line is the analytical systematic error on mode height derived assuming there is no systematic error on the linewidth derived from Eq.~(\ref{ratio_sys}).  (Top left) The deviations to the reference mode visibilities are identical for $l=1$ and $l=2$.  (Bottom left) The deviations to the reference mode visibilities are {\it of opposite sign} for $l=1$ and $l=2$.  (Top right) The deviation to the reference mode visibilities is zero for $l=2$.  (Bottom right) The deviation to the reference mode visibilities is zero for $l=1$.}
\label{mode_ratio}
\end{figure*}

Some fitters also allow the mode height ratio to vary, which will also lead, when compared to the results obtained with a notional mode height ratio, to some systematic error in the measured linewidth and mode height. Neglecting the systematic error on mode linewidth, we analytically computed the systematic error on mode height (see Appendix B.4).  We also applied the one-fit method of \citet{TT2005b} for computing the systematic errors.  Figure~\ref{mode_ratio} shows the results of the analytical and numerical calculations of the systematic errors on mode linewidth, mode height and mode amplitudes biases as a function of deviations to the reference mode height ratio for a set of $l=0,1,2$ modes with a 2-$\mu$Hz linewidth, and a signal-to-noise ratio of 10 for the $l$=0 mode.

To first order, the impact of a different mode height ratio on the linewidth is rather small and typically of the order of 10\%.  On the other hand, the impact of a different mode height ratio is much larger on the mode height, being close to 40\% in the worst case.  In addition, the systematic error  on the mode height ratio will produce a non-negligible systematic error on the mode amplitude.  \citet{Ballot2011} showed, using adiabatic computations, that the mode height ratio varies with the effective temperature, surface gravity and metallicity.  For the stars listed in this study, using the work of \citet{Ballot2011} and the stellar parameters provided by \citet{Bruntt2011} we derived that the mode height ratio for $l=1$, $l=2$, $l=3$ is 1.5, 0.52, 0.025 varying by 0.02, 0.02, 0.004, respectively.  

As shown by \citet{Salabert2011} for the solar case, the measured ratios agree very well with the theoretical ratios to within 0.06, and do not vary with time.  Such a deviation of 0.06 would result in systematic errors not larger than 4\% for the mode height, a value that can also be derived from Eq.~(\ref{ratio_sys}).  The values derived above from \citet{Ballot2011} would provide even a smaller systematic error not larger than 2\%.  Larger deviations of stellar mode ratio with respect to the value given above were also observed (D.Salabert, private communication, 2013).  At the time of writing, given that the variation of the mode height ratio is primarily geometrical, there is very little variation with mode frequency. Then, in order to achieve a repeatable measurement, it is advised to keep the mode height ratio at reference values of 1.5, 0.5, 0.05 for $l=1$, $l=2$ and $l=3$, respectively.


\subsection{Effect of a different definition of mean frequency}
By definition of the model, the linewidth is set to be common for modes that are within half a large separation.  There are two possible configurations defining for which modes the linewidth will be common: configuration A for which the $l=0,2$ mode pair is at the {\it left} of the $l=1$ mode; and configuration B for which the $l=0,2$ mode pair is at the {\it right} of the $l=1$ mode.
As a result, the mean frequency will differ for the different configurations.  
For configuration A, the mean frequency is $\nu_{n,0}+\Delta \nu / 4$, while for configuration B, it is $\nu_{n,0}-\Delta \nu / 4$.  Therefore depending on how the fits are performed there could be up to a $\Delta \nu / 2$ difference in the mean frequency.  Since it is assumed that the mode linewidth only depends upon mode frequency, this would directly produce a systematic error that would be proportional to the linewidth frequency derivative and the frequency difference.  

\begin{figure*}[htbp]
\centering
\hspace{1.5cm}
\hbox{\includegraphics[width=5.5 cm,angle=90]{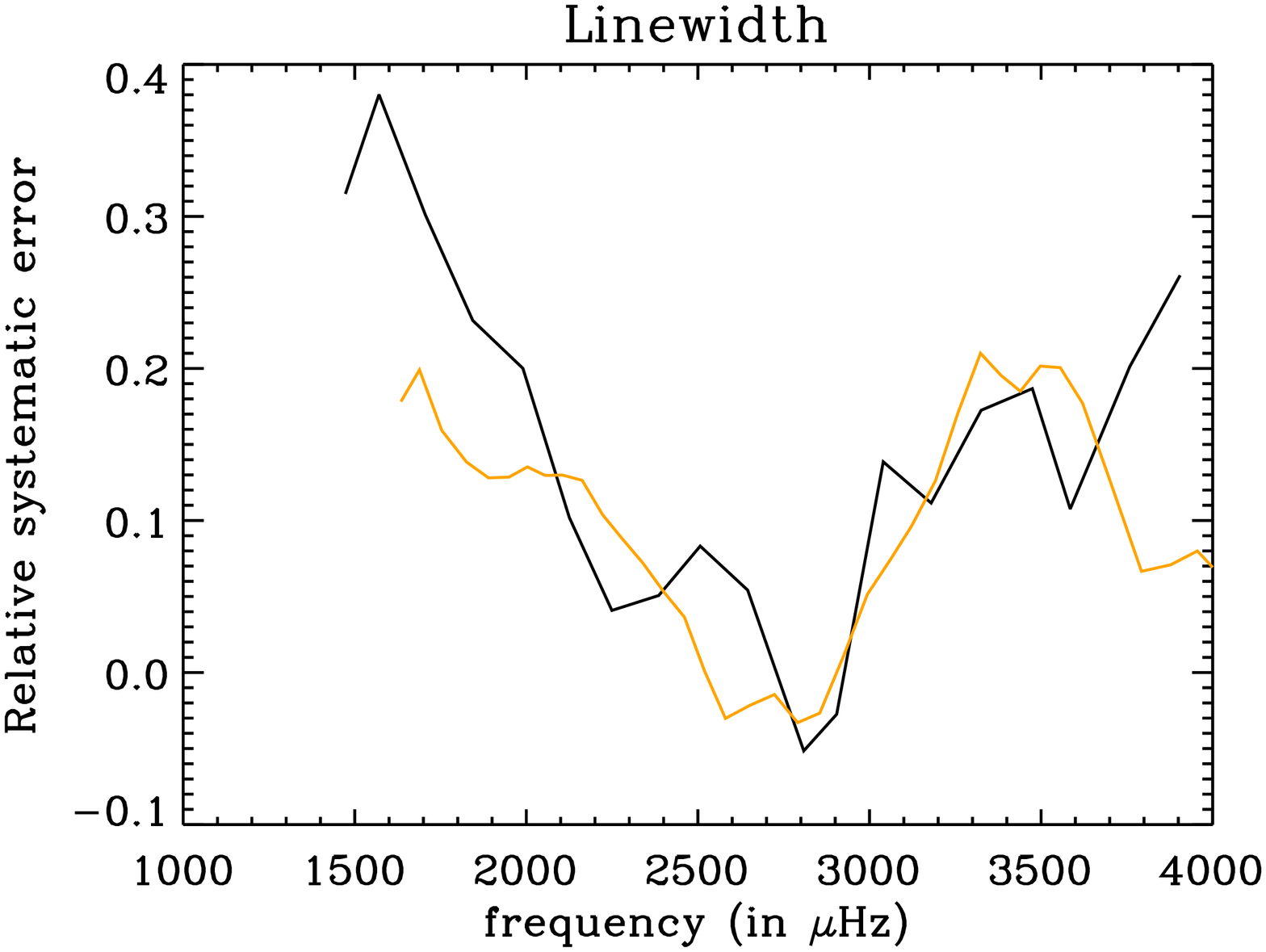}
\hspace{2.0cm}
\includegraphics[width=5.5 cm,angle=90]{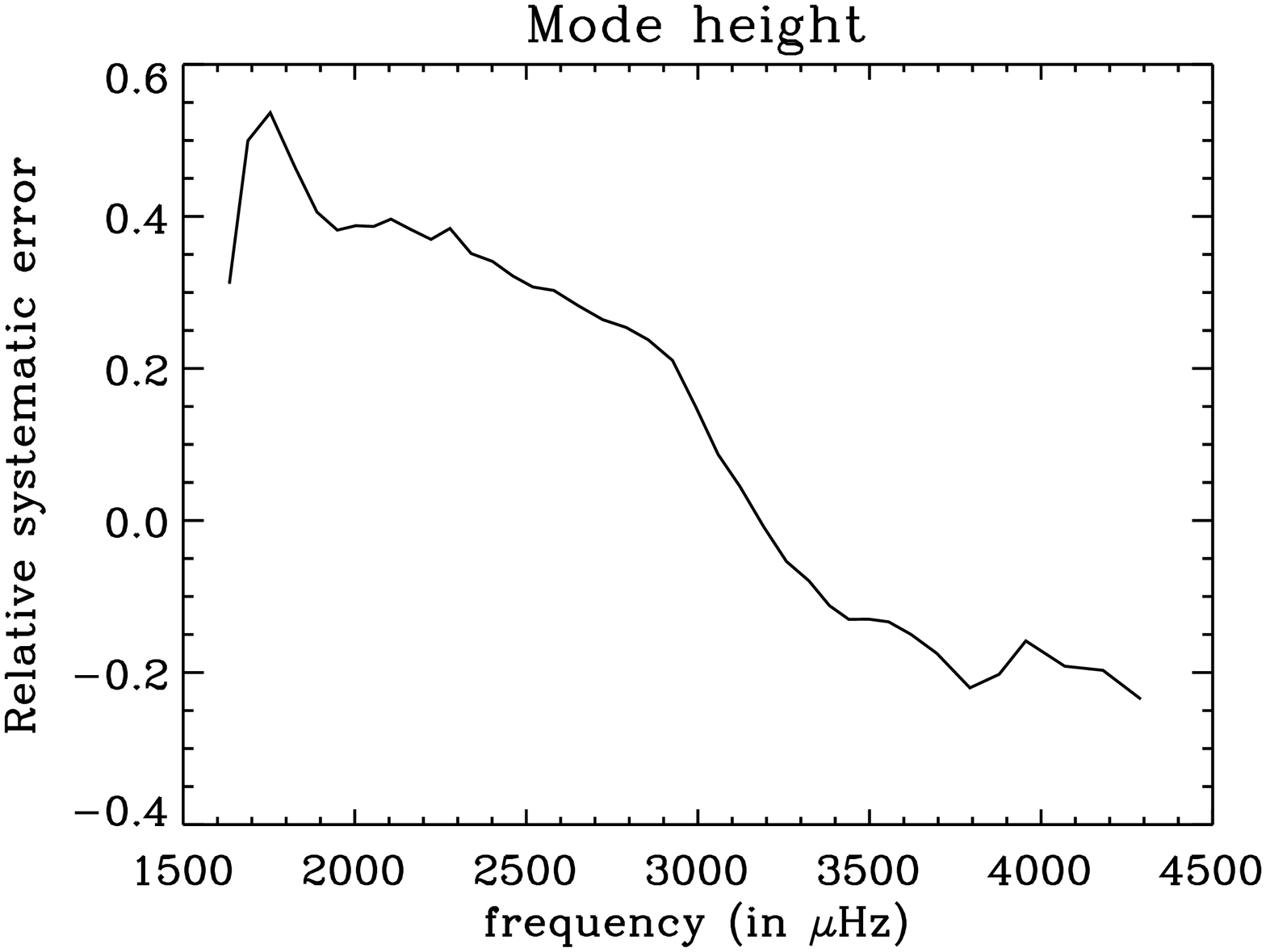}
}
\caption{Relative systematic error as a function of frequency for a mean frequency shifted by half the large separation for the solar case for mode linewidth (left) and for mode height (right) using BiSON data \citep[black line,][]{Chaplin1997} and GONG data  \citep[orange line,][]{Komm2000}.}
\label{shift}
\end{figure*}

Figure~\ref{shift} shows the resulting relative systematic error on mode linewidth and mode height in the solar case using BiSON and GONG data.  The relative error could be up to 40\% especially at the low frequencies.  Fortunately, the signal-to-noise ratio of the \textit{Kepler} observations allows the observation of low order modes only for stars more massive than the Sun having higher mode amplitude.  These stars then have smaller large separation than the Sun, typically about the third of the solar case.  Therefore the relative error due to the different mean frequency is never larger than 10\% for either the mode linewidth or the mode height.

Finally, only one fitter (Howe) used the formal order definition for which the $l$=2 mode is separated by a large separation of the $l=0$ mode of the same order $n$ (configuration {\it a} in Table A.1).  The problem of using this formal definition for fitting across a large separation is that the $l=2$ mode of order $n$ is very close to the $l=0$ mode of order $n+1$.  In practice, then the formal order definition is almost never used for mode fitting.  In any case, the difference of mean frequency of configuration A with respect to the formal order definition is then only $\Delta \nu/4$.  It means that the systematics, resulting from the different definition in that case, are twice as small as those resulting from using configuration B, or less than 5\%.

\subsection{Implementation of the correction of the systematic errors}
For each fit, the correction was computed by modelling the $l=0-2$ and $l=1-3$ modes over a window as large as the large separation ($\Delta\nu$) with the pair of modes separated by half of that.  The large separation is 240 $\mu$Hz.  The $l=2$ and $l=3$ modes are assumed to be 6 $\mu$Hz and 10 $\mu$Hz to the left of their adjacent $l=0$ or $l=1$ mode.  The correction could have been implemented with the proper large separation for each star but the results shows that the correction works even with the default large separation we used.  The mode height ratios are assumed to be either 1.0 / 1.5 / 0.5 / 0.05 or 1.0 / 2.0 / 1.0 / 0.0, depending on the star or the fitter.  The latter mode height ratio were typical results obtained by some fitters when freeing the ratios.  The linewidth is common to the set of mode frequencies following configuration A.  Note that the differences in the mean frequency between the different configurations were not implemented in the correction because of the rather small biases introduced.  The stellar background is flat.  The profile is modelled using the reference fit with the measured linewidth, mode height, splitting, inclination angle and stellar background.  Since the value of the background is taken at the $l=0$ mode, we neglect the frequency dependence of the stellar background within a large separation.  In the worst case the differential variation of the fitted stellar background with respect to the reference fit is typically less than 0.5\% across a large separation which is negligible compared to other systematic errors.   The fit is performed for only the linewidth and mode height as free parameters with a fixed splitting, inclination angle and stellar background provided by each fitter using the one-fit approach of \citet{TT2005b}.  Finally, the correction of the systematic errors is computed by taking the ratio of the fitted linewidth or mode height to that of the reference fit.  

We emphasize that the correction scheme could have been improved by using the mode-frequency configuration of each fitter, the \textit{true} fitted mode height ratio of each mode or of the ensemble of modes and a non-flat stellar background.  The {\it simple} correction scheme is sufficiently efficient to argue against making the correction model more complex.  

Figures B.1 to B.22 show for the remaining 22 stars the mode linewidth, height and amplitude corrected for systematic errors using the procedure described above.

\section{Analytical derivation of the biases leading to systematic errors}

\subsection{Assumptions}
The mode parameters are usually extracted from the power spectra using MLE which consists in finding the maximum of the following
figure of merit \citep[See][for details]{TTTA94}:
\begin{equation}
l = -\sum_{k=1}^{N} 
\left(\ln S_{1}(\nu_k,\lambda)+{S(\nu_k)\over{S_{1}(\nu_k,\lambda)}}\right)
\label{mle}
\end{equation}
where $S_1$ is the model of the power spectra, $S(\nu_k)$ represents the spectral data at the frequency bin $\nu_k$, $\lambda$ is the vector of parameters (frequency, splitting, linewidth, etc...).
In order to find the maximum of Eq.~(\ref{mle}), one needs to compute the derivative with respect to $\lambda$.  At the maximum, the derivative should be zero for all values of the element of the vector $\lambda$:

\begin{equation}
\frac{{\partial} l}{\partial {\lambda_i}}=-\sum_{k=1}^{N} \left(\frac{S_{1}(\nu_k,\lambda)-S(\nu_k)}{S_{1}^2(\nu_k,\lambda)}\right)\frac{{\partial} S_{1}}{\partial {\lambda_i}}=0
\end{equation}
which can be approximated, as per \citet{TTTA94}, by:
\begin{equation}
\frac{{\partial} l}{\partial {\lambda_i}}\approx-T\int_{-W/2}^{W/2} \left(\frac{S_{1}(\nu,\lambda)-S_{0}(\nu,\lambda)}{S_{1}^2(\nu,\lambda)}\right)\frac{{\partial} S_{1}}{\partial {\lambda_i}}{\rm d}\nu
\label{b3}
\end{equation}
where $\nu$ is the frequency, $W$ is the frequency range over which the maximization is performed, $T$ is the observation duration and $S_{0}$ is the asymptotic mean spectra.

The procedure used for analytically finding  the systematic errors is then simply to use the fact that $\frac{{\partial} l}{\partial {\lambda_i}}=0$ for all $\lambda_i$.  The analytical derivation can then be compared with the one-fit approach of \citet{TT2005b}, which explicitly assumes that the bias obtained by fitting an observed spectrum $S(\nu)$ with a profile $S_1(\nu,\lambda)$ is the same as fitting the asymptotic noiseless profile $S_0(\nu,\lambda)$.  The procedure used by \citet{TT2005b} is then simply to fit {\ $S_0(\nu,\lambda)$ once}  instead of doing Monte-Carlo simulations of many realization of $S(\nu)$.  With this procedure, this is exactly the asymptotic expression of Eq (B.1) that one maximizes, which results in finding asymptotically the zero of $\frac{{\partial} l}{\partial {\lambda_i}}$ as a function of the parameters $\lambda_i$. 

In all of the following subsections, we assumed a symmetrical Lorentzian profile for the asymptotic and fitted profiles.

\subsection{Effect of an incorrect estimation of the stellar background}
When the fitted spectra $S_1$ is different from $S_0$, we can write to first order:
\begin{equation}
S_{0}(\nu,\lambda)-S_{1}(\nu,\lambda)=\Delta h \frac{{\partial} S_{1}}{\partial {h}}+\Delta \gamma \frac{{\partial} S_{1}}{\partial {\gamma}}+\Delta b \frac{{\partial} S_{1}}{\partial {b}}
\label{b4}
\end{equation}
where $\gamma=\ln \Gamma$, $h=\ln H$, $b=\ln B$ with $\Gamma$, $H$ and $B$ being the mode linewidth, the mode height and the stellar background, respectively.  Note that if we assume that the model of the mode profile includes a white noise background, we have $ \frac{{\partial} S_{1}}{\partial {b}}=B$.
Inserting Eq.~(\ref{b4}) in Eq.~(\ref{b3}), one gets for the derivative with respect to $\gamma$ and $h$:
\begin{equation}
\frac{{\partial} l}{\partial {\gamma}} \approx {\cal H}_{h \gamma}\, \Delta h+ {\cal H}_{\gamma \gamma}\,\Delta \gamma +   i_\gamma \, \Delta b
\label{b5}
\end{equation}
\begin{equation}
\frac{{\partial} l}{\partial {h}} \approx {\cal H}_{h h} \, \Delta h +{\cal H}_{h \gamma}\, \Delta \gamma  +   i_h \, \Delta b
\label{b6}
\end{equation}
where ${\cal H}_{h \gamma}$, ${\cal H}_{\gamma \gamma}$ and ${\cal H}_{h h}$ are the elements of the Hessian ${\cal H}$ given by \citet{TTTA94} as:
\begin{equation}
{\cal H}_{h \gamma}=T  \int_{-W/2}^{W/2}   \frac{{\partial} S_{1}}{\partial {h}} \frac{{\partial} S_{1}}{\partial {\gamma}} \frac{1}{S_{1}^2(\nu,\lambda)}{\rm d}\nu
\end{equation}
\begin{equation}
{\cal H}_{\gamma \gamma}=T  \int_{-W/2}^{W/2}  \left( \frac{{\partial} S_{1}}{\partial {\gamma}}\right)^2 \frac{1}{S_{1}^2(\nu,\lambda)}{\rm d}\nu
\end{equation}
\begin{equation}
{\cal H}_{h h}=T  \int_{-W/2}^{W/2}  \left( \frac{{\partial} S_{1}}{\partial {h}}\right)^2 \frac{1}{S_{1}^2(\nu,\lambda)}{\rm d}\nu
\end{equation}
and with:
\begin{equation}
i_\gamma=T B  \int_{-W/2}^{W/2}    \frac{{\partial} S_{1}}{\partial {\gamma}} \frac{1}{S_{1}^2(\nu,\lambda)}{\rm d}\nu
\end{equation}
\begin{equation}
i_h=T B  \int_{-W/2}^{W/2}    \frac{{\partial} S_{1}}{\partial {h}} \frac{1}{S_{1}^2(\nu,\lambda)}{\rm d}\nu
\end{equation}
Since $\frac{{\partial} l}{\partial {\gamma}}=\frac{{\partial} l}{\partial {h}}=0$ by construction of the MLE solution, Eqs.~(\ref{b5}) and (\ref{b6}) will provide the solution for $\Delta h$ and $\Delta \gamma$ as a function of $\Delta b$ as:
\begin{equation}
\Delta \gamma= r_{b}^{\gamma} \Delta b
\label{b12}
\end{equation}
\begin{equation}
\Delta h=r_{b}^{h}\Delta b
\label{b13}
\end{equation}
with
\begin{equation}
 r_{b}^{\gamma}=\frac{{\cal H}_{h\gamma}i_h-{\cal H}_{\gamma \gamma} i_{\gamma}}{{\cal H}_{\gamma \gamma}h_{h h}-{\cal H}_{{\cal H} \gamma}^2}
\end{equation}
\begin{equation}
r_{b}^{h}=\frac{{\cal H}_{h \gamma} i_{\gamma}-{\cal H}_{\gamma\gamma}i_h}{{\cal H}_{\gamma \gamma}{\cal H}_{h h}-{\cal H}_{h \gamma}^2}
\end{equation}
The bias on the mode amplitude A=$\sqrt{\pi \Gamma H/2}$ can be expressed as
\begin{equation}
\Delta a=r_{b}^{a}b=\frac{\Delta \gamma+\Delta h}{2}=\left( \frac{r_{b}^{\gamma}+r_{b}^{h}}{2}\right)b
\end{equation}
where $a=\ln A$.  Eqs. (\ref{b12}) and (\ref{b13}) are the ratios for a notional bias in $B$. Since the calculation provided above was done assuming that the frequency window was large with respect to the mode linewidth, we must point out that the stellar background itself may be substantially over- or underestimated when the linewidth is not small compared to the window of interest.  As a consequence, even though the sensitivity to the background bias may be small, it is likely that the effective biases on mode linewidth and mode height may be larger because the bias on the stellar background may be large.

\subsection{Effect of a different estimated splitting and inclination angle}
As shown by \citet{TTTA94}, there is no correlation between the splitting error and the mode linewidth error or the mode height error.  It implicitly means that at least to first order there should not be any systematic errors in the mode linewidth and mode height measurements resulting from an incorrectly estimated splitting.  We study here is the second order effect.   

When the fitted spectra $S_1$ is different from $S_0$, we can write to the second order :
\begin{eqnarray}
S_{0}(\nu,\lambda)-S_{1}(\nu,\lambda)&=&\sum_j \Delta \lambda_j \frac{{\partial} S_{1}}{\partial {\lambda_j}}+\sum_{j \ne k} \Delta \lambda_j \Delta \lambda_k \frac{{\partial} S_{1}}{\partial {\lambda_j}}\frac{{\partial} S_{1}}{\partial {\lambda_k}}+\nonumber\\
&+&\sum_{j} \frac{\Delta \lambda_j^2}{2}\frac{{\partial}^2 S_{1}}{\partial {\lambda_j^2}}
\label{b17}
\end{eqnarray}
where $\lambda_j$ are $\gamma, h$, $b$ and $\nu_s$ (where $\nu_s$ is the mode splitting of the multiplet).  Inserting Eq.~(\ref{b17}) in Eq.~(\ref{b3}) one gets:
\begin{eqnarray}
\frac{{\partial} l}{\partial {\lambda_i}} &\approx & T  \int_{-W/2}^{W/2} \left[ \sum_j \Delta \lambda_j \frac{{\partial} S_{1}}{\partial {\lambda_j}}+\sum_{j \ne k} \Delta \lambda_j \Delta \lambda_k \frac{{\partial} S_{1}}{\partial {\lambda_j}}\frac{{\partial} S_{1}}{\partial {\lambda_k}}\right] \frac{{\partial} S_{1}}{\partial {\lambda_i}}  {\rm d}\nu +\nonumber\\
&+&T \int_{W/2}^{W/2}\left[\sum_{j} \frac{\Delta \lambda_j^2}{2}\frac{{\partial}^2 S_{1}}{\partial {\lambda_j^2}}\right]  \frac{{\partial} S_{1}}{\partial {\lambda_i}} {\rm d}\nu
\label{b18}
\end{eqnarray}
Here we note that the parity with respect to frequency ($\nu$) will play a key role in whether some terms disappears or not.  For instance, all derivatives of $S_1$ with respect to $\gamma$, $h$ and $b$ will be even in ($\nu-\nu_0$), where $\nu_0$ is the central mode frequency of the multiplet.  The derivatives of $S_1$ with respect to $\nu_0$ are odd in ($\nu-\nu_0$) while the derivative of $S_1$ with respect to $\nu_s$ is even in ($\nu-\nu_0$).  Therefore of all the cross terms involving $\gamma, h, b$ and $\nu_0$ are null.  At first sight all of the cross terms with $\nu_s$ would not necessarily be null.  But since the multiplet is composed of a superposition of several peaks, there is a local parity in the function around the mode frequency $\nu_{lm}$, i.e. around ($\nu-\nu_{lm}$) which \citet{TTTA94} termed as locally odd (or even) function, in other words:
\begin{equation}
  \int_{-W/2}^{W/2} \frac{{\partial} S_{1}}{\partial {\lambda_j}} \frac{{\partial} S_{1}}{\partial {\nu_s}}  {\rm d}\nu \approx 0
  \end{equation}
with $\lambda_j \ne \nu_s$.   We can then rewrite Eq.~(\ref{b18}) for the splitting $\nu_s$ as
\begin{equation}
\frac{{\partial} l}{\partial {\nu_s}} \approx \left[\Delta \nu_s {\cal H}_{\nu_s \nu_s}\right]+\left(\sum_{j \ne k}\Delta \lambda_j \Delta \lambda_k g_{jk}^{\nu_s} +  \sum_{j} \frac{\Delta \lambda_j^2}{2} g_{jj}^{\nu_s}\right)
\label{b20}
\end{equation}
and then for $\gamma, h$ and $b$ (=$\lambda_{i'}$) as:
\begin{equation}
\frac{{\partial} l}{\partial \lambda_{i'}} \approx \left[ \sum_{\lambda_j \ne \nu_s} \Delta \lambda_{j} {\cal H}_{ji'}\right]+\left(\sum_{j \ne k}\Delta \lambda_j \Delta \lambda_k g_{jk}^{i'} +  \sum_{j} \frac{\Delta \lambda_j^2}{2} g_{jj}^{i'}\right)
\label{b21}
\end{equation}
where $g_{jk}^{i}$ is written as:
\begin{equation}
g_{ij}^{k} = T  \int_{-W/2}^{W/2}  \frac{{\partial} S_{1}}{\partial {\lambda_j}}\frac{{\partial} S_{1}}{\partial {\lambda_k}} \frac{{\partial} S_{1}}{\partial {\lambda_i}}  {\rm d}\nu 
\end{equation}
The terms in brackets in Eqs.~(\ref{b20}) and (\ref{b21}) represent the first order terms, while the terms in parentheses are the higher order terms.  Since $\frac{{\partial} l}{\partial {\lambda_i}}=0$ by construction of the MLE solution, it is clear that the solution to the system of equation is non-linear.  As a consequence, the systematic errors on mode linewidth and mode height strongly depend to the second order to deviation to the nominal splitting. 

This simple analysis shows that the derivation of the biases due to splitting is not straightforward.  This is why we resorted to the single-fit approach of \citet{TT2005b} in order to have some clues as to the magnitude of the effect of an incorrect splitting.  Figures~\ref{splitting_1} and \ref{splitting_2} shows the results of the numerical calculation of the systematic errors on mode linewidth, mode height and mode amplitude biases as a function of the fitted splitting, for two different nominal splitting and mode linewidth.  Figure~\ref{splitting_3} shows the results of the numerical calculation of the systematic errors on mode linewidth, mode height and mode amplitude biases as a function of the fitted inclination angle, for a single nominal splitting and mode linewidth, and for two nominal inclination angles.

\subsection{Effect of different assumption on mode height ratio}
Here we assume that the fitted mode profile is given by a superposition of a profile for each $l$ such that we can write
\begin{equation}
S_{1}(\nu,\lambda)=\sum_{l} \alpha'_{l}Hs_{l}(\nu,\lambda) + B
\end{equation}
where $\alpha'_l$ is the fixed (or fitted) mode height ratio normalized to the $l=0$ degree, i.e. $\alpha'_0=1$.  We have a similar expression for the reference profile $S_0$ introducing the mode height $H$ and the mode height ratio $\alpha_l$.  Therefore we can write that:
\begin{eqnarray}
S_{1}(\nu,\lambda)-S_{0}(\nu,\lambda)&=&\sum_{l}^{} (H'\alpha'_{l}-H\alpha_{l})s_{l}(\nu,\lambda)
\label{b24}
\end{eqnarray}
Using Eq.~(\ref{b3}) we can write for the fitted mode height:
\begin{equation}
\frac{{\partial} l}{\partial H'}\approx-T\int_{-W/2}^{W/2} \left(\frac{S_{1}(\nu,\lambda)-S_{0}(\nu,\lambda)}{S_{1}^2(\nu,\lambda)}\right)\frac{{\partial} S_{1}}{\partial H'}{\rm d}\nu
\label{b25}
\end{equation}
Deriving Eq. (B.23) with respect to $H'$, we have:
\begin{equation}
\frac{{\partial} S_{1}}{\partial H'}=\sum_{l} \alpha'_{l}s_{l}(\nu,\lambda)
\label{b26}
\end{equation}
{\ Since} $\frac{{\partial} l}{\partial H'}=0$, using Eqs.~(\ref{b24}), (\ref{b25}) and (\ref{b26}) we can derive that we have:
\begin{equation}
J_{H}=J_{H'}
\label{b27}
\end{equation}
with
\begin{equation}
J_{H'}=H'\int_{-W/2}^{W/2} \left(\frac{\sum_{l} \alpha'_{l}s_{l}(\nu,\lambda)}{S_{1}^2(\nu,\lambda)}\right)\left(\sum_{l} \alpha'_{l}s_{l}(\nu,\lambda)\right){\rm d}\nu
\label{b28}
\end{equation}
\begin{equation}
J_H=H\int_{-W/2}^{W/2} \left(\frac{\sum_{l} \alpha'_{l}s_{l}(\nu,\lambda)}{S_{1}^2(\nu,\lambda)}\right)\left(\sum_{l} \alpha'_{l}s_{l}(\nu,\lambda)\right){\rm d}\nu
\label{b29}
\end{equation}
assuming that the cross terms between different $l$ are negligible, Eqs.~(\ref{b28}) and (\ref{b29}) can be approximated as:
\begin{equation}
J_{H'} \approx H'\sum_{l} \int_{-W/2}^{W/2} \left(\frac{(\alpha'_{l}s_{l}(\nu,\lambda))^2}{(H' \alpha'_{l}s_{l}(\nu,\lambda)+B)^2}\right){\rm d}\nu
\end{equation}
\begin{equation}
J_{H} \approx H \sum_{l} \frac{\alpha_{l}}{\alpha'_{l}}\int_{-W/2}^{W/2} \left(\frac{(\alpha'_{l}s_{l}(\nu,\lambda))^2}{(H' \alpha'_{l}s_{l}(\nu,\lambda)+B)^2}\right){\rm d}\nu
\end{equation}
both equations can be simplified as
\begin{equation}
J_{H'} \approx \frac{1}{H'}\sum_{l} \int_{-W/2}^{W/2} \left(\frac{s_{l}(\nu,\lambda)^2}{\left(s_{l}(\nu,\lambda)+\frac{B}{H' \alpha'_{l}}\right)^2}\right){\rm d}\nu
\end{equation}
and
\begin{equation}
J_{H} \approx \frac{H}{H'^2}\sum_{l} \frac{\alpha_{l}}{\alpha'_{l}} \int_{-W/2}^{W/2} \left(\frac{s_{l}(\nu,\lambda)^2}{\left(s_{l}(\nu,\lambda)+\frac{B}{H' \alpha'_{l}}\right)^2}\right){\rm d}\nu
\end{equation}
We can calculate these integrals using Eq. (A1) of \citet{TTTA94} by replacing $\beta$ by $\beta'/\alpha'$, where $\beta'=B/H'$ and assuming that the {\ frequency} window is large compared to the linewidth.   Omitting the common factors, we finally obtain:
\begin{equation}
J_{H'} \propto \frac{1}{H'} \sum_{l}  \frac{\sqrt{\alpha'_{l}}}{\sqrt{\beta'}(1+\beta'/\alpha'_{l})^{3/2}}
\end{equation}

\begin{equation}
J_{H}  \propto \frac{H}{H'^2}\sum_{l}  \frac{\alpha_{l}}{\alpha'_{l}} \frac{\sqrt{\alpha'_{l}}}{\sqrt{\beta'}(1+\beta'/\alpha'_{l})^{3/2}}
\end{equation}
then using Eq.~(\ref{b27}), we have:
\begin{equation}
\frac{H'}{H}=\frac{\sum_{l}  \frac{\alpha_{l}}{\alpha'_{l}} \frac{\sqrt{\alpha'_{l}}}{(1+\beta'/\alpha'_{l})^{3/2}}}{\sum_{l}  \frac{\sqrt{\alpha'_{l}}}{(1+\beta'/\alpha'_{l})^{3/2}}}
\end{equation}
Then to first order in $\Delta H=H'-H$ and in $\delta \alpha_l=\alpha'_l-\alpha_l$, we can write
\begin{equation}
\frac{\Delta H}{H}=-\frac{\sum_{l}  \frac{\Delta \alpha_{l}}{\alpha'_{l}} \frac{\sqrt{\alpha'_{l}}}{(1+\beta/\alpha'_{l})^{3/2}}}{\sum_{l}  \frac{\sqrt{\alpha'_{l}}}{(1+\beta/\alpha'_{l})^{3/2}}}+O(\Delta H)g(\Delta \alpha_l)
\end{equation}
such that we have the following approximation:
\begin{equation}
\frac{\Delta H}{H} \approx -\frac{\sum_{l}  \frac{\Delta \alpha_{l}}{\alpha'_{l}} \frac{\sqrt{\alpha'_{l}}}{(1+\beta/\alpha'_{l})^{3/2}}}{\sum_{l}  \frac{\sqrt{\alpha'_{l}}}{(1+\beta/\alpha'_{l})^{3/2}}}
\label{ratio_sys}
\end{equation}


\clearpage

\Online




\begin{figure*}[htbp]
\centering
\includegraphics[width=10.75 cm,angle=90]{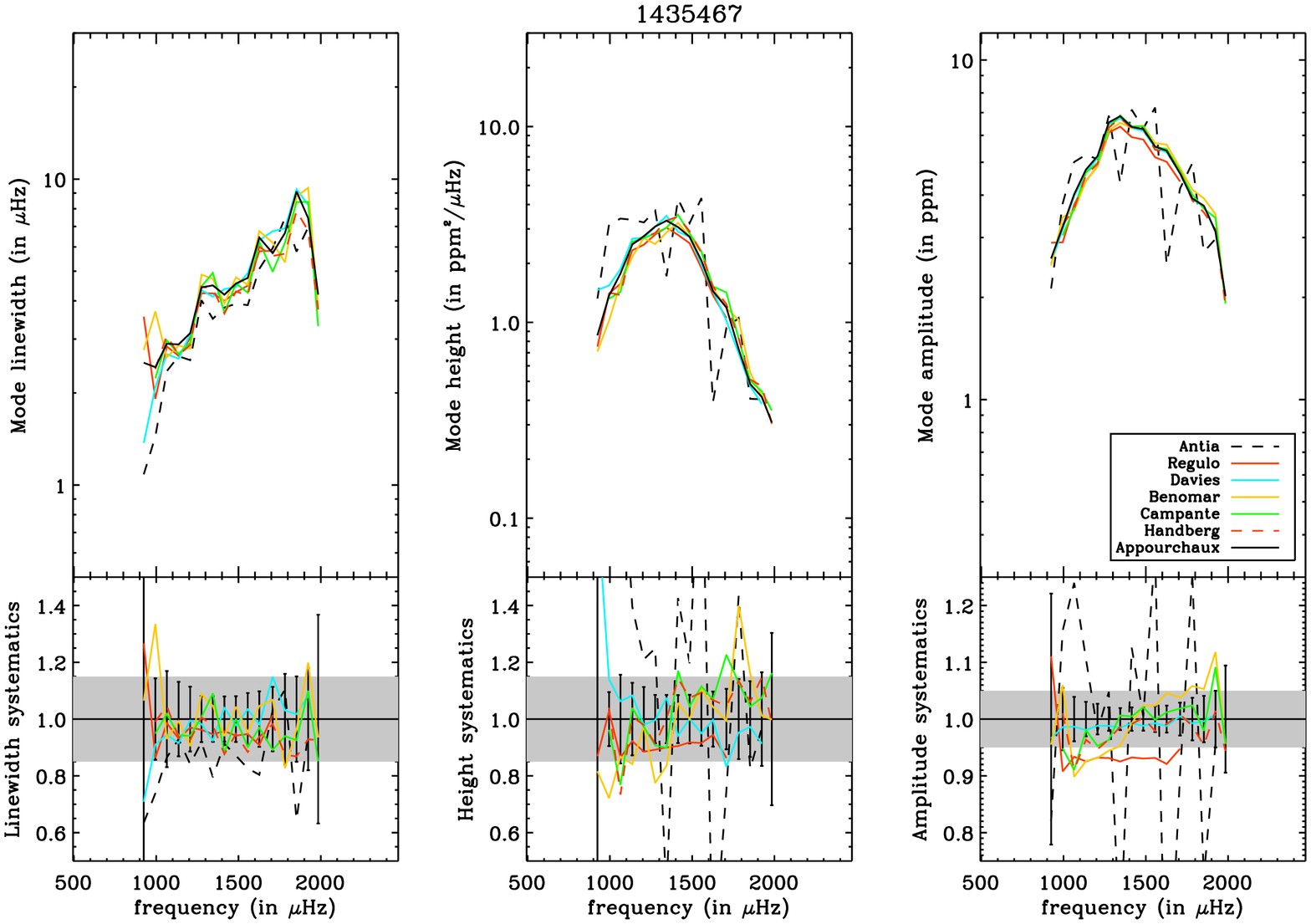}
\caption{Corrected mode linewidth, mode height and mode amplitude (Top) and relative values of these parameters with respect to the reference fit (Bottom) as a function of mode frequency for KIC 1435467. The grey band indicates the range of systematic error around the reference fit values of $\pm$ 15\% for mode linewidth and mode height, of $\pm$ 5\% for mode amplitude.  The error bars are those of the reference fit.}
\label{}
\end{figure*}

\begin{figure*}[htbp]
\centering
\includegraphics[width=10.75 cm,angle=90]{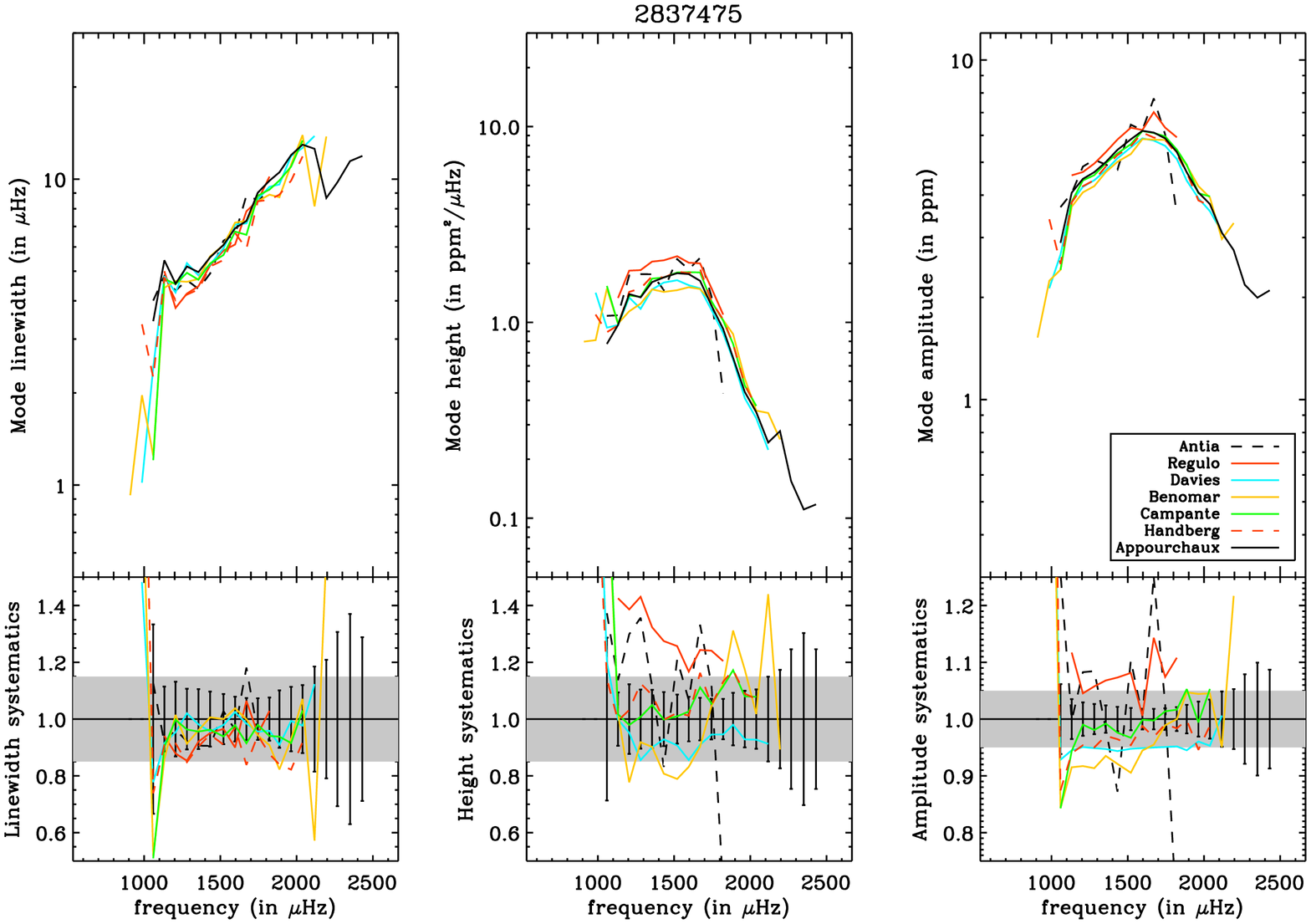}
\caption{Corrected mode linewidth, mode height and mode amplitude (Top) and relative values of these parameters with respect to the reference fit (Bottom) as a function of mode frequency for KIC 2837475. The grey band indicates the range of systematic error around the reference fit values of $\pm$ 15\% for mode linewidth and mode height, of $\pm$ 5\% for mode amplitude.  The error bars are those of the reference fit.}
\label{}
\end{figure*}

\begin{figure*}[htbp]
\centering
\includegraphics[width=10.75 cm,angle=90]{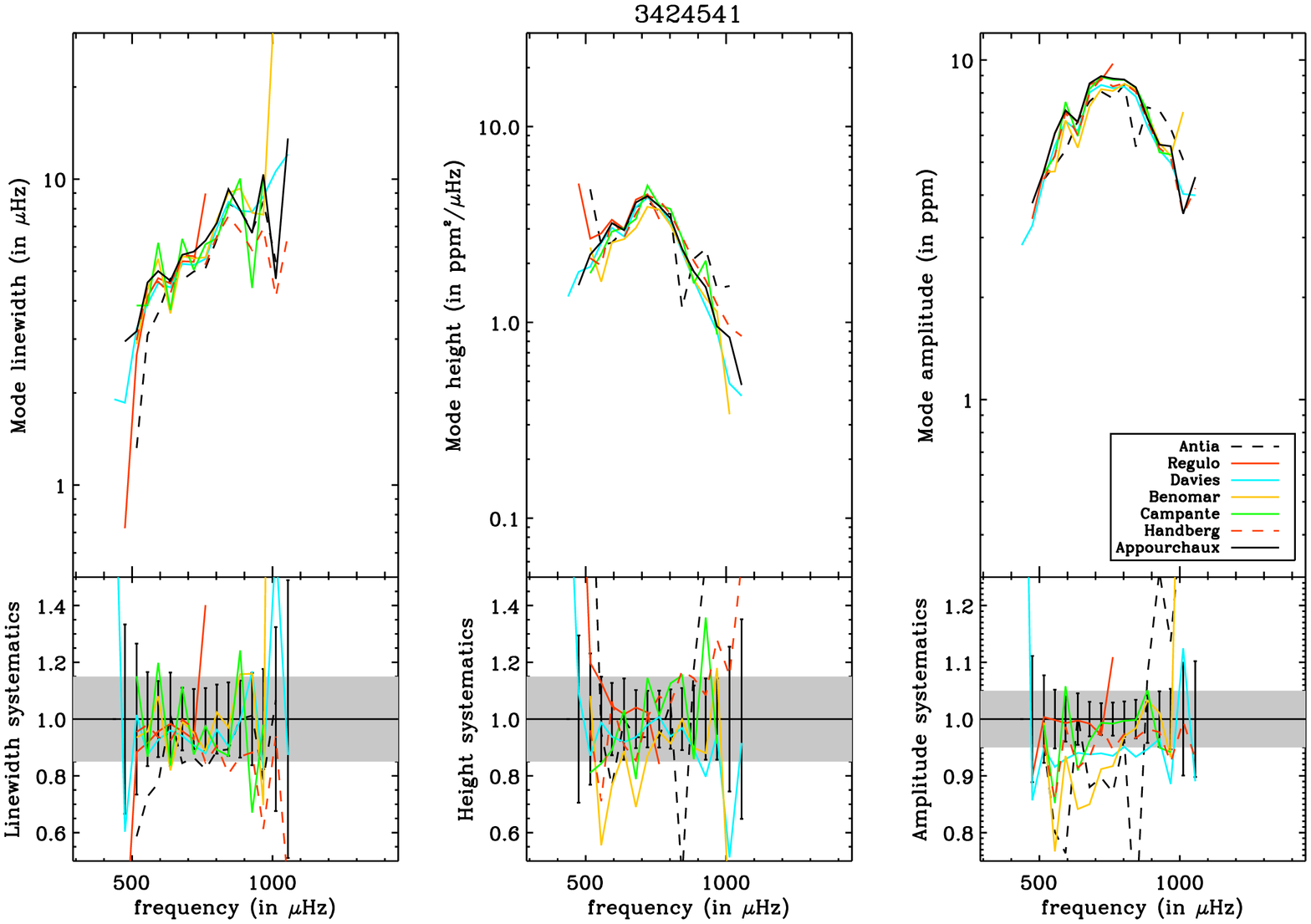}
\caption{Corrected mode linewidth, mode height and mode amplitude (Top) and relative values of these parameters with respect to the reference fit (Bottom) as a function of mode frequency for KIC 3424541. The grey band indicates the range of systematic error around the reference fit values of $\pm$ 15\% for mode linewidth and mode height, of $\pm$ 5\% for mode amplitude.  The error bars are those of the reference fit.}
\label{}
\end{figure*}

\begin{figure*}[htbp]
\centering
\includegraphics[width=10.75 cm,angle=90]{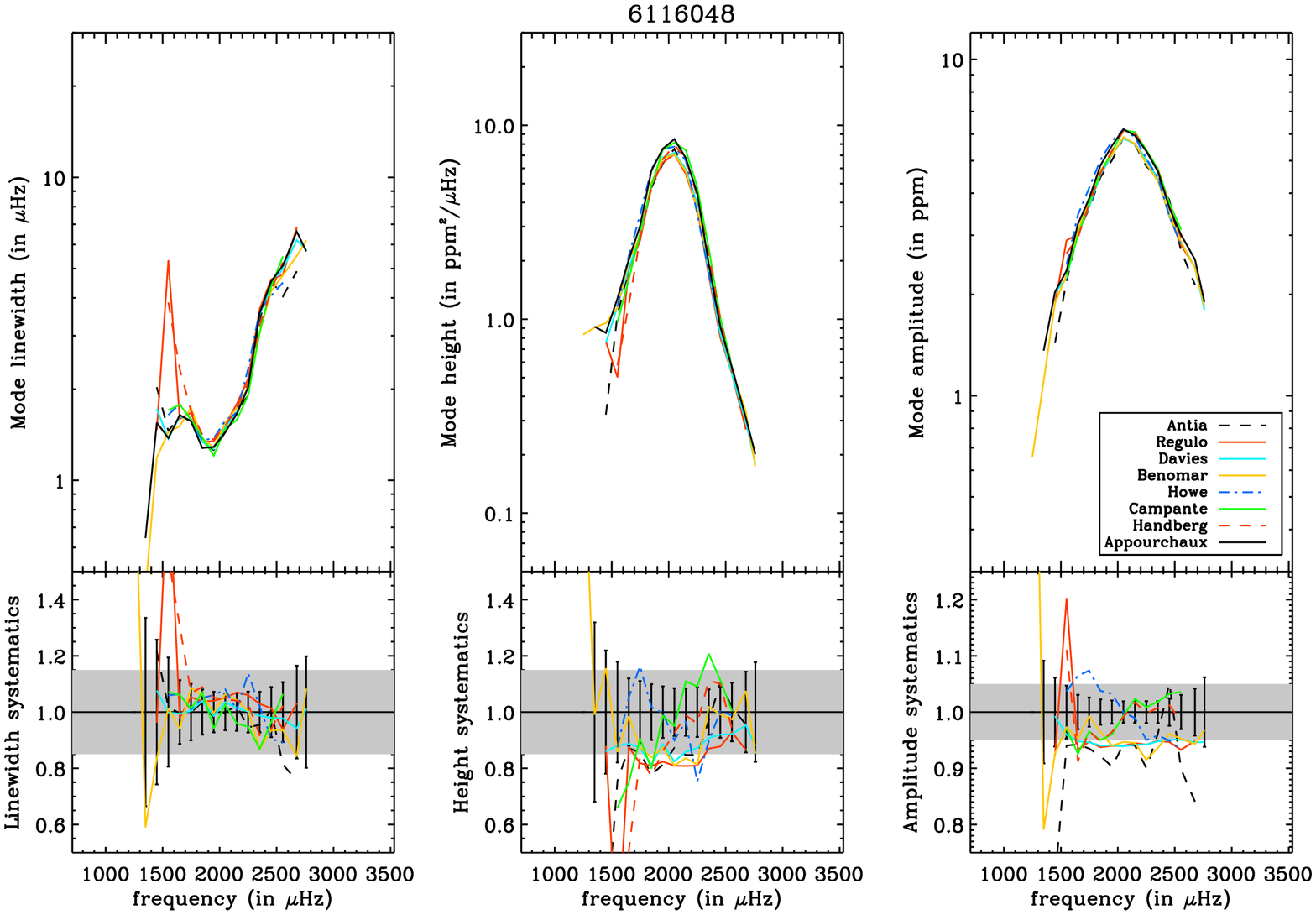}
\caption{Corrected mode linewidth, mode height and mode amplitude (Top) and relative values of these parameters with respect to the reference fit (Bottom) as a function of mode frequency for KIC 6116048. The grey band indicates the range of systematic error around the reference fit values of $\pm$ 15\% for mode linewidth and mode height, of $\pm$ 5\% for mode amplitude.  The error bars are those of the reference fit.}
\label{}
\end{figure*}

\begin{figure*}[htbp]
\centering
\includegraphics[width=10.75cm,angle=90]{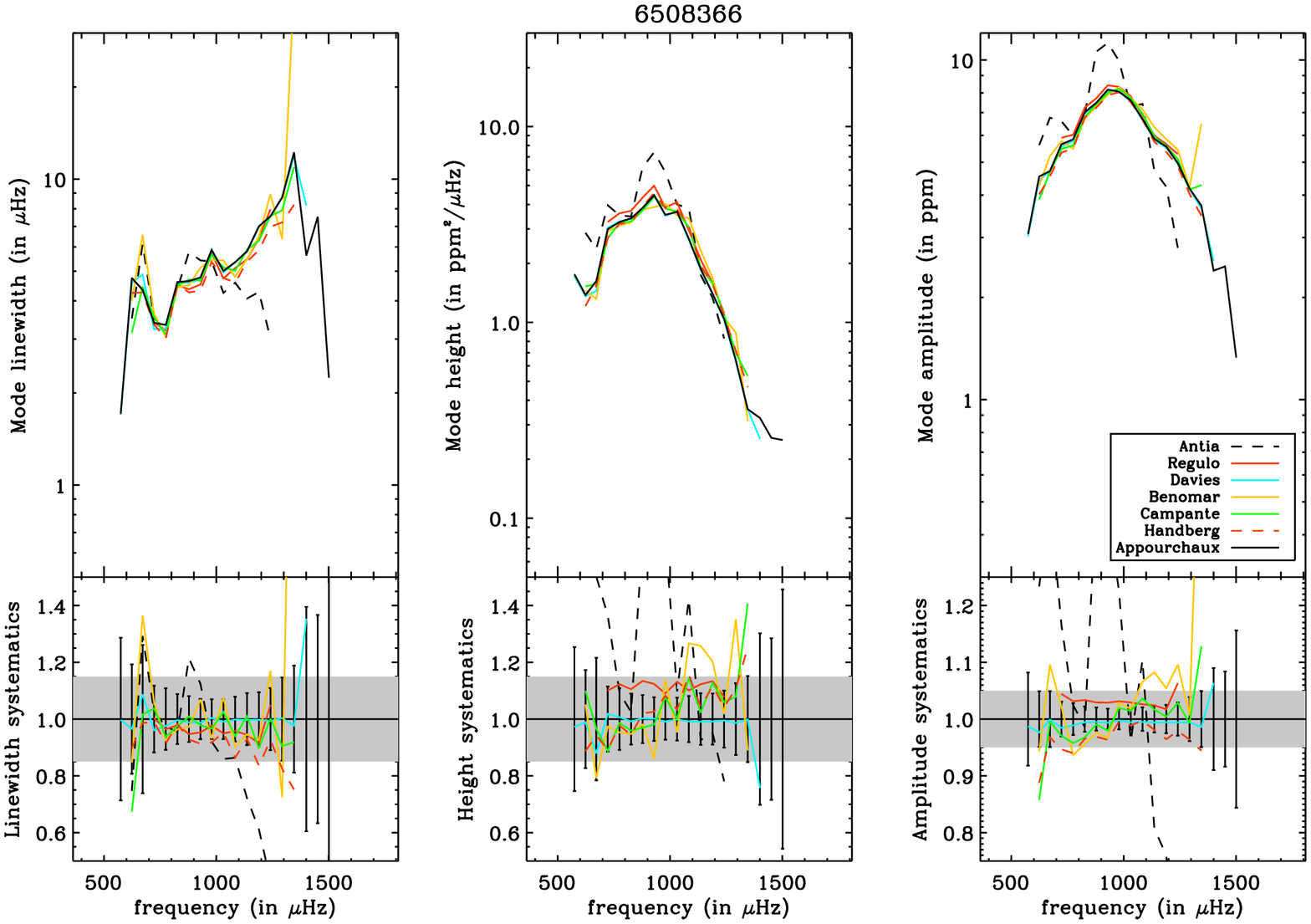}
\caption{Corrected mode linewidth, mode height and mode amplitude (Top) and relative values of these parameters with respect to the reference fit (Bottom) as a function of mode frequency for KIC 6508366. The grey band indicates the range of systematic error around the reference fit values of $\pm$ 15\% for mode linewidth and mode height, of $\pm$ 5\% for mode amplitude.  The error bars are those of the reference fit.}
\label{}
\end{figure*}

\begin{figure*}[htbp]
\centering
\includegraphics[width=10.75 cm,angle=90]{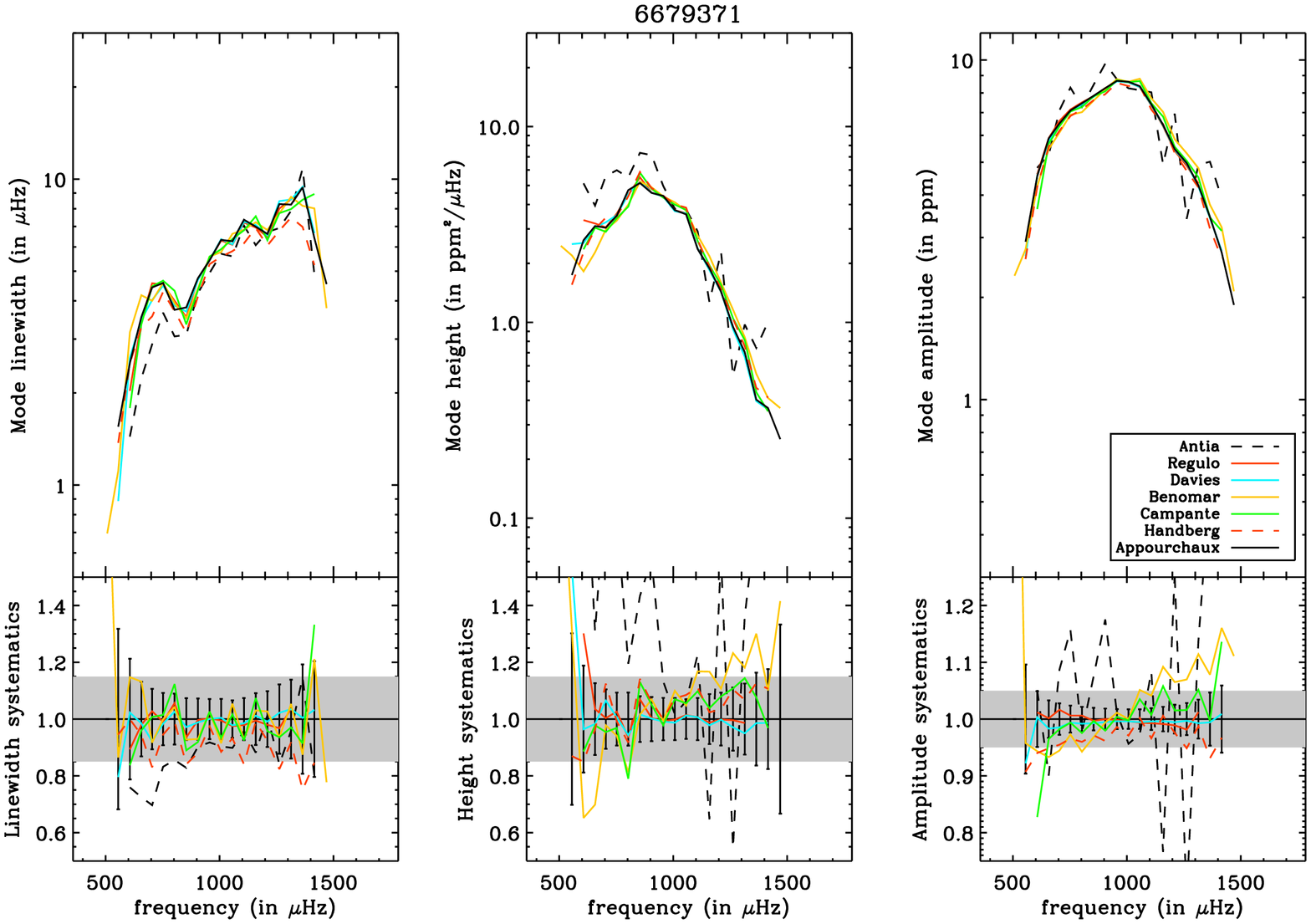}
\caption{Corrected mode linewidth, mode height and mode amplitude (Top) and relative values of these parameters with respect to the reference fit (Bottom) as a function of mode frequency for KIC 6679371. The grey band indicates the range of systematic error around the reference fit values of $\pm$ 15\% for mode linewidth and mode height, of $\pm$ 5\% for mode amplitude.  The error bars are those of the reference fit.}
\label{}
\end{figure*}

\begin{figure*}[htbp]
\centering
\includegraphics[width=10.75 cm,angle=90]{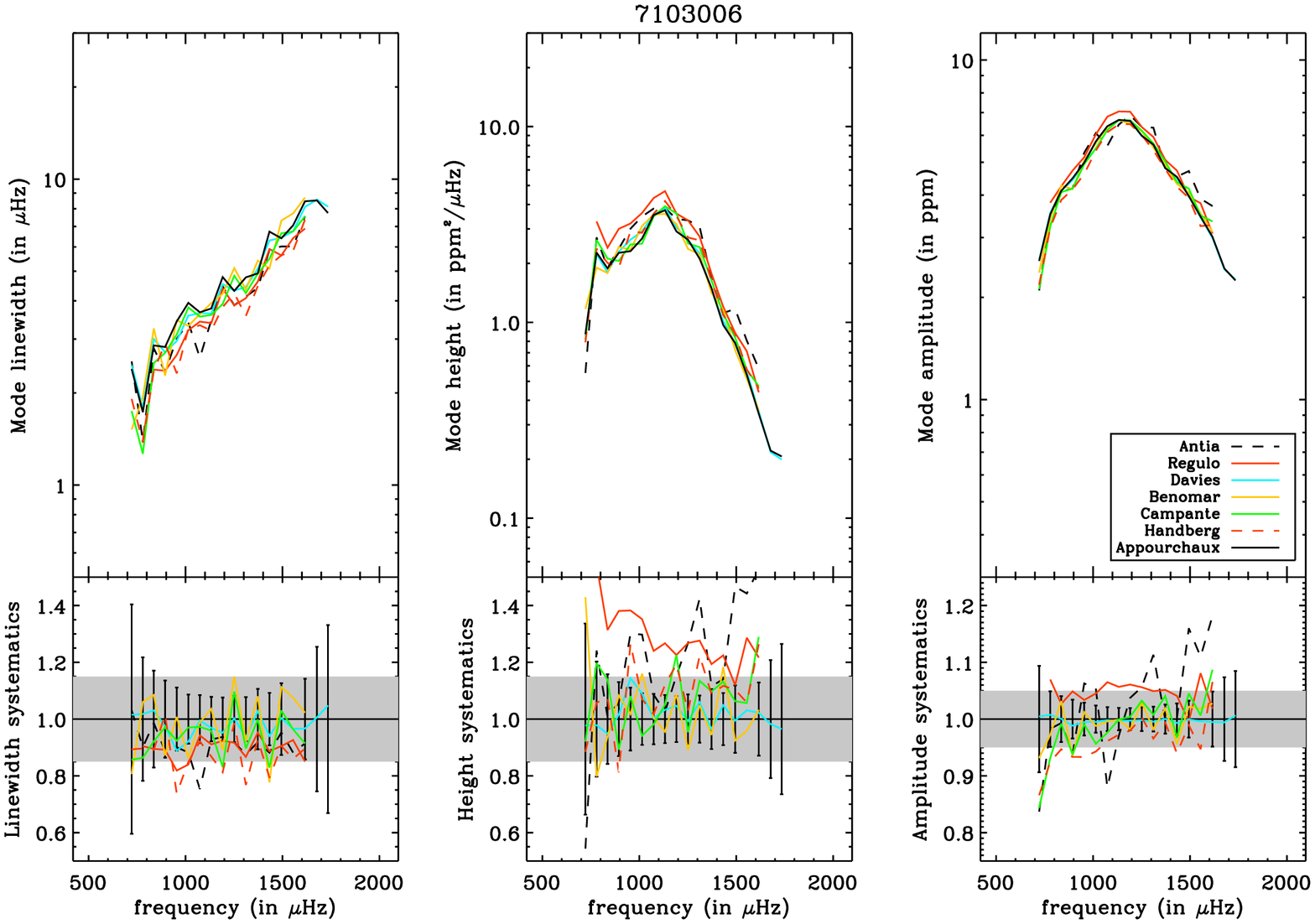}
\caption{Corrected mode linewidth, mode height and mode amplitude (Top) and relative values of these parameters with respect to the reference fit (Bottom) as a function of mode frequency for KIC 7103006. The grey band indicates the range of systematic error around the reference fit values of $\pm$ 15\% for mode linewidth and mode height, of $\pm$ 5\% for mode amplitude.  The error bars are those of the reference fit.}
\label{}
\end{figure*}

\begin{figure*}[htbp]
\centering
\includegraphics[width=10.75 cm,angle=90]{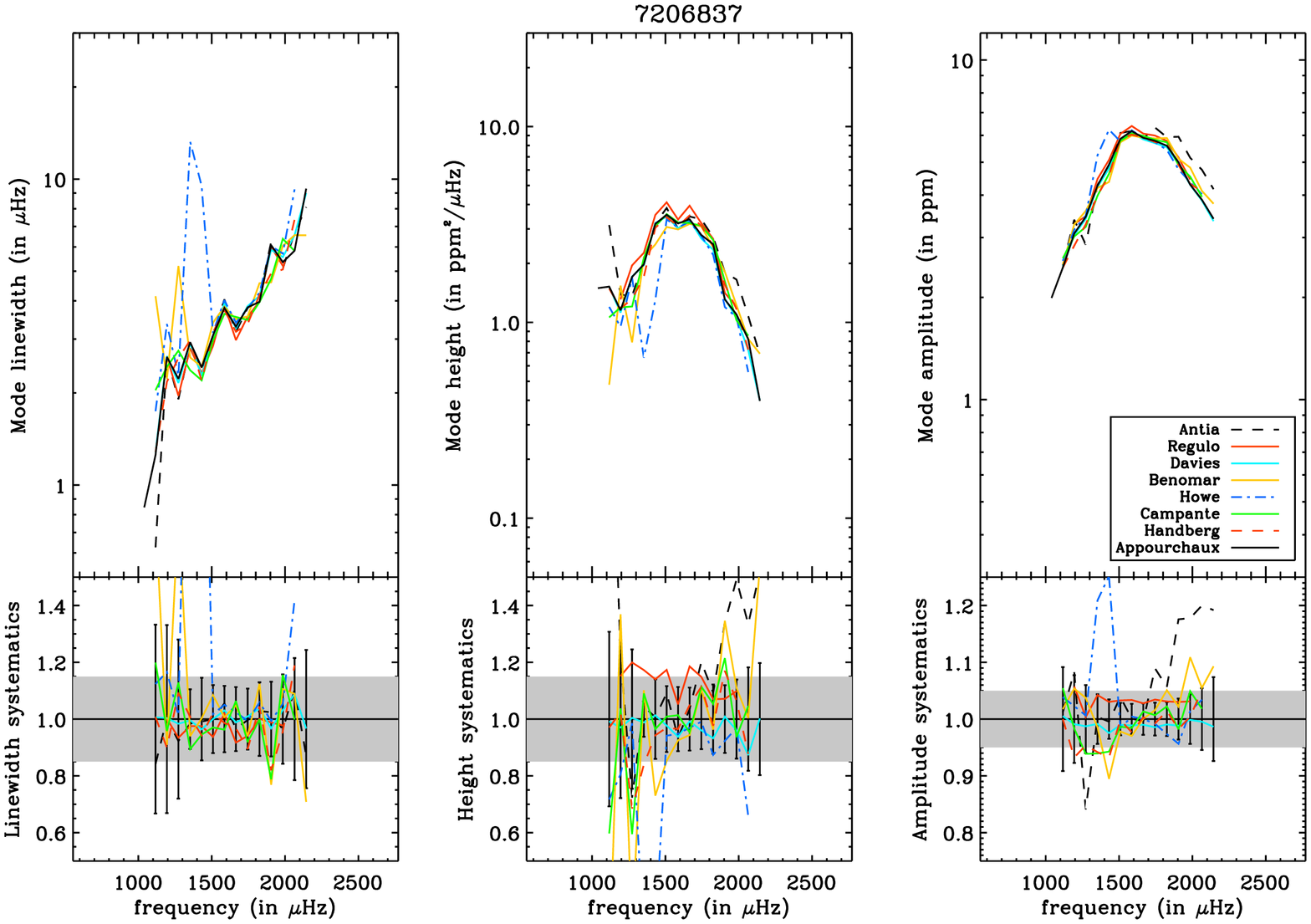}
\caption{Corrected mode linewidth, mode height and mode amplitude (Top) and relative values of these parameters with respect to the reference fit (Bottom) as a function of mode frequency for KIC 7206837. The grey band indicates the range of systematic error around the reference fit values of $\pm$ 15\% for mode linewidth and mode height, of $\pm$ 5\% for mode amplitude.  The error bars are those of the reference fit.}
\label{}
\end{figure*}

\begin{figure*}[htbp]
\centering
\includegraphics[width=10.75 cm,angle=90]{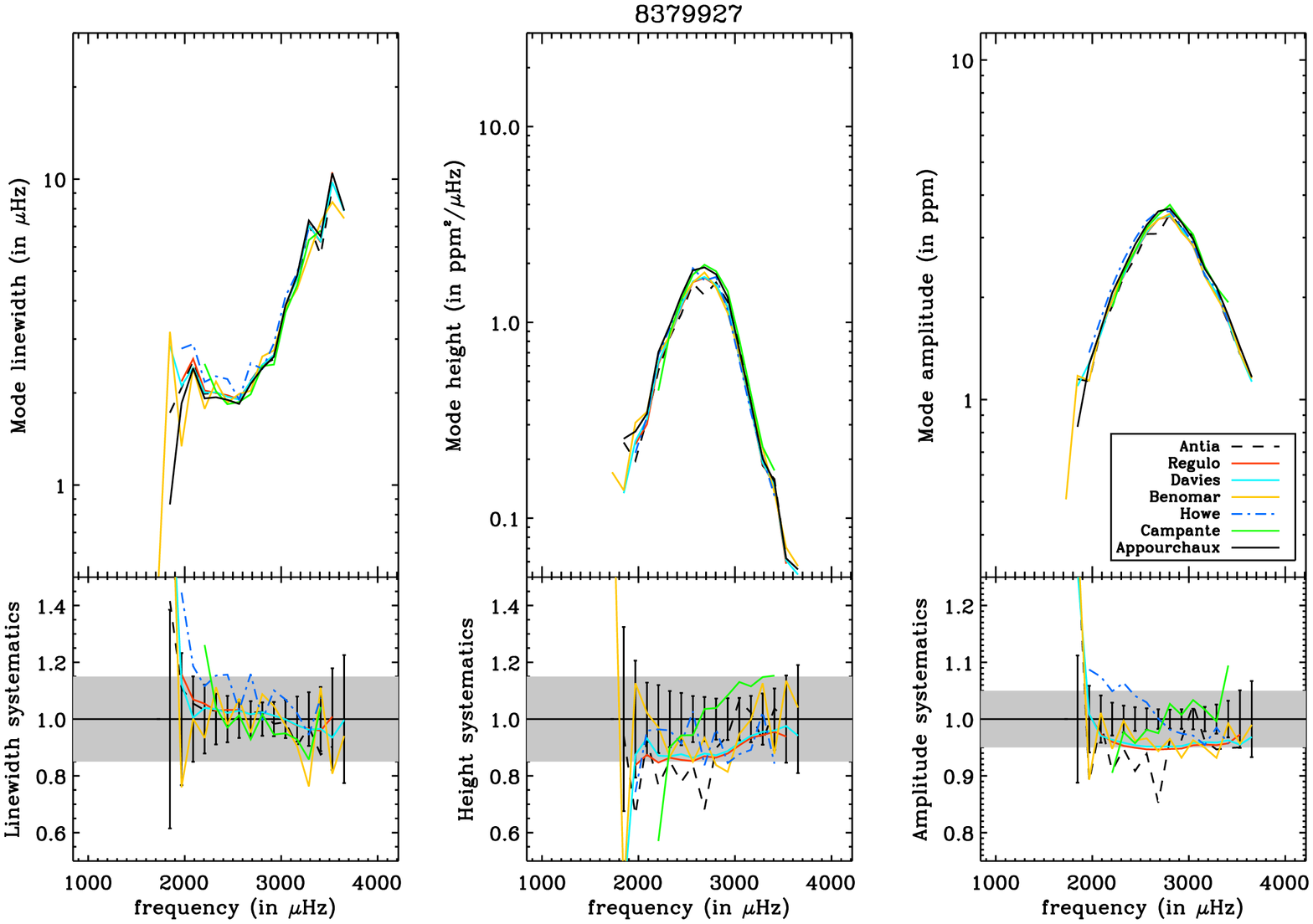}
\caption{Corrected mode linewidth, mode height and mode amplitude (Top) and relative values of these parameters with respect to the reference fit (Bottom) as a function of mode frequency for KIC 8379927 . The grey band indicates the range of systematic error around the reference fit values of $\pm$ 15\% for mode linewidth and mode height, of $\pm$ 5\% for mode amplitude.  The error bars are those of the reference fit.}
\label{8379927}
\end{figure*}

\begin{figure*}[htbp]
\centering
\includegraphics[width=10.75 cm,angle=90]{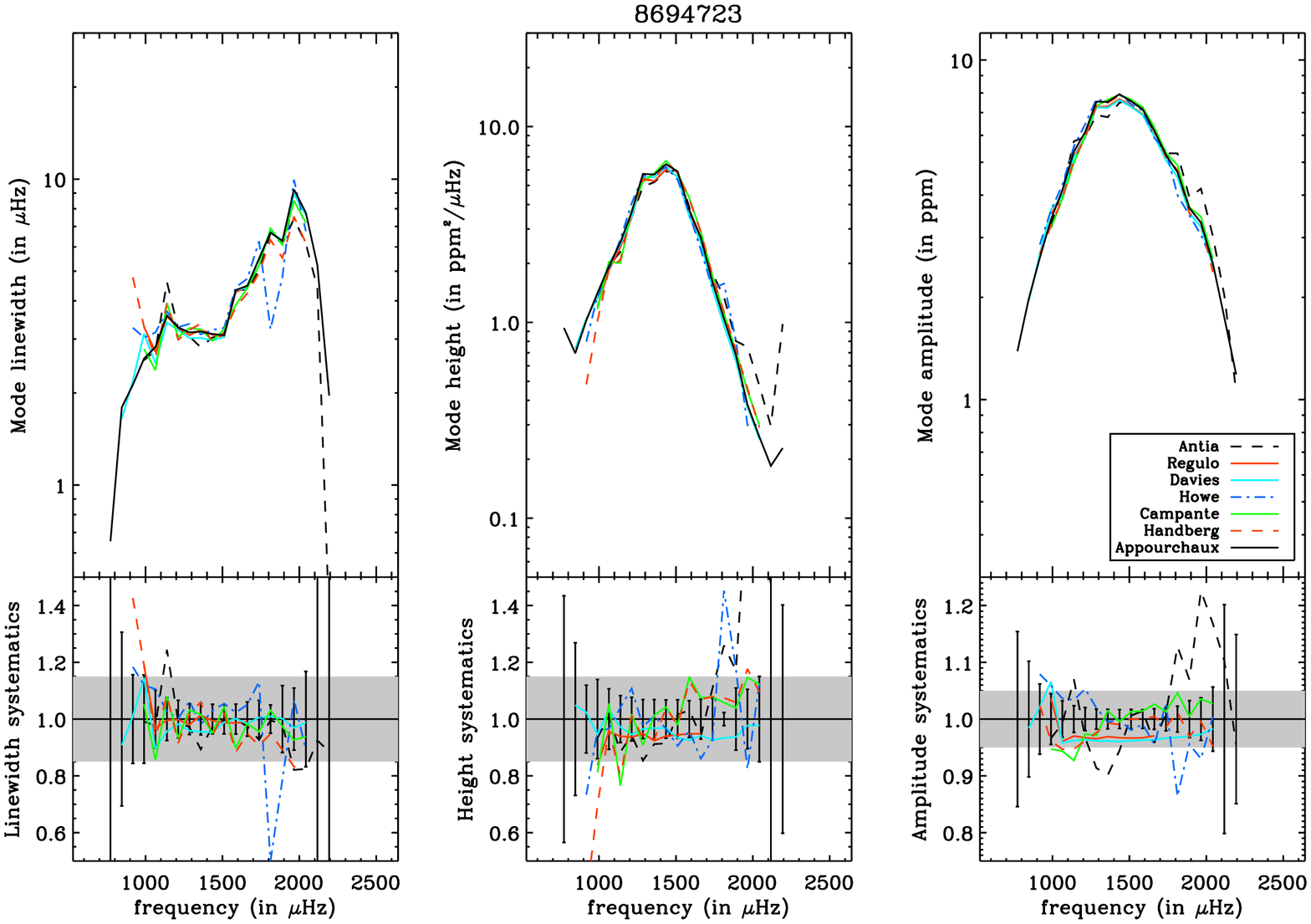}
\caption{Corrected mode linewidth, mode height and mode amplitude (Top) and relative values of these parameters with respect to the reference fit (Bottom) as a function of mode frequency for KIC 8694723. The grey band indicates the range of systematic error around the reference fit values of $\pm$ 15\% for mode linewidth and mode height, of $\pm$ 5\% for mode amplitude.  The error bars are those of the reference fit.}
\label{}
\end{figure*}

\begin{figure*}[htbp]
\centering
\includegraphics[width=10.75 cm,angle=90]{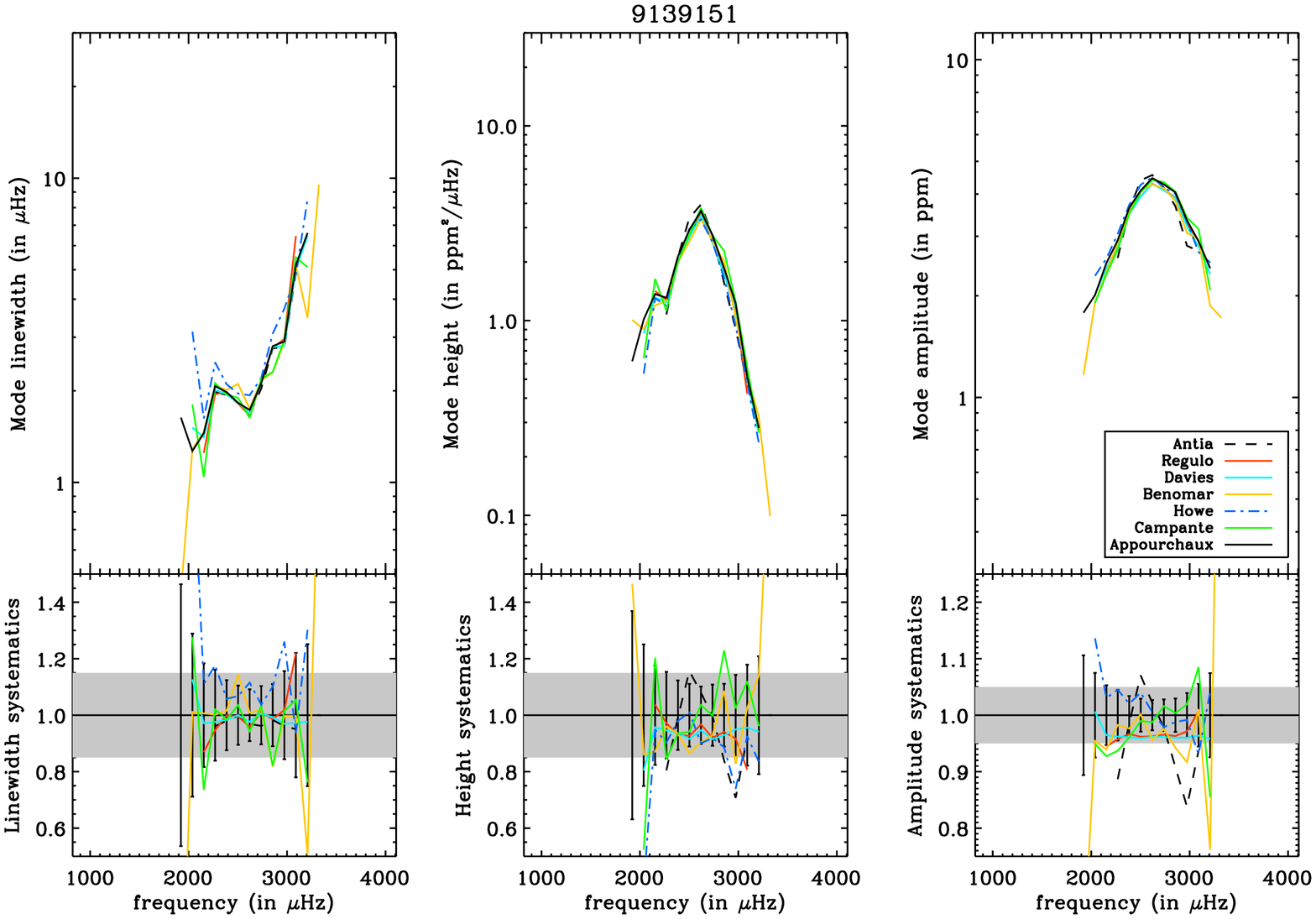}
\caption{Corrected mode linewidth, mode height and mode amplitude (Top) and relative values of these parameters with respect to the reference fit (Bottom) as a function of mode frequency for KIC 9139151. The grey band indicates the range of systematic error around the reference fit values of $\pm$ 15\% for mode linewidth and mode height, of $\pm$ 5\% for mode amplitude.  The error bars are those of the reference fit.}
\label{}
\end{figure*}

\begin{figure*}[htbp]
\centering
\includegraphics[width=10.75 cm,angle=90]{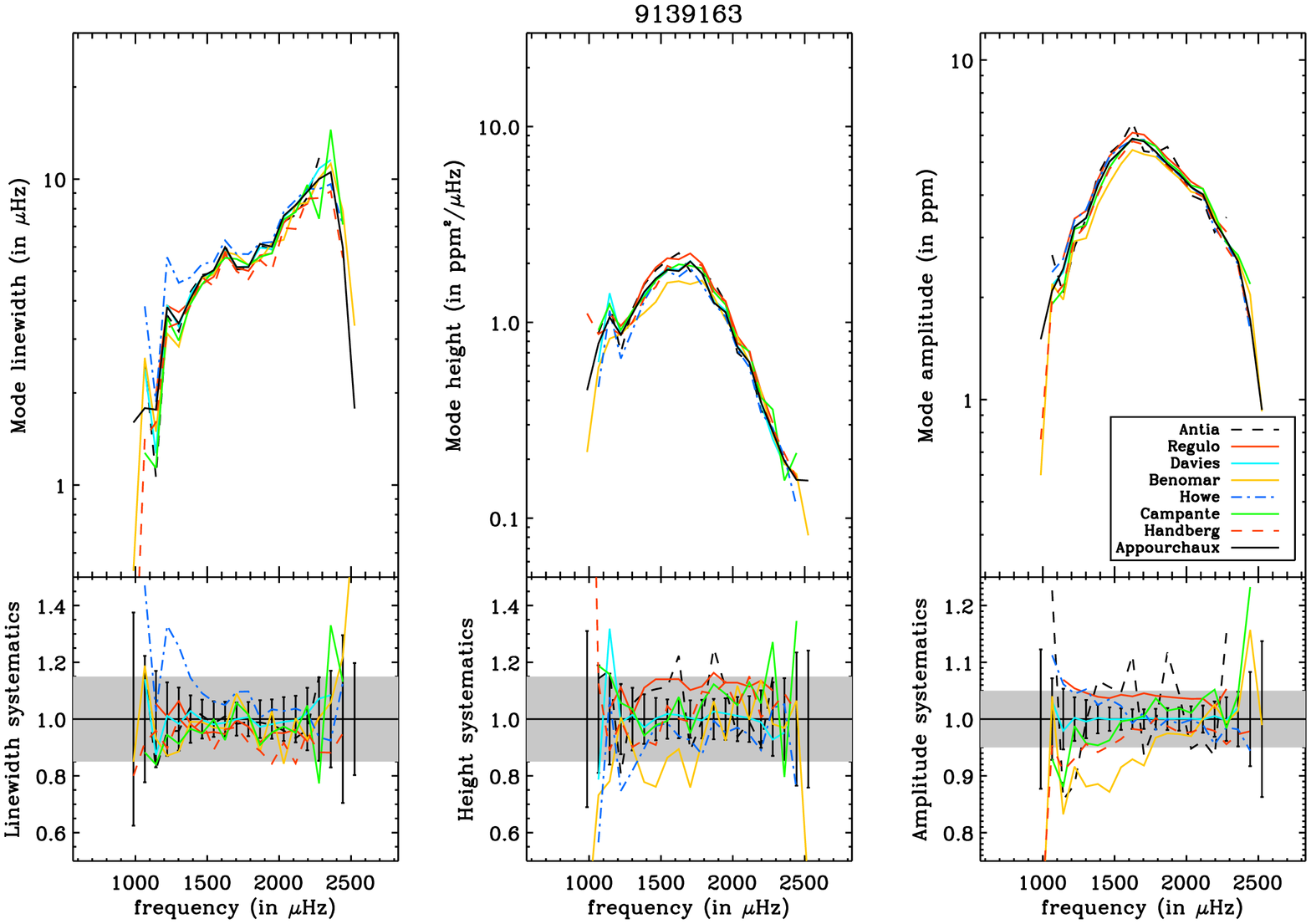}
\caption{Corrected mode linewidth, mode height and mode amplitude (Top) and relative values of these parameters with respect to the reference fit (Bottom) as a function of mode frequency for KIC 9139163. The grey band indicates the range of systematic error around the reference fit values of $\pm$ 15\% for mode linewidth and mode height, of $\pm$ 5\% for mode amplitude.  The error bars are those of the reference fit.}
\label{}
\end{figure*}

\clearpage

\begin{figure*}[htbp]
\centering
\includegraphics[width=10.75 cm,angle=90]{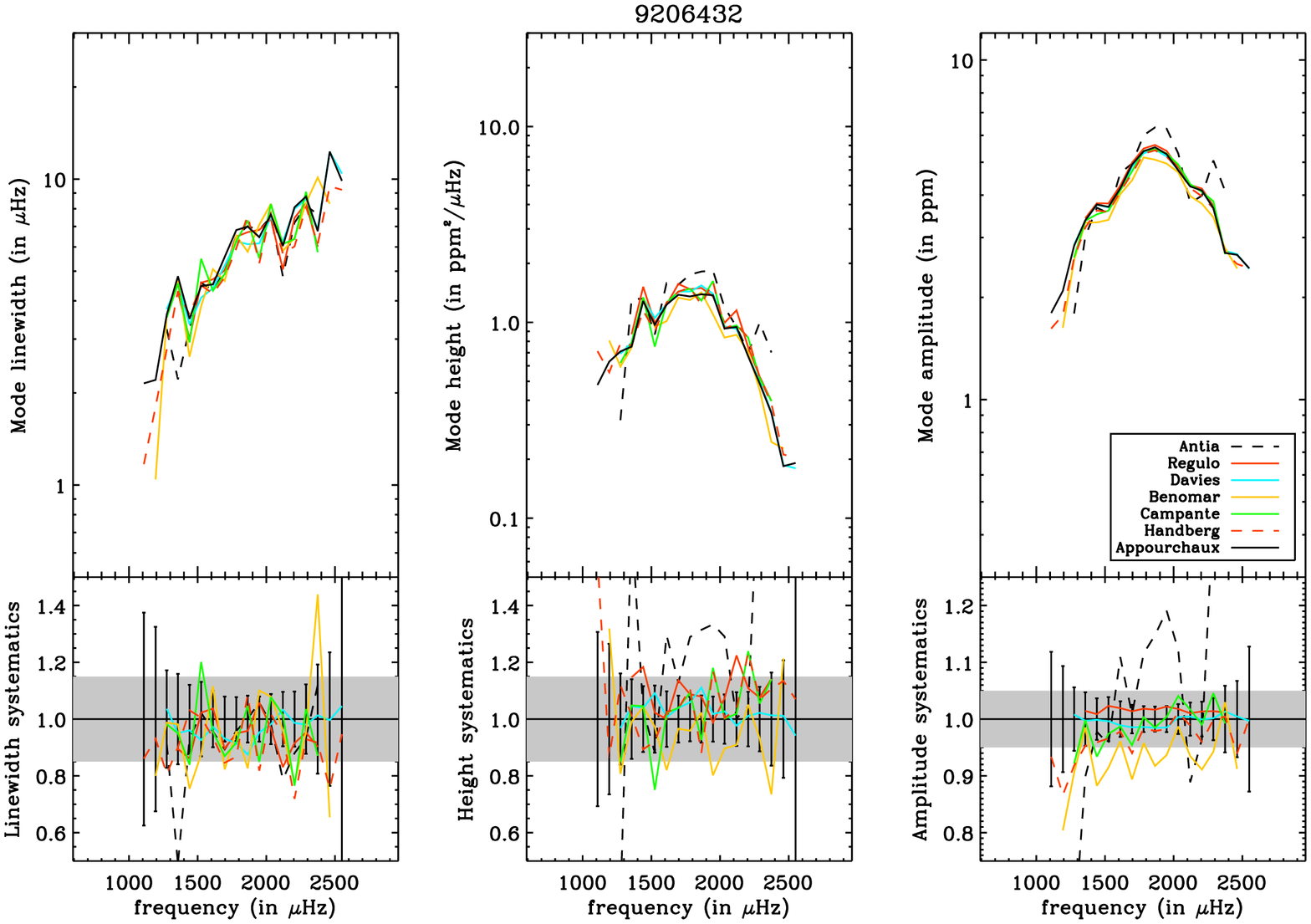}
\caption{Corrected mode linewidth, mode height and mode amplitude (Top) and relative values of these parameters with respect to the reference fit (Bottom) as a function of mode frequency for KIC 9206432. The grey band indicates the range of systematic error around the reference fit values of $\pm$ 15\% for mode linewidth and mode height, of $\pm$ 5\% for mode amplitude.  The error bars are those of the reference fit.}
\label{}
\end{figure*}

\begin{figure*}[htbp]
\centering
\includegraphics[width=10.75 cm,angle=90]{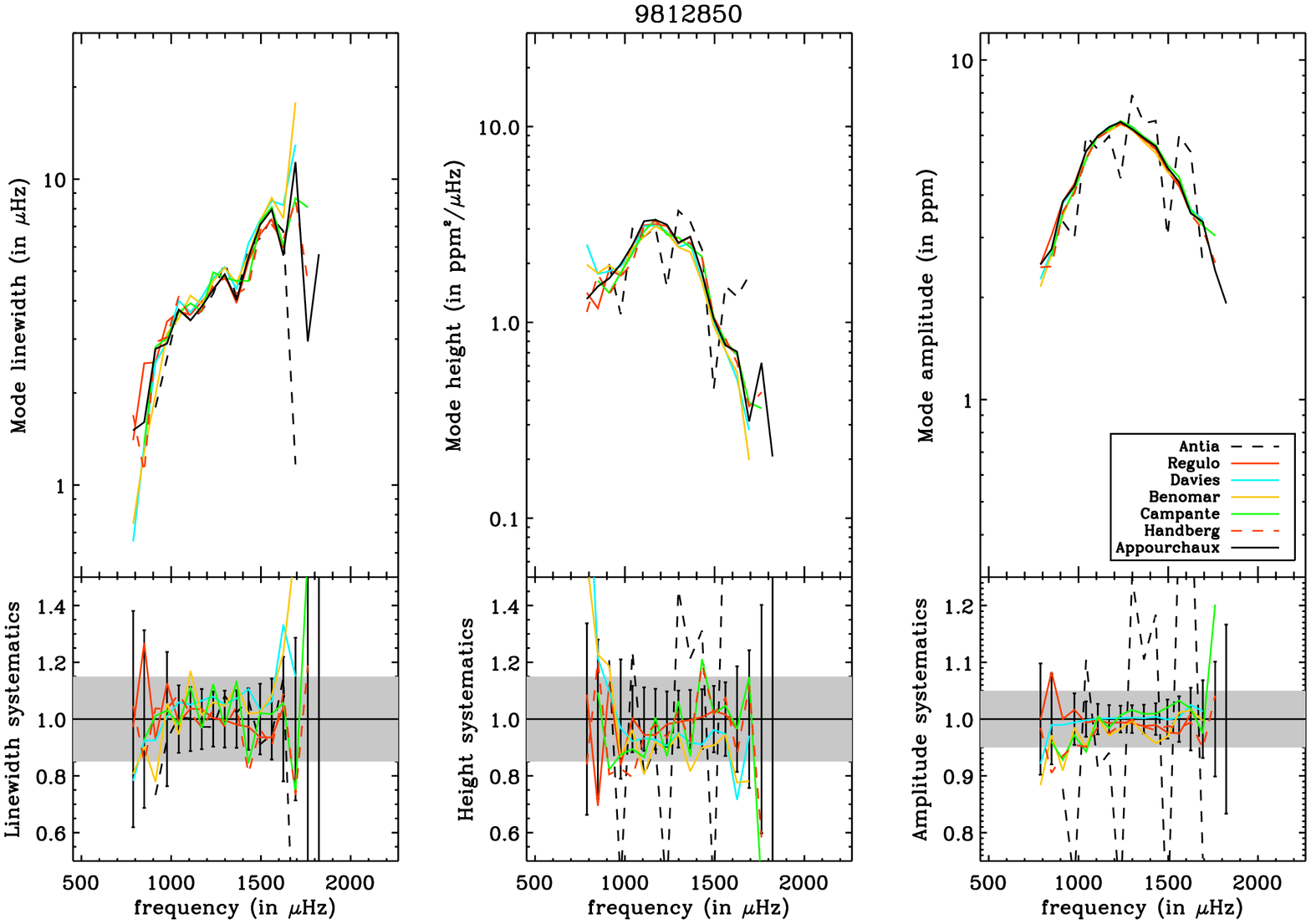}
\caption{Corrected mode linewidth, mode height and mode amplitude (Top) and relative values of these parameters with respect to the reference fit (Bottom) as a function of mode frequency for KIC 9812850. The grey band indicates the range of systematic error around the reference fit values of $\pm$ 15\% for mode linewidth and mode height, of $\pm$ 5\% for mode amplitude.  The error bars are those of the reference fit.}
\label{}
\end{figure*}

\begin{figure*}[htbp]
\centering
\includegraphics[width=10.75 cm,angle=90]{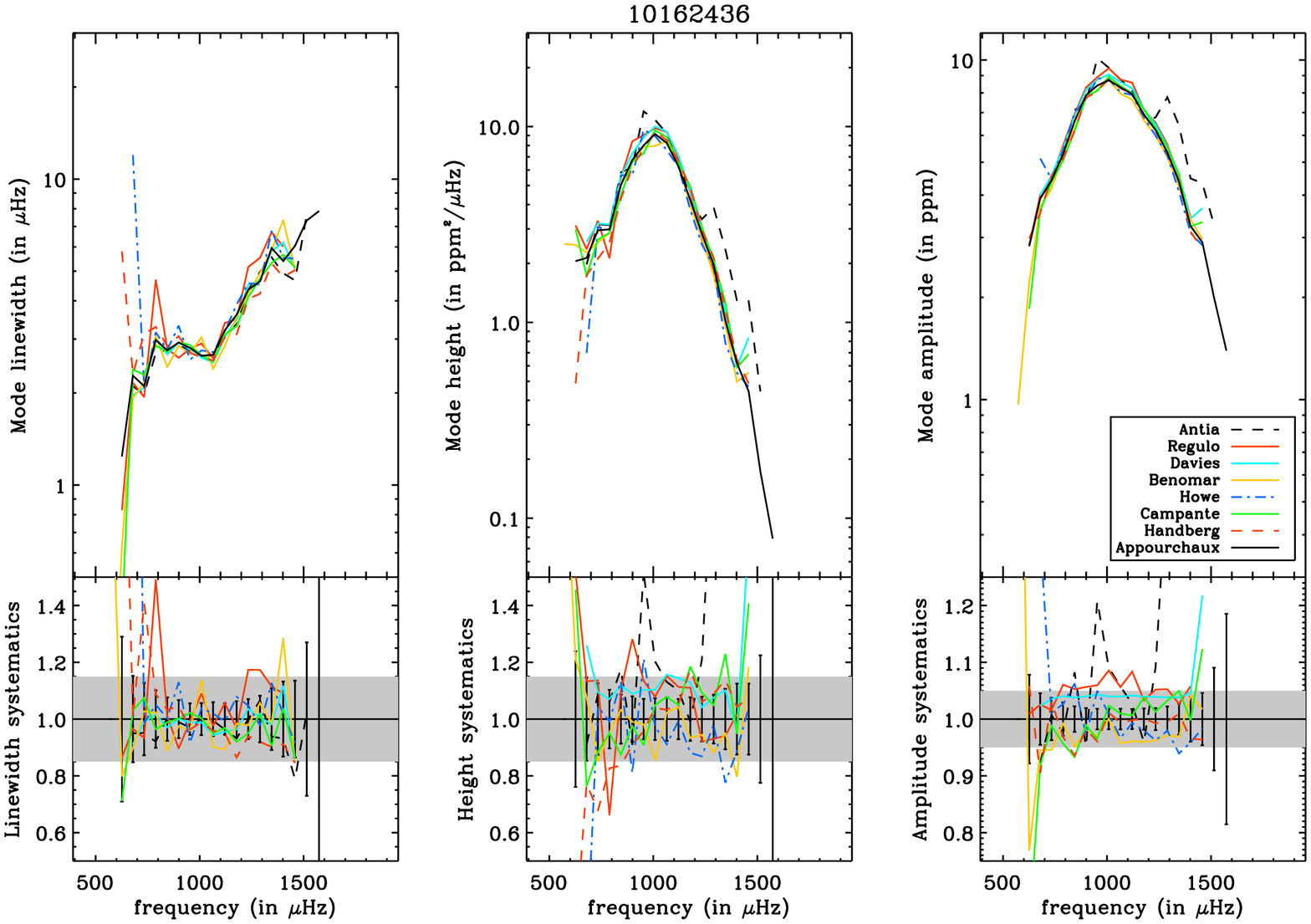}
\caption{Corrected mode linewidth, mode height and mode amplitude (Top) and relative values of these parameters with respect to the reference fit (Bottom) as a function of mode frequency for KIC 10162436. The grey band indicates the range of systematic error around the reference fit values of $\pm$ 15\% for mode linewidth and mode height, of $\pm$ 5\% for mode amplitude.  The error bars are those of the reference fit.}
\label{}
\end{figure*}

\begin{figure*}[htbp]
\centering
\includegraphics[width=10.75 cm,angle=90]{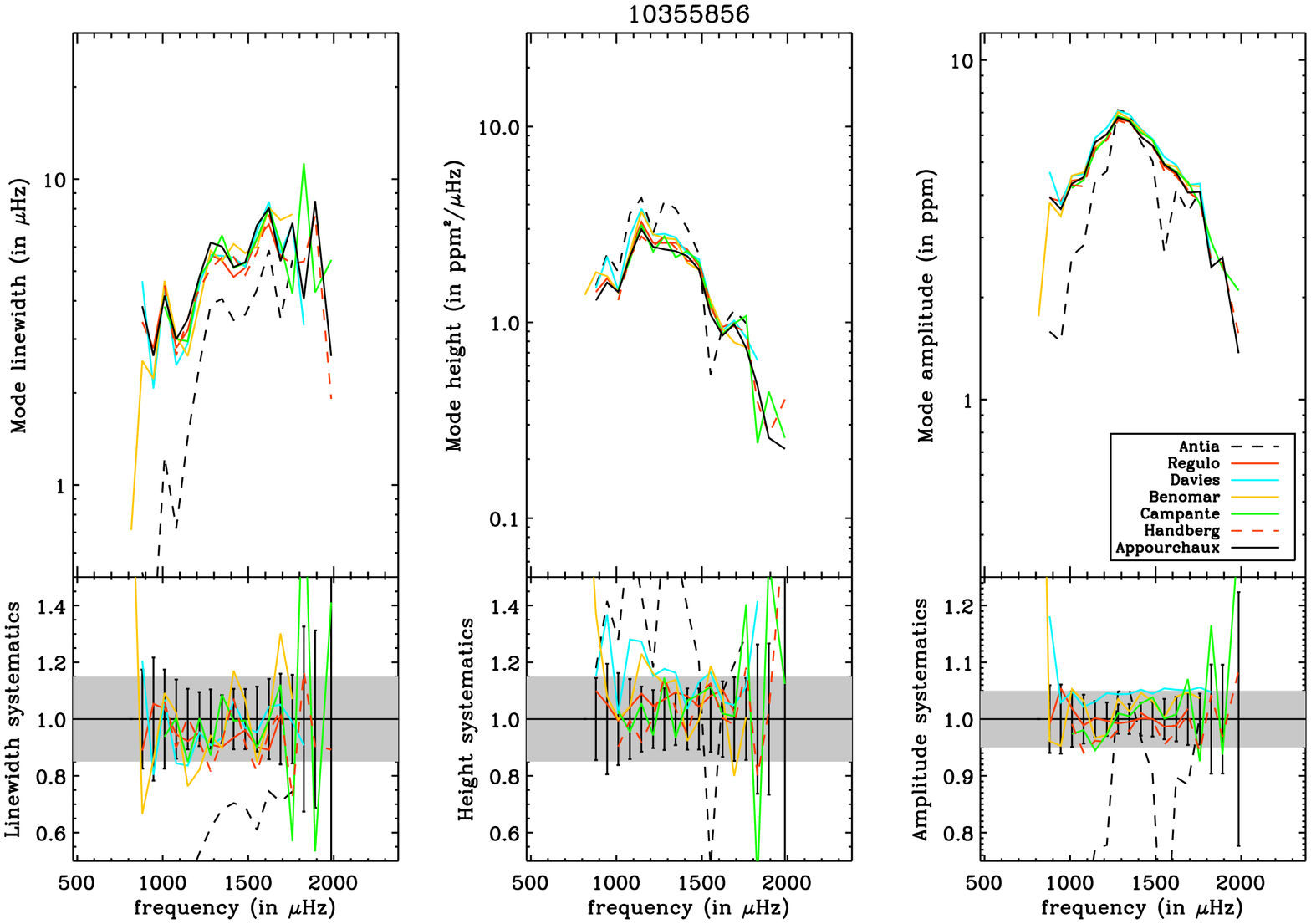}
\caption{Corrected mode linewidth, mode height and mode amplitude (Top) and relative values of these parameters with respect to the reference fit (Bottom) as a function of mode frequency for KIC 10355856. The grey band indicates the range of systematic error around the reference fit values of $\pm$ 15\% for mode linewidth and mode height, of $\pm$ 5\% for mode amplitude.  The error bars are those of the reference fit.}
\label{}
\end{figure*}

\begin{figure*}[htbp]
\centering
\includegraphics[width=10.75 cm,angle=90]{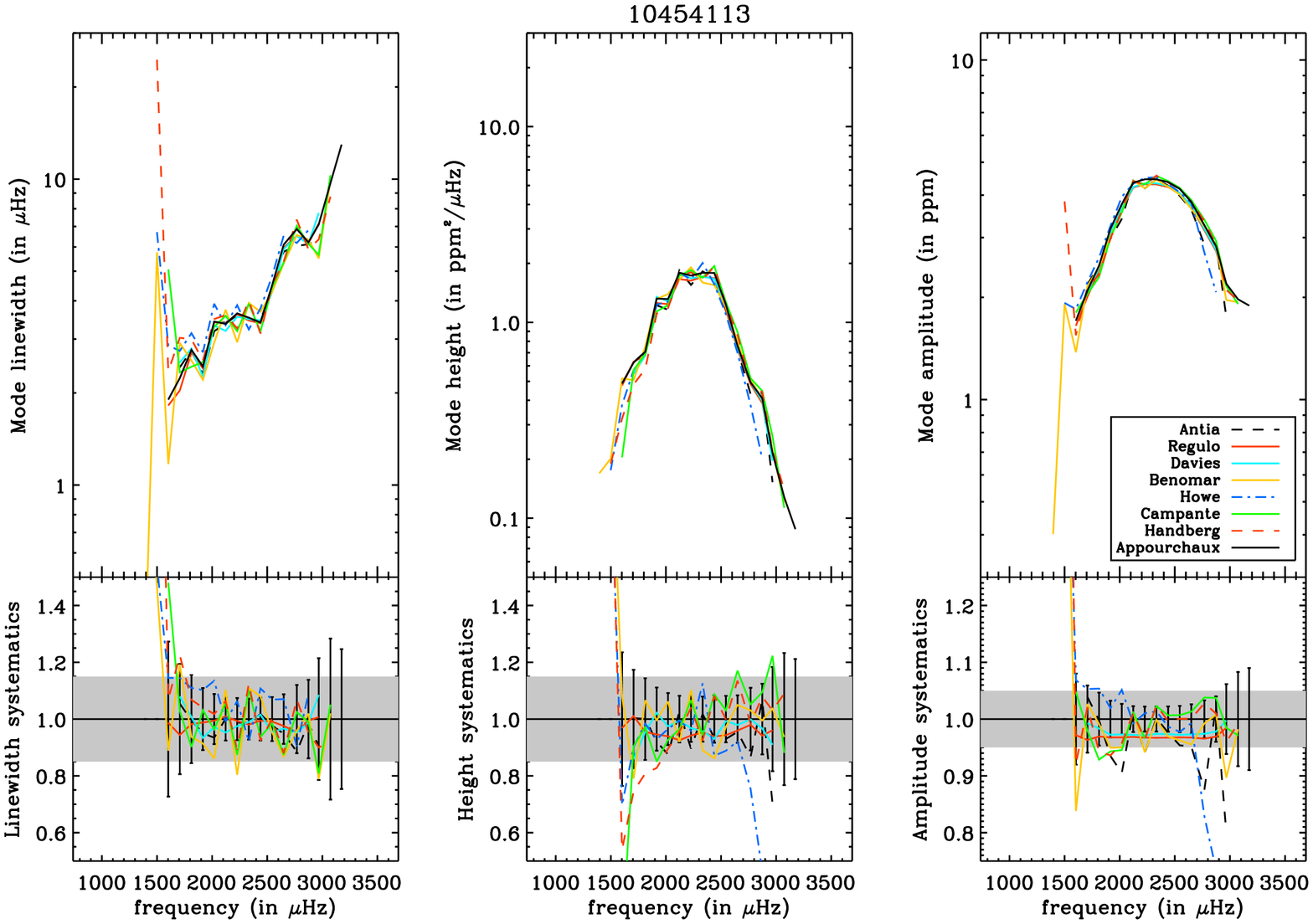}
\caption{Corrected mode linewidth, mode height and mode amplitude (Top) and relative values of these parameters with respect to the reference fit (Bottom) as a function of mode frequency for KIC 10454113. The grey band indicates the range of systematic error around the reference fit values of $\pm$ 15\% for mode linewidth and mode height, of $\pm$ 5\% for mode amplitude.  The error bars are those of the reference fit.}
\label{}
\end{figure*}

\begin{figure*}[htbp]
\centering
\includegraphics[width=10.75 cm,angle=90]{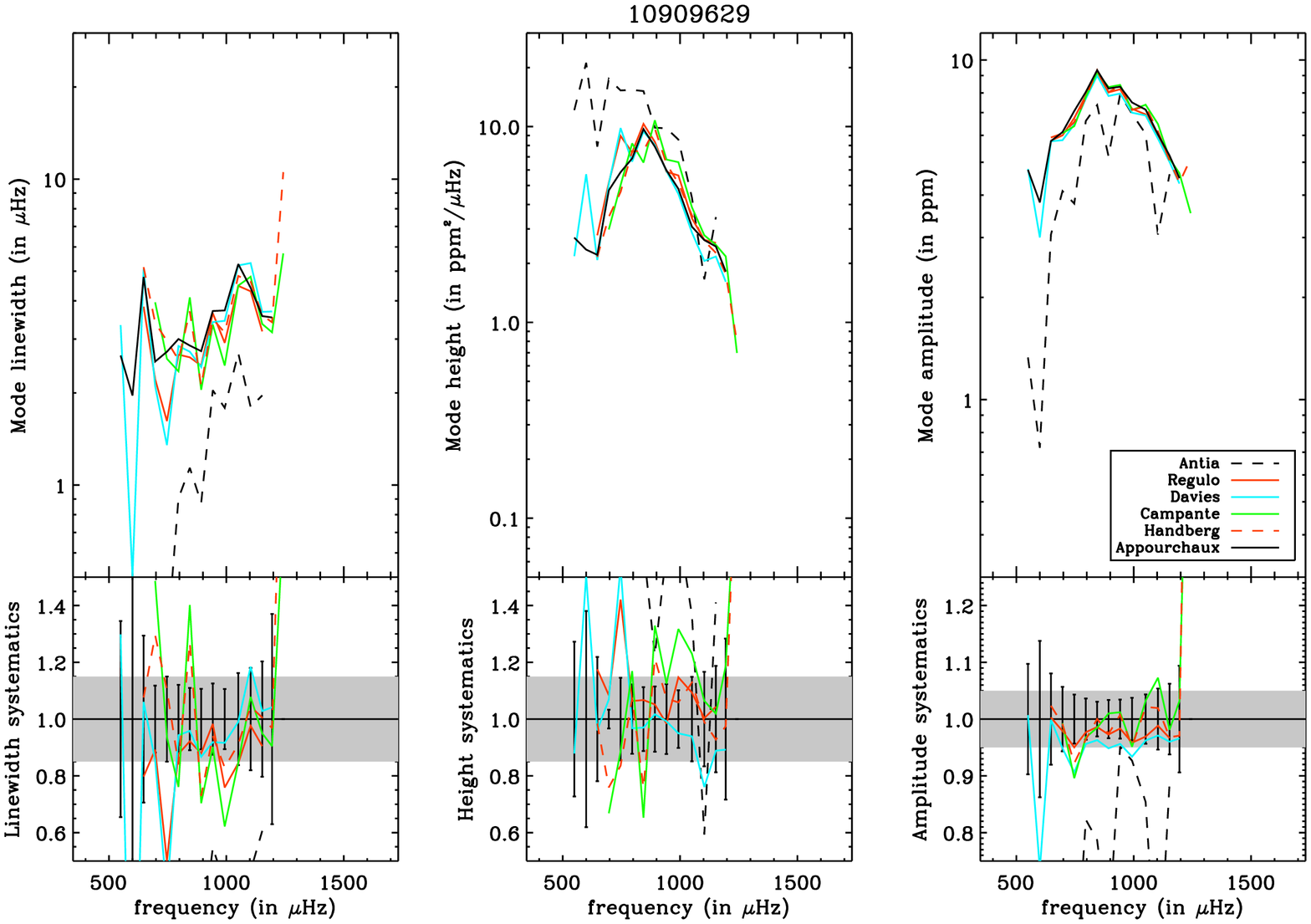}
\caption{Corrected mode linewidth, mode height and mode amplitude (Top) and relative values of these parameters with respect to the reference fit (Bottom) as a function of mode frequency for KIC 10909629. The grey band indicates the range of systematic error around the reference fit values of $\pm$ 15\% for mode linewidth and mode height, of $\pm$ 5\% for mode amplitude.  The error bars are those of the reference fit.}
\label{}
\end{figure*}

\begin{figure*}[htbp]
\centering
\includegraphics[width=10.75 cm,angle=90]{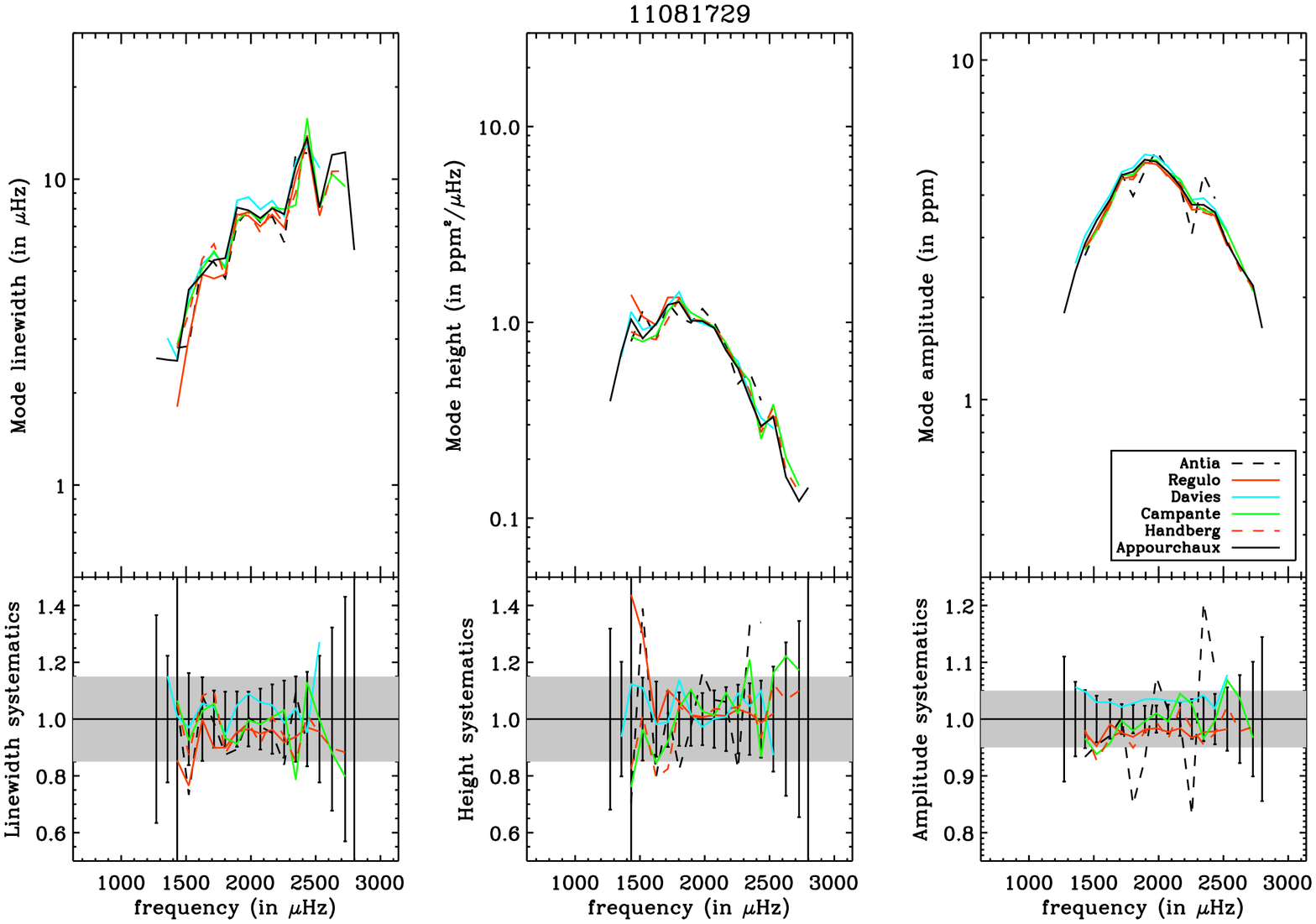}
\caption{Corrected mode linewidth, mode height and mode amplitude (Top) and relative values of these parameters with respect to the reference fit (Bottom) as a function of mode frequency for 11081729. The grey band indicates the range of systematic error around the reference fit values of $\pm$ 15\% for mode linewidth and mode height, of $\pm$ 5\% for mode amplitude.  The error bars are those of the reference fit.}
\label{}
\end{figure*}

\begin{figure*}[htbp]
\centering
\includegraphics[width=10.75 cm,angle=90]{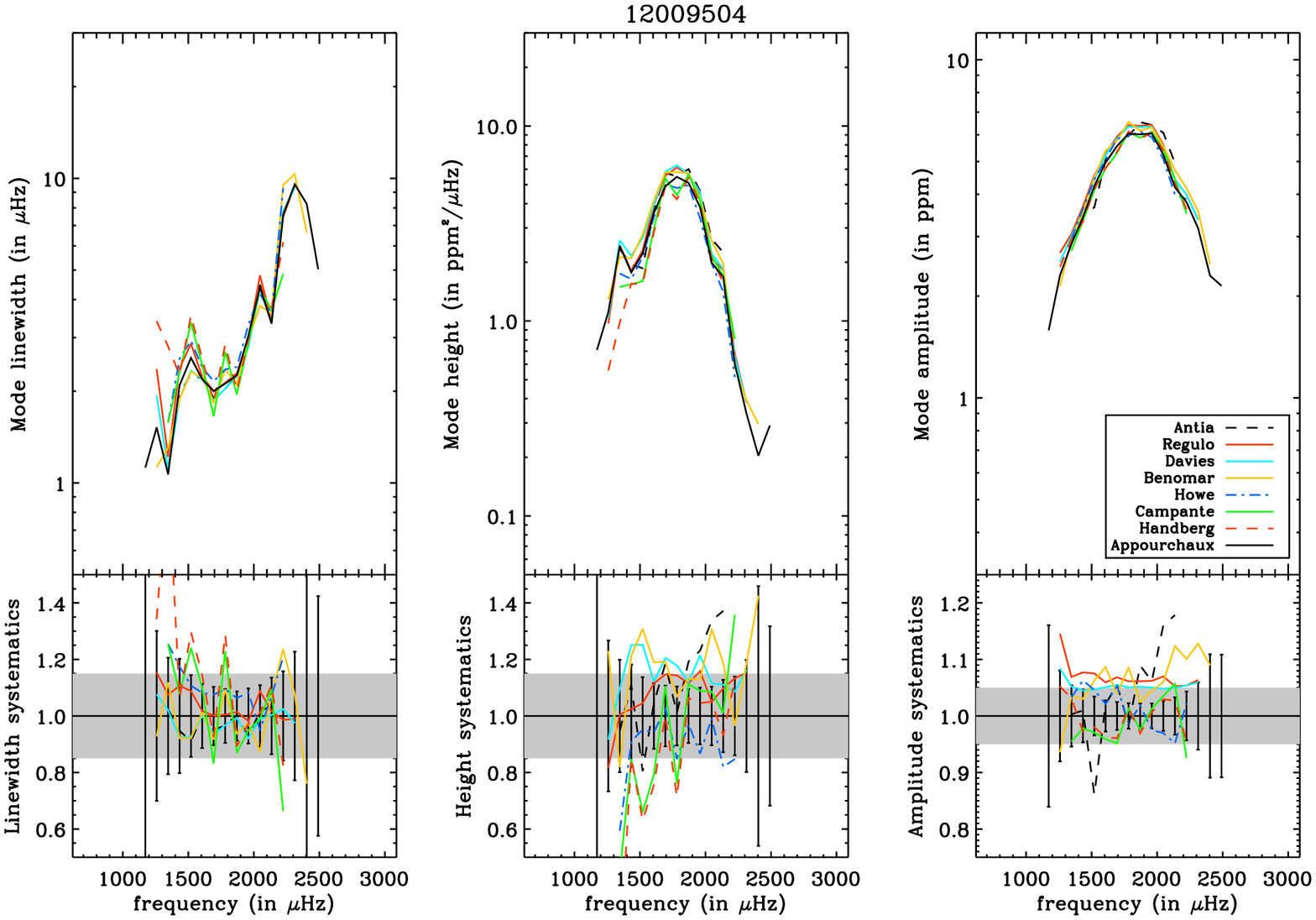}
\caption{Corrected mode linewidth, mode height and mode amplitude (Top) and relative values of these parameters with respect to the reference fit (Bottom) as a function of mode frequency for 12009504. The grey band indicates the range of systematic error around the reference fit values of $\pm$ 15\% for mode linewidth and mode height, of $\pm$ 5\% for mode amplitude.  The error bars are those of the reference fit.}
\label{}
\end{figure*}

\begin{figure*}[htbp]
\centering
\includegraphics[width=10.75 cm,angle=90]{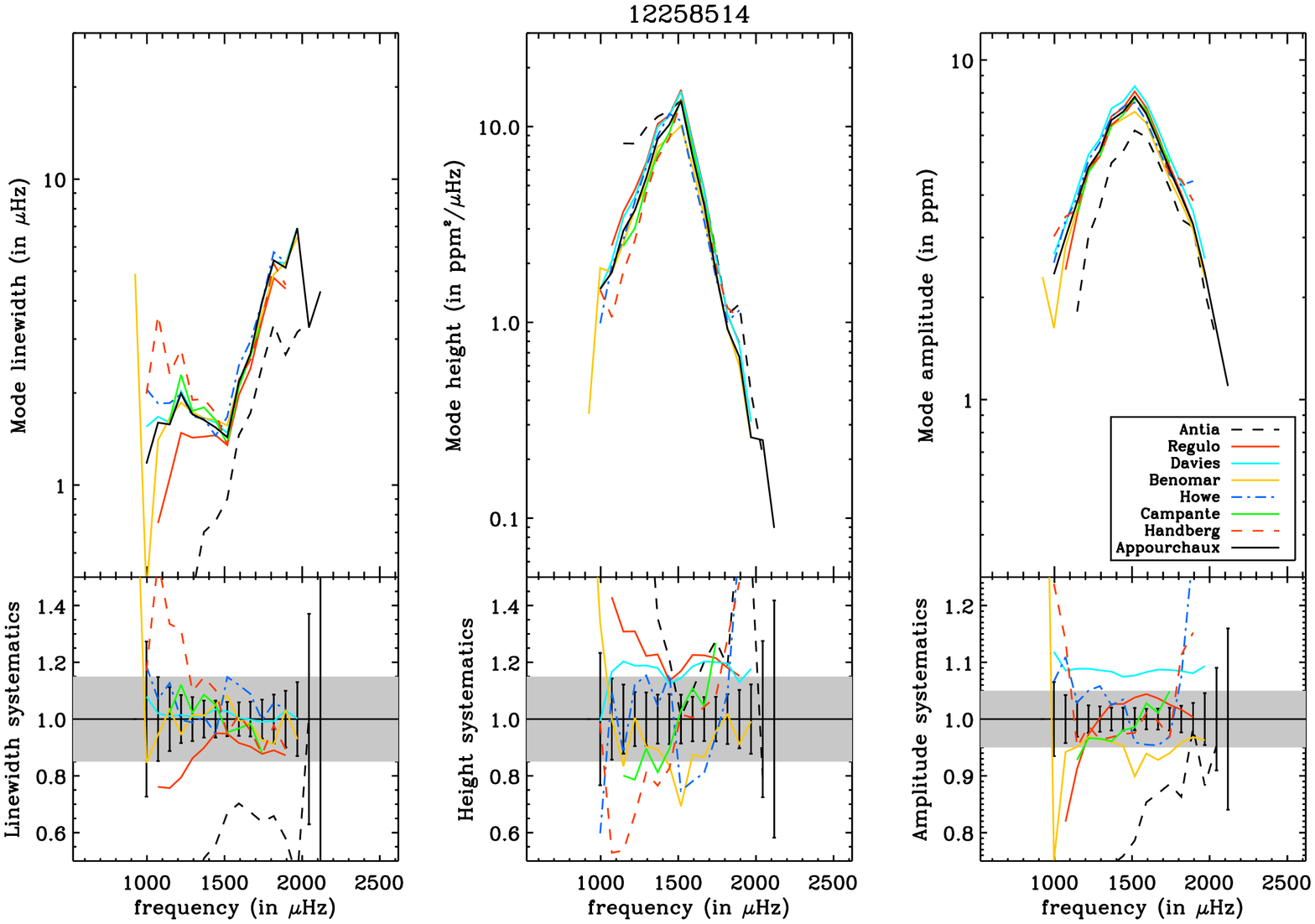}
\caption{Corrected mode linewidth, mode height and mode amplitude (Top) and relative values of these parameters with respect to the reference fit (Bottom) as a function of mode frequency for 12258514. The grey band indicates the range of systematic error around the reference fit values of $\pm$ 15\% for mode linewidth and mode height, of $\pm$ 5\% for mode amplitude.  The error bars are those of the reference fit.}
\label{}
\end{figure*}

\begin{figure*}[htbp]
\centering
\includegraphics[width=10.75 cm,angle=90]{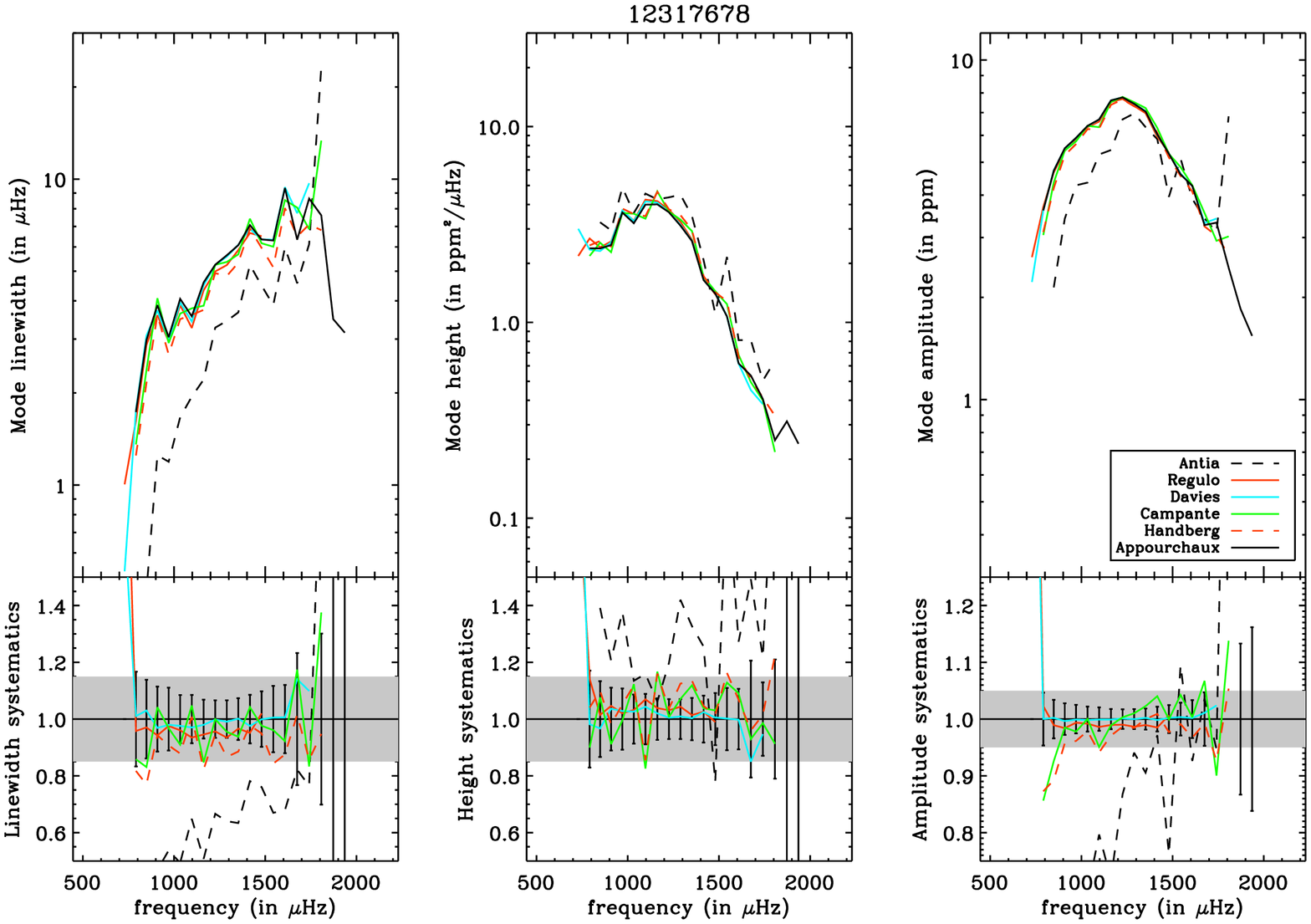}
\caption{Corrected mode linewidth, mode height and mode amplitude (Top) and relative values of these parameters with respect to the reference fit (Bottom) as a function of mode frequency for 12317678. The grey band indicates the range of systematic error around the reference fit values of $\pm$ 15\% for mode linewidth and mode height, of $\pm$ 5\% for mode amplitude.  The error bars are those of the reference fit.}
\label{}
\end{figure*}

\end{document}